\documentclass[10pt,a4paper]{article}
\usepackage{graphicx} 

\usepackage{hyperref}
\usepackage{stmaryrd}
\usepackage{framed}
\usepackage{mathtools}
\usepackage{physics}
\usepackage{pgf}

\usepackage{subfigure}
\usepackage{caption}

\usepackage{amsmath}

\usepackage{amssymb, bm}
\usepackage{amsfonts}
\usepackage{bbm}
\usepackage{amsthm}
\usepackage{amstext}
\usepackage{array}
\usepackage{amsfonts}

\usepackage[linesnumbered,ruled,vlined]{algorithm2e}
\SetKwComment{Comment}{/* }{ */}
 
\newcommand{\vertiii}[1]{{\left\vert\kern-0.25ex\left\vert\kern-0.25ex\left\vert #1 
   \right\vert\kern-0.25ex\right\vert\kern-0.25ex\right\vert}}

\newtheorem{definition}{Definition}[section]
\newtheorem{theorem}{Theorem}[section]

\newtheorem{theorem2}{Theorem}[section]
\newtheorem{proposition}[theorem]{Proposition}

\newtheorem{remark}[theorem2]{Remark}

\usepackage{geometry}
\date{\today}

\title{Parametric vector flows for registration  fields in bounded domains with applications to nonlinear interpolation of shock-dominated flows}
\author{Jon Labatut,  Jean-Baptiste Chapelier, Angelo Iollo,  Tommaso Taddei}

\begin{document}

\maketitle

\begin{abstract}
We present  a  registration procedure for parametric model order reduction (MOR)  in two- and three-dimensional bounded domains.
In the MOR framework, registration methods exploit  solution snapshots to identify a parametric coordinate transformation that improves the approximation  of the solution set through linear subspaces.
For each training parameter,  optimization-based (or variational) 
registration methods minimize a target function that measures the alignment of the    coherent structures of interest  (e.g., shocks, shear layers, cracks) for different parameter values, over a family of bijections of the computational domain $\Omega$.
We consider diffeomorphisms $\Phi$ that are vector flows of given velocity fields $v$ with vanishing normal
component on $\partial \Omega$;
we rely on a sensor to extract appropriate point clouds from the solution snapshots and 
we  develop an expectation-maximization procedure to simultaneously solve the point cloud matching problem and to determine the velocity $v$ (and thus the bijection $\Phi$); finally, 
we combine our registration method with the nonlinear interpolation  technique  of [Iollo, Taddei, J. Comput. Phys., 2022]
to perform accurate interpolations of fluid dynamic fields in the presence of shocks.
Numerical results for a two-dimensional inviscid transonic flow past a NACA airfoil and a three-dimensional viscous transonic flow past an ONERA M6 wing illustrate the many elements of the methodology and demonstrate the effectiveness of nonlinear interpolation for shock-dominated fields.
\end{abstract}

\section{Introduction}
\label{sec:intro}

The past decade has witnessed a surge in the development of nonlinear approximation strategies for  model order reduction (MOR) to tackle parametric problems that feature a slow-decaying Kolmogorov $n$-width
\cite{barnett2023neural,iollo2014advection,ohlberger2016reduced}.
In this regard, nonlinear approximation methods 
based on coordinate transformations have established themselves as 
a prominent technique to handle parametric systems with sharp 
parameter-dependent coherent structures with compact support such as shocks, shear layers and cracks.
If we denote by $\mathcal{M}$ 
the solution set 
associated with a given parametric partial differential equation (PDE), the problem of 
finding a  parametric transformation
$\Phi$ that  improves the approximation of 
$\mathcal{M}$ through linear subspaces
is referred to as \emph{registration problem}
\cite{taddei2020registration}:
these methods  are inspired by registration techniques from image processing \cite{brown1992survey,ma2015robust,myronenko2010point} and are also related to   morphing techniques in mesh adaptation
\cite{budd2009adaptivity,mcrae2018optimal} and shock-tracking methods in CFD \cite{shubin1982steady,zahr2018optimization}.
The aim of this work is to present a new registration method for model reduction  and to apply it to nonlinear interpolation of shock-dominated fields.

We denote by  $\Omega$ a Lipschitz bounded domain in $\mathbb{R}^d$ with $d=2$ or $d=3$; we also denote by $\mu$ a vector of model parameters in the compact 
parameter space $\mathcal{P}$;
we further define the parametric field $U:\Omega \times \mathcal{P} \to \mathbb{R}^D$ that solves the parametric PDE model of interest; we use notation
$U_\mu = U(\cdot; \mu)$ to refer to the solution to the PDE for the parameter $\mu \in \mathcal{P}$.
We further   introduce the  set of diffeomorphisms  
${\rm Diff}(\overline{\Omega})$ that contains all vector fields in $\Omega$ that are differentiable up to the boundary and are bijective (one-to-one and onto) in $\Omega$.
For MOR applications, given the parameter space $\mathcal{P}$, we are interested in parametric diffeomorphisms $\Phi: \overline{\Omega} \times \mathcal{P} \to \overline{\Omega}$ such that 
  (i) $\Phi_\mu\in {\rm Diff}(\overline{\Omega})$ for all $\mu\in \mathcal{P}$,  
  (ii) $\Phi$ is continuous with respect to the parameter, and 
  (iii) for all $\mu \in \mathcal{P}$,  
  $\Phi_\mu$ approximately minimizes
\begin{equation}
\label{eq:optimization_based_registration_ideal}
\min_{\Phi \in {\rm Diff}(\overline{\Omega})} 
\mathfrak{f}_\mu^{\rm tg}(\Phi),
\end{equation}
where $\mathfrak{f}_\mu^{\rm tg}$ should measure the alignment  between the  coherent structures associated with the parameter $\mu$ and the deformed coherent structures associated with a reference parameter $\mu_{\rm ref}\in \mathcal{P}$.
Registration techniques should hence take as input a set of snapshots $\{U_k:=U_{\mu_k} : \mu_1,\ldots,\mu_{n_{\rm train}} \in \mathcal{P} \}$ and return a parametric diffeomorphism that satisfies (i)-(ii)-(iii). 

The development of registration techniques in bounded domains requires to address three major challenges: 
first,
    the choice of the target function in \eqref{eq:optimization_based_registration_ideal};
second,
the choice of a finite-dimensional parameterization $\texttt{N}:\mathbb{R}^{N_h}$ --- the subscript $(\cdot)_h$ refers to a finite-dimensional spatial discretization that  shall be formally introduced  in section \ref{sec:methodology} --- of diffeomorphisms (or Lipschitz bijections) in $\Omega$; and
third,
the generalization of $\Phi$ to unseen parameters.
As regards the  second challenge, we can show that
${\rm Diff}(\overline{\Omega})$ is infinite-dimensional, and bijectivity is not stable under small perturbations in the $C^1(\Omega)$ norm
(cf. \cite[Lemma 3.1]{taddei2025compositional}):
formally, given two bijections of a curved domain $\Omega$ that do not coincide everywhere on the boundary, we can show that the convex combination is not a bijection for any intermediate value of the interpolation parameter $t\in (0,1)$.
The choice of the parameterization is hence crucial to devise tractable finite-dimensional counterparts of the optimization problem \eqref{eq:optimization_based_registration_ideal}.

In this work, we introduce vector flows \cite{lee2003smooth} for parametric registration in bounded domains. Given the domain $\Omega$, we introduce a finite element (FE) discretization $V_h \subset {\rm Lip}(\Omega; \mathbb{R}^d)$; next, we define $\texttt{N}(\mathbf{v}): \Omega \to \mathbb{R}^d$ as the flow of the vector field $v\in V_h$ associated with the FE vector of coefficients $\mathbf{v}\in \mathbb{R}^{N_h}$. 
Vector flows are diffeomorphisms in $\Omega$ under mild assumptions on the velocity $v$
(cf. Theorem \ref{th:theoryVFs})
and can be readily applied in the optimization context.

Following
\cite{cucchiara2024model,iollo2022mapping}, we rely on a scalar testing function 
to extract a point cloud
$Y_\mu : = \{  y_j(\mu) \}_{j=1}^{Q_\mu} \subset \overline{\Omega}$ from the state $U_\mu$ that is associated with the solution features that we want to track. Next, we define the target function $\mathfrak{f}^{\rm tg}(\Phi)$ as a weighted distance between a reference point cloud $\{  \xi_i \}_{i=1}^N$ and $Y_\mu$: since  in general we cannot anticipate a one-to-one correspondence between the two point clouds, we introduce an additional matrix of unknowns $\mathbf{P}\in \mathbb{R}^{N \times Q_\mu}$ whose entries $P_{i,j}$  describe the likelihood of the point $\xi_i$ to be mapped in the point $y_j(\mu)$. To simultaneously learn the velocity $\mathbf{v}\in \mathbb{R}^{N_h}$ and the matrix $\mathbf{P}$, we adapt the expectation maximization (EM) procedure of \cite{myronenko2010point} that iteratively chooses $\mathbf{P}^\star$ based on the current estimate $\mathbf{v}^\star$ and then updates 
$\mathbf{v}^\star$ by approximately solving a minimization problem of the form
\begin{equation}
\label{eq:tractable_optimization_based_registration}
\min_{   \mathbf{v}  \in \mathbb{R}^{N_h}}
 \mathfrak{f}_{\mu}^{\rm obj}(\mathbf{v} ):=
 \mathfrak{f}^{\rm tg}(\texttt{N}( \mathbf{v} ) ; 
 \mathbf{P}^\star, 
 \mu)
\; + \lambda  \mathfrak{f}_{\rm pen}(\mathbf{v} ),
\end{equation}
where  $ \mathfrak{f}_{\rm pen}$ is a regularization term that is designed to promote the smoothness of the map, and $\lambda>0$ is a weighting parameter.

To demonstrate the impact of our registration method
to MOR, we consider the application to shock-dominated flows in two and three dimensions.
In particular, we consider the transonic flow past a NACA0012 airfoil for varying inflow Mach numbers and the three-dimensional viscous flow past a wing (Onera M6) for varying angles of attack. Towards this end, we resort to the convex displacement interpolation (CDI) method introduced in \cite{iollo2022mapping} and further developed in \cite{cucchiara2024model}: as described below, CDI can be interpreted as a convex interpolation along the characteristics associated with the vector field $v$, which is determined using the registration procedure.

Our method is related to several previous works.
\begin{itemize}
\item 
The registration problem \eqref{eq:optimization_based_registration_ideal} is related  to the Monge's problem in optimal transportation theory \cite{santambrogio2015optimal}, for a suitable choice of the target function and the penalty function;
the analogy with optimal transportation theory enables to guarantee the existence and the uniqueness of the solution to \eqref{eq:optimization_based_registration_ideal} for specific target functions.
In this work, we pursue an optimization-based approach: we introduce a   problem-dependent target function and a parameterization $\texttt{N}$ that is amenable for computations; then, we we use standard optimization tools to estimate a (possibly local) minimum of 
 \eqref{eq:optimization_based_registration_ideal}.
 Optimization-based (or variational) methods for registration were previously considered in \cite{taddei2020registration,taddei2025compositional} for MOR,
and also \cite{beg2005computing,christensen1996deformable} for image registration.
\item
Thanks to the seminal contributions by Benamou and Brenier
\cite{benamou2000computational} (see also 
\cite[Chapter 6]{santambrogio2015optimal}),
vector flows have been extensively studied in optimal transportation theory;
furthermore, 
they have also been 
widely adopted  for image registration tasks
 \cite{beg2005computing,christensen1996deformable},
 mesh adaptation  \cite{alauzet2014changing,de2016optimization},  
parametric  geometry  reduction 
\cite{kabalan2025elasticity}, and also
 model reduction  \cite{iollo2014advection}.
 In this regard, the main feature of the present work is the focus on bounded domains: the accurate preservation of the boundary is  of critical importance for computational mechanics applications, due to the need to accurately resolve sharp boundary features (e.g., boundary layers) and also to ensure the satisfaction of boundary conditions. 
\item
Since  the work of \cite{taddei2020registration}, 
several authors
have proposed to directly parameterize the displacement field rather than the velocity field  for MOR applications \cite{mirhoseini2023model,razavi2025registration,taddei2025compositional}:
a thorough comparison of these two classes of methods is beyond the scope of the present work and is the subject of the preprint \cite{iollo2026mathematical}.
\item 
We can distinguish between two classes of nonlinear approximation methods based on coordinate transformation for MOR: Lagrangian and Eulerian methods.
\emph{Lagrangian methods} rely on the definition of a reference configuration where coherent structures of interest are approximately fixed \cite{barral2024registration,mirhoseini2023model,mojgani2021low,ohlberger2013nonlinear,razavi2025registration}; on the other hand,
\emph{Eulerian methods} 
\cite{iollo2014advection,iollo2022mapping}
 do not require the definition of a reference configuration and can in principle cope with multiple transport fields
\cite{bergmann2025model,krah2025robust,reiss2018shifted}.
As stated above, we here rely on the CDI method for nonlinear interpolation: this method  is inherently  Eulerian. Nevertheless, we envision that our registration method can be applied in conjunction with other MOR techniques based on coordinate transformation.
\item 
We (Iollo, Taddei) already considered the problem of point cloud matching in \cite{iollo2022mapping} and \cite{cucchiara2024model}: first, we applied a point set registration method for unbounded geometries to determine the matching and then we used the results to inform the bounded registration problem.
The approach of this work can be interpreted as a generalization of the one in 
\cite{cucchiara2024model,iollo2022mapping} and, in our experience, it  is significantly more robust, especially for geometric configurations with slender bodies.
\item 
Finally, we remark that several alternative target functions have been proposed in the literature.
In the MOR framework, 
similarly to \cite{beg2005computing},
the author of \cite{taddei2020registration}  minimizes the distance between a parameter-dependent mapped function $u_\mu$   and a template field $\bar{u}$, that is
$\mathfrak{f}^{\rm tg}(\Phi; \mu)=
\| u_\mu \circ \Phi - \bar{u}  \|_{L^2(\Omega)}^2$;
the authors of \cite{taddei2021space} generalize the latter to measure the $L^2$ distance from a low-dimensional template space. 
In the image processing literature, alternative objective functions can be found in \cite{han2023diffeomorphic} and the references therein.
\end{itemize}

The outline of this paper is as follows.
In section \ref{sec:formulation},
we introduce vector flows    and we review some critical properties that justify its application to 
registration in bounded domain;  we introduce the target function that is used in the computations;
and we introduce the CDI method of \cite{iollo2022mapping}.
In section \ref{sec:methodology}, we present the main elements of the  methodology --- namely, the FE discretization, the evaluation of the target function and its gradient, the choice of the penalty function, 
the choice of the sensor, and the deployment of the EM procedure.
In section \ref{sec:numerics}, we present the numerical results for the two model problems introduced above.
Section \ref{sec:conclusions} concludes the paper.

\section{Formulation}
\label{sec:formulation}

\subsection{Vector flows}
Given the Lipschitz connected domain $\Omega\subset \mathbb{R}^d$, we denote by $\mathbf{n}: \partial \Omega \to \mathbb{S}^2 = \{x \in \mathbb{R}^d: \|  x \|_2=1  \}$ the outward unit normal, and by $C^k(\overline{\Omega}; \mathbb{R}^d)$ the space of $k$-times differentiable functions up to the boundary, with values in $\mathbb{R}^d$. 
We   introduce the space of time-dependent vector fields whose normal component vanishes on $\partial \Omega$ at all times,
 $$
 \mathcal{V}_0:=  \{ v\in C^1(\overline{\Omega} \times [0,1]; \mathbb{R}^d) \,: \, v(\cdot,t) \cdot \mathbf{n} \big|_{\partial \Omega} = 0 \;\;  \forall \, t\in [0,1] \}
 $$ 
Below, $\texttt{id}:\Omega \to \Omega$ denotes the identity map,
$\texttt{id}(x)=x$, while $\mathbbm{1}\in \mathbb{R}^{d\times d}$ is the identity matrix.

Let 
 $v:\overline{\Omega} \times [0,1]\to \mathbb{R}^d$ 
satisfy  $v \in C(\overline{\Omega} \times [0,1]; \mathbb{R}^d)$, $v(\cdot,t) \in C^1(\overline{\Omega}; \mathbb{R}^d)$ for all $t\in [0,1]$, and
$v(\cdot,t) \cdot \mathbf{n} \big|_{\partial \Omega} = 0$.
Then, we say that $X:\overline{\Omega} \times [0,1]\to \overline{\Omega}$ is the flow of the vector field $v$ if
\begin{equation}
\label{eq:flow_diffeomorphisms}
\left\{
\begin{array}{ll}
\frac{\partial X}{\partial t}(\xi,t) = v( X(\xi,t) , t) & t\in (0,1], \\[3mm]
X(\xi,0) = \xi, & \\
\end{array}
\right. 
\quad
\forall \, \xi \in \overline{\Omega}.
 \end{equation}
 Furthermore, we define 
the vector flow (VF)
 $F[v]: \overline{\Omega} \to \overline{\Omega}$ 
as the end point of the flow $X$, that is  
 $F[v](\xi):=X(\xi,t=1)$. 
If we restrict ourself to a finite-dimensional velocity space, we obtain the family of diffeomorphisms
 \begin{equation}
 \label{eq:VB_maps}
 \texttt{N}(\xi; \mathbf{v}) := F[v(\mathbf{v})](\xi), \quad
 {\rm where} \;\;
 v(x, t; \mathbf{v})
 =\sum_{i=1}^{N_h} \, (\mathbf{v})_i \phi_i(x,t),
 \end{equation}
 where $\mathbf{v}\in \mathbb{R}^{N_h}$ is a vector of coefficients,
 and 
 $\phi_1 \ldots,\phi_{N_h} \in  \mathcal{V}_0(\Omega)$ 
 are linearly-independent. In view of the analysis, we introduce a relevant subset of diffeomorphisms.
 
\begin{definition}
\label{def:Diff0}
(\cite[Chapter 1, page 3]{banyaga2013structure})
Let $\Phi$ be a diffeomorphism in the Lipschitz domain $\Omega\subset \mathbb{R}^d$. We say that $\Phi$ is isotopic to the identity if there exists a smooth field (isotopy) $X \in C^1(\overline{\Omega} \times[0,1]; \mathbb{R}^d)$ such that
(i) \emph{$X(\cdot, 0)=\texttt{id}$}, (ii) $X(\cdot, 1)=\Phi$, 
(iii) the map $X(\cdot, t) \in {\rm Diff}(\overline{\Omega})$ for all $t\in [0,1]$.
We denote by ${\rm Diff}_0(\overline{\Omega})$ the subset of diffeomorphisms that are isotopic to the identity.  
\end{definition}

We can easily construct diffeomorphisms that do not belong to 
 ${\rm Diff}_0(\overline{\Omega})$ (see, e.g., \cite{iollo2026mathematical}); nevertheless,  we observe that ${\rm Diff}_0(\overline{\Omega})$ is the subset of diffeomorphisms of interest for model reduction applications.
Given the   parameter domain $\mathcal{P}$, 
in model order reduction, we seek parametric mappings
$\Phi: \overline{\Omega} \times \mathcal{P} \to \overline{\Omega}$ such that
$\Phi(\cdot,\mu) :\overline{\Omega} \to \overline{\Omega}$ is a diffeomorphism for all $\mu\in \mathcal{P}$ \cite{taddei2020registration}.
Exploiting Definition \ref{def:Diff0}, we can readily show
that
 if 
(i) 
 $\mathcal{P}$ is connected,
 (ii) $\Phi(\cdot,\mu')$ is equal to the identity for some $\mu'\in \mathcal{P}$, and
 (iii) $\Phi$ is continuous with respect to the parameter,  we must have that 
 $\Phi(\cdot,\mu)\in {\rm Diff}_0(\overline{\Omega})$  for all $\mu\in \mathcal{P}$.
Below, we summarize key properties of vector flows that show their relevance for registration tasks.
We refer to 
\cite{iollo2026mathematical}
(see also \cite{younes2010shapes}) for the proofs.

\begin{proposition}
\label{th:theoryVFs}
Let $\Omega\subset \mathbb{R}^d$ be  a Lipschitz domain;
we denote by 
$\mathcal{W}_h:= {\rm span} \{\phi_i \}_{i=1}^{N_h} \subset \mathcal{V}_0(\Omega)$ 
the approximation space 
associated to \eqref{eq:VB_maps}.
Then, the  following hold.
\begin{itemize}
\item
\textbf{Bijectivity.}
Let 
$v \in \mathcal{V}_0(\Omega)$. Define $\Phi: \overline{\Omega} \to \mathbb{R}^d$ such that
$\Phi=F[v]$.
 Then, $\Phi$ is a diffeomorphism (one-to-one and onto) in $\overline{\Omega}$.
 \item
 \textbf{Approximation.}
Let   $\Phi \in {\rm Diff}_0(\overline{\Omega})$. Then, there exists 
$v \in \mathcal{V}_0(\Omega)$ 
such that   $\Phi=F[v]$. Furthermore, if we denote by $L$ the Lipschitz constant of $v$,  we have that
\emph{
\begin{equation}
\label{eq:approximation_result_explained}
\inf_{\mathbf{v}\in \mathbb{R}^{N_h}} 
\| \Phi - \texttt{N}(\mathbf{v})  \|_{L^\infty(\Omega)}
\leq
\frac{e^L - 1}{L}
\inf_{\phi \in \mathcal{W}_{h}} 
\| v - \phi \|_{L^\infty(\Omega\times (0,1))}.
\end{equation}
}
 \item
 \textbf{Derivative.}
 The derivative of the VF \eqref{eq:VB_maps} with respect to the coefficients $\mathbf{v}$ is given by:
\emph{
\begin{equation}
\label{eq:derivative_VB}
\frac{\partial  \texttt{N} }   {\partial a_i}
(\xi; \mathbf{v})
=
\nabla X(\xi, 1; \mathbf{v}) 
\int_0^1 \, 
\left(
\nabla X(\xi, \tau; \mathbf{v}) 
\right)^{-1}
\phi_i \left( X(\xi, \tau; \mathbf{v}), \tau \right) \, d\tau,
\quad
{\rm for} \;\; i=1,\ldots,M;
\end{equation}
}
where the gradient $t\mapsto \nabla X(\xi, t; \mathbf{v}) $ satisfies the differential equation:
\begin{equation}
\label{eq:gradX_odes}
\left\{
\begin{array}{ll}
\frac{\partial \nabla X}{\partial t}(\xi,t; \mathbf{v}) = \nabla_x v( X(\xi,t; \mathbf{v}) , t; \mathbf{v})  \nabla X(\xi,t; \mathbf{v}) & t\in (0,1], \\[3mm]
\nabla  X (\xi,0; \mathbf{v}) = \mathbbm{1}. & \\
\end{array}
\right.     
\end{equation}
\item
\textbf{Jacobian determinant.}
Let $\Phi=F[v]$ for some $v\in \mathcal{V}_0(\Omega)$. Then, the Jacobian determinant $J :={\rm det} (\nabla \Phi)$ satisfies
\begin{equation}
\label{eq:jacobian_determinant}
J(\xi) = {\rm exp}
\left(
\int_0^1 \nabla_x \cdot v(X(\xi,s),s) \, ds
\right),
\qquad
\xi \in \Omega, t\in [0,1].
\end{equation}
\end{itemize}
\end{proposition}

Some comments are in order.
\begin{itemize}
\item
The VF \eqref{eq:VB_maps}
is guaranteed to be  a diffeomorphism of $\overline{\Omega}$  for all $\mathbf{v}\in \mathbb{R}^{N_h}$; therefore,  the penalty term in \eqref{eq:tractable_optimization_based_registration} should solely be designed to promote smoothness rather than  to ensure bijectivity.
\item
Eq. \eqref{eq:approximation_result_explained}
shows that 
any element of ${\rm Diff}_0(\overline{\Omega})$  
can be expressed in the form \eqref{eq:flow_diffeomorphisms} and that we can explicitly control the approximation error.
\item
It is easy to show that the practical computation of 
$\frac{\partial \texttt{N}}{\partial \mathbf{v}}  (\mathbf{v})$  at the discrete level 
for typical (e.g., Runge Kutta, linear multi-step) time-integration schemes   involves the application of the chain rule over all time steps: it is hence  computationally expensive.
On the other hand,  Eq.  \eqref{eq:derivative_VB} provides a constructive way to evaluate the
continuous derivative of the mapping 
--- and ultimately of the target function ---
with respect to the generalized coordinates $\mathbf{v}$ at roughly the same cost of evaluating $\texttt{N}$.
\item
Eq. \eqref{eq:jacobian_determinant} signifies that the Jacobian determinant of VF maps is strictly positive, which implies that VF maps preserve orientation.
\end{itemize}

\subsection{Target function}
In this work, we resort to a discrete matching function, which is  broadly used in image registration to align unmatched point clouds
\cite{myronenko2010point}.
Given the parameter $\mu\in \mathcal{P}$, 
we introduce the reference (parameter-independent) point cloud $\{  \xi_i \}_{i=1}^{N}$ and the target (parameter-dependent) cloud
$\{  y_j(\mu) \}_{j=1}^{Q_\mu}$.
Next, we define
 the matrix of weights 
$\mathbf{P}  = [P_{i,j}]_{i,j} \in \mathbb{R}^{N\times Q_\mu}$, which was already introduced in section
\ref{sec:intro}, 
such that
$P_{i,j}\in [0,1]$, and
$\sum_{j} P_{i,j}\leq 1$ and 
$\sum_{i} P_{i,j} \leq 1$ for all $i=1,\ldots,Q_\mu$ and $j=1,\ldots,N$.
As discussed in \cite{myronenko2010point} the weights
 $\{ P_{i,j} \}_{i,j}$ describe the likelihood  of the point $\xi_j$ to be mapped in the point $y_i(\mu)$.
Finally, 
we introduce the 
\emph{pointwise objective}:
 \begin{equation}
 \label{eq:pointwise_sensor}
 \mathfrak{f}^{\rm tg}(\Phi;  \mathbf{P}, \mu) = 
 \frac{1}{2} 
\sum_{i=1}^{N}  \sum_{j=1}^{Q_\mu}
P_{i,j} \, \big\|
 \Phi(\xi_i) - y_j(\mu)
\big\|_2^2.
 \end{equation}

We notice that the target function \eqref{eq:pointwise_sensor} is bounded from below; provided that
 $\lim_{\| \mathbf{v} \|_2 \to \infty}
\mathfrak{f}_{\rm pen}(\mathbf{v}) = +\infty$, we hence find that  the minimization statement 
\eqref{eq:tractable_optimization_based_registration}  has a (non-necessarily unique) minimum in $\mathbb{R}^{N_h}$ for any $\lambda>0$ (cf. \cite{iollo2026mathematical}).
The point clouds 
in \eqref{eq:pointwise_sensor}
are extracted from the solution field  using a properly-chosen point-set sensor (cf. section \ref{sec:sensors}); 
on the other hand, the weights 
$\mathbf{P}$ are chosen using 
a variant of the   expectation-maximization (EM)
procedure of 
\cite{myronenko2010point}
(cf. section \ref{sec:EM}).
We postpone the discussion on the 
practical construction of 
the weights $\mathbf{P}$ and 
the point-set sensor 
 for the model problems  of this work to  section \ref{sec:methodology}.

We can exploit 
\eqref{eq:derivative_VB} and 
the chain rule to compute the derivative of the objective
$E:\mathbf{v} \mapsto 
 \mathfrak{f}^{\rm tg}(\texttt{N}(\mathbf{v}) ;  \mathbf{P}, \mu)$.
  In more detail, we find
\begin{subequations}
 \begin{equation}
\label{eq:chain_rule}
\frac{\partial E}{\partial v_i} (\mathbf{v}; \mathbf{P}, \mu)
\; = \; 
 D \mathfrak{f}^{\rm tg} 
 \left[  \texttt{N}(\mathbf{v}),  \mathbf{P}, \mu\right] 
\left( 
  \frac{\partial \texttt{N}}{\partial v_i} 
(\mathbf{v})
\right),
\quad
i=1,\ldots,M,
\end{equation} 
 where the expression of 
 $ \frac{\partial \texttt{N}}{\partial v_i} 
(\mathbf{v})$ is provided in \eqref{eq:derivative_VB} and 
\begin{equation}
\label{eq:sensors_der}
 D \mathfrak{f}^{\rm tg}[\Phi; \mathbf{P}, \mu]  \,    (h)
\,  = \,
\sum_{i=1}^{N}  \sum_{j=1}^{Q_\mu}
P_{i,j}
 \left(
  \Phi(\xi_i) - y_j(\mu)
 \right)\cdot h(\xi_i)
  = 
\sum_{i=1}^{N} 
\left(
 \Phi(\xi_i)   \, \, -
 \sum_{j=1}^{Q_\mu}
P_{i,j}
 \,  y_j(\mu)
\right)
\cdot  h(\xi_i) ,
 \end{equation}
 \end{subequations}
As discussed in the next section, the gradient of $E$ is employed by the gradient-descent technique to find local minima of \eqref{eq:tractable_optimization_based_registration}. 
 
 \subsection{Convex displacement interpolation}
In this work, we rely on convex displacement interpolation (CDI) to perform nonlinear interpolation of fluid dynamics fields. 
CDI was introduced in \cite{iollo2022mapping} for the interpolation of two snapshots, and further extended in \cite{cucchiara2024model} to deal with multiple snapshots and multi-dimensional parameterizations.
Below, we briefly review the CDI as introduced in 
\cite{iollo2022mapping}.

We denote by 
$\mu_0,\mu_1 \in \mathcal{P}$ two parameters and by 
$U_0,U_1:\Omega \to \mathbb{R}^D$ the corresponding solution fields, $U_i := U(\cdot; \mu_i)$, $i=0,1$; CDI provides a nonlinear  interpolation  $\widehat{U} : \Omega \times [0,1] \to \mathbb{R}^D$ of the solution
$s \in [0,1] \mapsto  U_s = U(\cdot; (1-s)\mu_0 + s \mu_1)$.
Towards this end, given the velocity field $v\in \mathcal{V}_0(\Omega)$ and $s\in [0,1]$, we define the mappings $\Phi_{s\to 0}$ and 
$\Phi_{s\to 1}$
such that
\begin{subequations}
\begin{equation}
\label{eq:flowsCDI}
\Phi_{s\to 0}(\xi) = Y(\xi,t=0),
\quad
\Phi_{s\to 1}(\xi) = X(\xi,t=1),
\end{equation}
where $X$ and $Y$ are the backward and the forward flows associated with the velocity $v$,
\begin{equation}
\left\{
\begin{array}{ll}
\frac{\partial X}{\partial t}(\xi,t) = v( X(\xi,t) , t) & t\in (s,1], \\[3mm]
X(\xi,s) = \xi, & \\
\end{array}
\right. 
\qquad
\left\{
\begin{array}{ll}
\frac{\partial Y}{\partial t}(\xi,t) = -v( Y(\xi,t) , t) & t\in (0,s], \\[3mm]
Y(\xi,s) = \xi, & \\
\end{array}
\right. 
\end{equation}
for all $\xi \in \Omega$.
Then, we introduce the CDI of the parametric field $U$ as
\begin{equation}
\label{eq:CDI}
\widehat{U}(\xi, s)
= (1-s) U_0
\circ \Phi_{s\to 0}(\xi)
 + s U_1 \circ \Phi_{s\to 1}(\xi)  
\quad
\xi\in \Omega, \quad
s\in [0,1].
\end{equation}
\end{subequations}

As discussed in \cite{iollo2022mapping}, 
we can interpret the CDI
as a convex interpolation along the characteristics associated with the vector field $v$. Similarly, we can also view \eqref{eq:CDI} as a symmetrized McCann (or displacement) interpolation \cite{peyre2019computational}:
note that the latter is defined only for probability measures, while  \eqref{eq:CDI} is defined for arbitrary vector-valued fields.
 
Recalling   Proposition \ref{th:theoryVFs}, we find that  
$\Phi_{s\to 0} $ and $\Phi_{s\to 1} $ are diffeomorphisms for all $s\in [0,1]$: CDI is hence well-defined for any $v\in \mathcal{V}_0(\Omega)$, provided that $U_0, U_1$ are measurable.
Furthermore,  CDI ensures interpolation at end points,  and satisfies  maximum and minimum principles \cite{cucchiara2024model}.
Finally, we notice that, if $v$ is time-independent,
$\Phi_{s\to 0} = X'(\cdot, s)$
where $X'$ is the flow associated with $-v$
 and $\Phi_{s\to 1} = X(\cdot, 1-s)$.

\section{Methodology}
\label{sec:methodology}
We present the  elements of the methodology.
\begin{itemize}
\item 
In section  \ref{sec:FEM}, we introduce the finite element (FE) discretization of the velocity field. The FE method provides a finite-dimensional discretization $V_h$ of the space of stationary admissible stationary velocities
$\mathcal{U}_0 = \{v\in C^1(\overline{\Omega} ; \mathbb{R}^d) : v\cdot \mathbf{n} \big|_{\partial \Omega} = 0   \}$.
\item 
In section  \ref{sec:sensors},
we define the 
scalar testing functions that are used to extract the point cloud $Y_\mu = \{y_j(\mu) \}_{j=1}^{Q_\mu}$ for the state $U_\mu$ (see \eqref{eq:pointwise_sensor}).
\item 
In section \ref{sec:eval_target}, we discuss the evaluation of the target 
\eqref{eq:pointwise_sensor}
$E(\mathbf{v}) = \mathfrak{f}^{\rm tg}(\texttt{N}(\mathbf{v}); \mathbf{P}, \mu)$ and its gradient for any choice of the weights $\mathbf{P}$ and the point clouds.
\item 
In section \ref{sec:penalty}, we comment on the choice of the penalty term in \eqref{eq:tractable_optimization_based_registration} that ensures the smoothness of the map and the satisfaction of the constraints $v\cdot \mathbf{n}\big|_{\partial \Omega} = 0$.
\item 
In section \ref{sec:EM}, we present the complete EM procedure for the computation of the velocity $v$ and the weights $\mathbf{P}$. The EM algorithm requires the introduction of a probabilistic model for the target points.
\end{itemize}
In this work we consider stationary velocity fields to reduce  computational costs.
Our implementation is based on the FE library FEniCS  
\cite{alnaes2015fenics}, and on the linear algebra and optimization software PETSC \cite{petsc-web-page}.

\subsection{Finite element discretization}
\label{sec:FEM}
We resort to a continuous FE discretization of the velocity field. Given the domain $\Omega \subset \mathbb{R}^d$, we introduce the mesh
$\mathcal{T}=(\{x_j\}_{j=1}^{N_{\text{v}}}, \texttt{T})$ 
with nodes $\{x_j\}_{j=1}^{N_{\text{v}}}$ and connectivity matrix 
$\texttt{T} \in \mathbb{N}^{N_{\rm e}\times n_{\text{lp}}}$, where $N_{\rm e}$ is the number of elements in the mesh and $n_{\rm lp}$ is the number of local degrees of freedom. In this work, we consider linear triangular meshes for  $d=2$-dimensional domains  and 
linear tetrahedral  meshes for  $d=3$-dimensional  domains: we denote by $\{ \texttt{D}_k \}_{k=1}^{N_{\rm e}}$ the elements of the mesh and by 
 $\{ \texttt{e}_j \}_{j=1}^{N_{\rm f}}$ the facets of the mesh. We further denote by $h>0$ the characteristic size of the mesh,
 $$
h:= \max_{k=1,\ldots,N_{\rm e}} \sqrt[d]{|\texttt{D}_k|},
 $$  
 and we introduce the approximate domain $\Omega_h:= \bigcup_{k=1}^{N_{\rm e}} \texttt{D}_k$ with outward normal $\mathbf{n}_h: \partial \Omega_h \to \mathbb{S}_2$.

Exploiting this notation, we introduce the FE space of degree $p$ associated with the mesh $\mathcal{T}$:
\begin{equation}
\label{eq:FEspace}
 V_h = \left\{ v  \in C^0(\overline{\Omega}_h; \mathbb{R}^d) \, :  \, v|_{\texttt{D}_k} \in [\mathbb{P}_p(\texttt{D}_k)]^d, k=1, ..., N_{\rm e} \right\},
\end{equation}
where $\mathbb{P}_p(\texttt{D}_k)$ is the space of polynomials in $\texttt{D}_k$ of degree less or equal to $p$. 
We further introduce the Lagrangian basis   
$\{ \ell_i  \}_{i=1}^{N_{\rm v}}$ 
associated with the nodes 
$\{x_j\}_{j=1}^{N_{\text{v}}}$ of $\mathcal{T}$
and the canonical basis $\mathbf{e}_1,\ldots,\mathbf{e}_d$ of $\mathbb{R}^d$. Given the FE field $v\in V_h$, we denote by $\mathbf{v}\in \mathbb{R}^{N_h}$ with $N_h=N_{\rm v} \, d$ the corresponding FE vector such that
\begin{equation}
\label{eq:FE2vec}
v(x) : = 
\sum_{j=1}^{N_h} 
\left(\mathbf{v} \right)_{j} \; \phi_j(x),
\quad
{\rm where} \;\; 
 \phi_{i + (\ell-1) N_{\rm v}}(x):=
\ell_i(x)
\; \mathbf{e}_\ell,
\;\;\;
i=1,\ldots,N_{\rm v}, \;
\ell=1,\ldots,d.
\end{equation}

We notice that the FE discretization introduces several sources of approximation that should be taken into account.
First,
the space $V_h$  does not embed the constraint $v\cdot \mathbf{n} \big|_{\partial \Omega} = 0 $; furthermore, the elements of $V_h$ do not belong to $C^1$. We should hence rely on the penalty function of \eqref{eq:tractable_optimization_based_registration} to impose these constraints.
Second, 
if the domain $\Omega$ is curved,
the use of linear meshes leads to a
geometric error ${\rm dist}_{\rm H}(\Omega_h, \Omega)$ of order $\mathcal{O}(h)$, where 
 the Hausdorff distance 
 ${\rm dist}_{\rm H}(\Omega_h, \Omega)$ 
between the domains $\Omega_h$ and $\Omega$ is given by
$$
{\rm dist}_{\rm H}(\Omega_h, \Omega) := \max\{\inf_{x\in \Omega_h} \sup_{y\in \Omega} \| x-y\|_2, \; 
\inf_{y\in \Omega} \sup_{x\in \Omega_h} \| x-y\|_2
\}.
$$
Furthermore, in the absence of a proper post-treatment,  the outward normal $\mathbf{n}_h$ to $\Omega_h$ differs from the outward normal 
 $\mathbf{n}$
to the domain $\Omega$: this discrepancy induces an approximation in the imposition of the constraint $v\cdot \mathbf{n} \big|_{\partial \Omega} = 0 $. In the numerical experiments, we assess the impact of these approximations to the results.

\subsection{Choice of the sensor}
\label{sec:sensors}
Given the parametric fluid state $\mu\in \mathcal{P} \mapsto U_\mu$, we should extract the appropriate point cloud $Y_\mu:= \{ y_j(\mu) \}_{j=1}^{Q_\mu} \subset \overline{\Omega}$ that is associated with the flow features  of interest (e.g., shocks, shear layers, or vortices). Following \cite{iollo2022mapping}, we perform this task by first  introducing a finite-dimensional discretization $\Omega_{\rm discr}$ of the domain $\Omega$ and a scalar testing function $\mathfrak{T}:\Omega\times \mathcal{P} \to \mathbb{R}$, and then by defining $Y_\mu$ as
\begin{equation}
\label{eq:point_cloud_definition}
Y_\mu:=
\left\{
x\in \Omega_{\rm discr} \, : \,
\mathfrak{T}(x,\mu) > 0
\right\}.
\end{equation}

In the numerical experiments,  we define $\Omega_{\rm discr}$ as the union of all the cell  centroids of the mesh which is used for CFD calculations; on the other hand, 
in order to define the function $\mathfrak{T}$, we first  introduce the sensor
\begin{subequations}
\label{eq:tecplot_scalar_testing}
\begin{equation}
\label{eq:tecplot_sensor}
\varsigma_{\rm s}(x; U_\mu) = \frac{1}{a_\mu(x) \|\nabla p_\mu(x)\|_2} \, \left( v_\mu(x)^\top \,  \nabla p_\mu(x) \right),
\end{equation}
where $a_\mu$ is the sound speed, $p_\mu$ is the pressure, and  $v_\mu$ is the fluid velocity. The sensor $\varsigma_{\rm s}$ is a broadly-used shock detector for compressible flows\footnote{See the User's manual of Tecplot 360, Release 1 (2013), page 580 of 598.}. The corresponding scalar testing function is given by 
\begin{equation}
\mathfrak{T}_{\rm s}(x,\mu) = 
\varsigma_{\rm s}(x; U_\mu)  - \texttt{tol},
\end{equation}
where $\texttt{tol} \in (0,1)$ is a tunable parameter.
\end{subequations}

In the numerical experiments, we also tested 
 the scalar testing function 
\begin{subequations}
\label{eq:iso_scalar_testing}
\begin{equation}
\mathfrak{T}_{\rm iso}^\pm(x,\mu) = 
\pm \varsigma_{\rm iso}(x; U_\mu)  - \texttt{tol}, 
\end{equation}
that aims to align selected isolines of the normalized density field,
\begin{equation}
\varsigma_{\rm iso}(x; U_\mu) 
:=  \begin{cases}
        \dfrac{\rho_\mu(x) - \overline{\rho}_\mu}{\rho_{\rm max,\mu} - \overline{\rho}_\mu}, & \text{if } \rho_\mu(x) - \overline{\rho}_\mu > 0, \\[3mm]
        \dfrac{\rho_\mu(x) - \overline{\rho}_\mu}{ \overline{\rho}_\mu-\rho_{\rm min,\mu}}, & \text{if } \rho_\mu(x) - \overline{\rho}_\mu < 0.
    \end{cases}    
\end{equation}
where $ \overline{\rho}_\mu$, $\rho_{\rm min,\mu}$, $\rho_{\rm max,\mu}$ are 
the mean, the minimum and the maximum of the density ${\rho}_\mu$ over $\Omega$.
This alternative  scalar testing function is designed to identify strong compression and expansion regions, for aerodynamic external flows.
\end{subequations}

\subsection{Evaluation of the target function}
\label{sec:eval_target}
As discussed in section \ref{sec:formulation},  the evaluation of the flow $X$ requires the integration of the system of $d$ nonlinear ODEs \eqref{eq:flow_diffeomorphisms}; furthermore, the evaluation of the derivatives of the flow with respect to the velocity degrees of freedom requires the solution to the linear system of $d\times d$ ODEs 
\eqref{eq:gradX_odes}. 
Algorithm \ref{alg:eval_target} summarizes the procedure
for the computation  of the target \eqref{eq:pointwise_sensor} and its derivative \eqref{eq:sensors_der}: for simplicity, in 
Algorithm \ref{alg:eval_target}, 
we consider the explicit Euler method for time integration; furthermore, we introduce the set of ``active''  degrees  of freedom,
\begin{equation}
\label{eq:active_dofs}
\texttt{I}_{\rm act} \left(   \{ x_j  \}_{j=1}^N   \right)
:= \left\{
i\in \{1,\ldots,N_h\} \, : \,
x_j \in 
{\rm spt}(\phi_i), \quad
 {\rm for \; some} \, j\in \{1,\ldots,N\}
\right\},
\end{equation}
where ${\rm spt}(\phi_i)$ denotes the support of the function $\phi_i$.

{\renewcommand{\baselinestretch}{1.5}\selectfont
\begin{algorithm}[H]
\caption{Evaluation of the target \eqref{eq:pointwise_sensor} and its derivative \eqref{eq:sensors_der}.}
\label{alg:eval_target}
\KwData{$\mathbf{P}\in \mathbb{R}_+^{N\times Q_\mu}$ sets of weights;
$\{ \xi_i \}_{i=1}^N$,   $\{ y_j \}_{j=1}^{Q_\mu}$ reference and target point clouds;
$\mathbf{v}\in \mathbb{R}^{N_h}$ velocity field;
 $(\Delta t , N_t)$ time grid.
}
\KwResult{$E(\mathbf{v})$, $\nabla E(\mathbf{v})$.}

\textit{Initialization}: set 
$X_i^{(0)} \gets \xi_i$, 
$\nabla X_i^{(0)} \gets \mathbbm{1}$, 
$\dfrac{\partial \texttt{N}_i}{\partial v_\ell} = 0$, 
for $i=1,\ldots,N$, $\ell=1,\ldots,N_h$.

\For{$k = 0,\ldots,N_t - 1$}{
    Compute $v_i^{(k)} := v(X_i^{(k)})$, 
    $\nabla v_i^{(k)} := \nabla v(X_i^{(k)})$ for $i=1,\ldots,N$ \;

    Update positions and gradients:  
    $X_i^{(k+1)} \gets X_i^{(k)} + \Delta t  \,  v_i^{(k)}$, 
    $\nabla X_i^{(k+1)} \gets \nabla X_i^{(k)} + \Delta t \, \nabla v_i^{(k)} \,  \nabla X_i^{(k)}$, for $i=1,\ldots,N$\;

    Identify active DOFs: 
    $\texttt{I}_{\rm act}^{(k)} := \texttt{I}_{\rm act}( \{X_i^{(k)}\}_i \cup \{X_i^{(k+1)}\}_i )$\;

    Update  
    $\dfrac{\partial \texttt{N}_i}{\partial v_\ell} = 
    \dfrac{\partial \texttt{N}_i}{\partial v_\ell}  +\dfrac{\Delta t}{2} \left(
(\nabla X_i^{(k+1)} )^{-1} \phi_\ell ( X_i^{(k+1)}   ) +
(\nabla X_i^{(k)} )^{-1} \phi_\ell ( X_i^{(k)}   )
   \right)    
    $ for all $\ell \in \texttt{I}_{\rm act}^{(k)}$.
    
}

 Return 
$\displaystyle{E  = \frac{1}{2} \sum_{i=1}^{N}  \sum_{j=1}^{Q_\mu} P_{i,j} \| X_i^{(N_t)}  - y_j(\mu)  \|_2^2 }$

$\displaystyle{\dfrac{\partial \texttt{N}_i}{\partial v_\ell} \gets
\nabla X_i^{(N_t)} \, \dfrac{\partial \texttt{N}_i}{\partial v_\ell}}$, 
for $i=1,\ldots,N$, $\ell\in \bigcup_k \texttt{I}_{\rm act}^{(k)}$.

 Return
$\displaystyle{
 \dfrac{\partial E }{\partial v_\ell} 
\,  = \,
\sum_{i=1}^{N} 
\left(
X_i^{(N_t)}   \, \, -
 \sum_{j=1}^{Q_\mu}
P_{i,j}
 \,  y_j(\mu)
\right)
\cdot  \dfrac{\partial \texttt{N}_i}{\partial v_\ell} 
}$, 
for  $\ell\in \bigcup_k \texttt{I}_{\rm act}^{(k)}$.
\end{algorithm}
}

Some comments are in order.
\begin{itemize}
\item 
At each time step, we evaluate the FE field $v\in V_h$ and its gradient at the points $\{  X_i^{(k)} \}_i$ (cf. Line 3). If $X_i^{(k)}  \in \Omega_h$, we first find the element of the mesh to which $X_i^{(k)}$ belongs using an octree search and then we use the functional expression of the velocity field to determine its pointwise value (cf. FEniCS documentation). If 
$X_i^{(k)}  \notin \Omega_h$, we consider a zero-th order extrapolation:  we replace $X_i^{(k)}$ with the nearest vertex of the mesh and then we use the same expression as before to evaluate $v$. We rely on $k$-dimensional (KD, \cite{luke2012fast}) trees to rapidly identify the nearest mesh node.
\item 
The approximation of $\{  \frac{\partial \texttt{N}_i}{\partial v_\ell} \}_{i,\ell}$ in Algorithm \ref{alg:eval_target} corresponds to a second-order approximation of the integral 
\eqref{eq:derivative_VB} based on the 
trapezoidal rule.
We notice that the procedure does not require the storage of the whole trajectories $\{ X_i^{(k)} \}_{i,k} $ and $\{ \nabla X_i^{(k)} \}_{i,k} $.
Furthermore, it 
requires exactly $N$ octree searches at each time iteration to determine the elements of the mesh
$\texttt{I}_{\rm el}^{(k)}\subset \{1,\ldots,N_{\rm e}\}$
where the points 
$\{ X_i^{(k)} \}_{i=1}^N$ lie: for this reason, to improve efficiency, we only update the estimate of 
$\{  \frac{\partial \texttt{N}_i}{\partial v_\ell} \}_{i,\ell}$ for the active degrees of freedom \eqref{eq:active_dofs} (cf. Line 9).
Note that the active degrees of freedom can be readily identified from   the sampled elements  $\texttt{I}_{\rm el}^{(k)}$, the definition of the Lagrangian basis and the connectivity matrix $\texttt{T}$.
\item 
In the numerical experiments, if not specified otherwise, we resort to the explicit midpoint method, which is a second-order two-stage Runge Kutta scheme: given the ODE system $\dot{y} = f(t,y)$ and the time grid $\{t^{(k)} = k \Delta t \}_{k=0}^{N_t}$, the method reads as
\begin{equation}
\label{eq:RK2}
y^{(k+1)}=y^{(k)} \, + \, \Delta t f
\left(
t^{(k)} + \frac{\Delta t}{2}  \, , \,
y^{(k)} \, + \, \frac{\Delta t}{2}
f \left(
t^{(k)}    \, , \,
y^{(k)} 
\right)
\right),
\quad
k=0,1,\ldots.
\end{equation}
We can readily adapt Algorithm 
\ref{alg:eval_target}
to cope with other Runge Kutta explicit schemes by properly modifying Line 4.
On the other hand, the development of effective adaptive methods is more involved due to the need to synchronize the $N$ independent systems of ODEs and  is beyond the scope of the present work.
\item 
As anticipated in section \ref{sec:formulation}, Algorithm \ref{alg:eval_target} returns an estimate of $\nabla E$ based on the continuous derivative of $\mathbf{v}\mapsto \texttt{N}(\mathbf{v})$  in \eqref{eq:derivative_VB}, which differs from the actual gradient of the discrete objective: in the numerical experiments, we assess the accuracy of this approximation for typical choices of the time step $\Delta t$.
\end{itemize}

\subsection{Choice of the penalty term}
\label{sec:penalty}
The penalty term $\mathfrak{f}_{\rm pen}$ should be designed to promote smoothness and to ensure the constraint $v\cdot \mathbf{n} \big|_{\partial \Omega} = 0$. Recalling the Rellich-Kondrachov embedding theorem (cf. \cite[Chapter 6.3]{adams2003sobolev}), we find that the Sobolev space $H^s(\Omega)$ is contained in $C^p(\Omega)$ if $s > \frac{d}{2}+p$.
It is hence natural to bound the  $H^s(\Omega)$ norm of the velocity for  $s=2$ or $s=3$ --- for $d=2$ or $d=3$, the former ensures continuity of the velocity, while the latter also ensures continuity of its derivatives. Since in this work we resort to a  standard FE discretization which is not $H^2$-conforming, the definition of the discrete counterpart of such norms requires care and is  outlined below. 

Towards this end, we introduce the bilinear forms $b_0, b, m: V_h \times V_h \to \mathbb{R}$ such that
\begin{equation}
\label{eq:bilinear_forms}
\left\{
\begin{array}{l}
\displaystyle{b_0(u,v) =
\int_{\Omega} 
\frac{1}{\kappa_0^2} \nabla u \, : \, \nabla v \, +  u\cdot v \, dx,
\quad
m(u,v) =
\int_{\Omega} 
u\cdot v \, dx;
}\\[3mm]
\displaystyle{
b(u,v) =
b_0(u,v)  +
\sum_{ \texttt{e}_h \in \mathcal{E}_h^{\rm bnd}}
\int_{\texttt{e}_h}
\left(
\dfrac{C_{\rm pen}}{|\texttt{e}_h|}
(u\cdot \mathbf{n}_h)
(v\cdot \mathbf{n}_h) 
\,
-
\,
(\mathbf{n}_h^\top \nabla u \, \mathbf{n}_h)
(v\cdot \mathbf{n}_h) 
-
\,
(
\mathbf{n}_h^\top \, \nabla v \, \mathbf{n}_h)
(u\cdot \mathbf{n}_h) 
\right) \, ds
}\\
\end{array}
\right.
\end{equation}
where $C_{\rm pen},\kappa_0>0$ are user-defined constants and $\mathcal{E}_h^{\rm bnd}$ denotes the set of boundary facets of the mesh $\mathcal{T}$. We further define the corresponding matrices $\mathbf{B}_0,\mathbf{B},\mathbf{M}$ such that 
$(\mathbf{B}_0)_{i,j} = b_0(\varphi_j, \varphi_i)$, 
$(\mathbf{B})_{i,j} = b(\varphi_j, \varphi_i)$ and
$(\mathbf{M})_{i,j} = m(\varphi_j, \varphi_i)$ for $i,j=1,\ldots,N_h$. It is straightforward to verify that $\mathbf{B}_0,\mathbf{M}$ are symmetric positive definite;   provided that $C_{\rm pen}$ is sufficiently large,
it can be shown that
$\mathbf{B}$ is also symmetric positive definite \cite{nitsche1971variationsprinzip}. Exploiting this notation, we   consider the quadratic penalty:
\begin{equation}
\label{eq:quadratic_penalty}
\mathfrak{f}_{\rm pen}(\mathbf{v})
=
\frac{1}{2} \mathbf{v}^\top \mathbf{K} \mathbf{v} \quad
{\rm where} \quad
\mathbf{K} = \left(  
\mathbf{B}  \mathbf{M}^{-1}
\right)^{s-1} \mathbf{B},
\end{equation}
and $s\in \mathbb{N}$, $s\geq 2$. 

To justify \eqref{eq:quadratic_penalty}, we provide a 
functional interpretation of the solution to the linear system 
$\mathbf{K} \mathbf{u}= \mathbf{F}$ where $\mathbf{F} \in \mathbb{R}^{N_h}$ is the FE vector associated with the functional $F\in V_h'$.
The solution $\mathbf{u}$ can be obtained using the iterative procedure:
first, 
for $k=1,\ldots,s-1$, we solve
\begin{equation}
\label{eq:smoothing_problem_a}
\mathbf{B} \mathbf{w}_k = \left\{
\begin{array}{ll}
 \mathbf{F}    &  {\rm if} \; k=1, \\[2mm]
 \mathbf{M} \mathbf{w}_{k-1}    &  {\rm if} \; k>1; \\
\end{array}
\right.
\end{equation}
then, we solve the system:
\begin{equation}
\label{eq:smoothing_problem_b}
\mathbf{B} \mathbf{u} =  \mathbf{M} \mathbf{w}_{s-1}.
\end{equation}
We observe that,
if $F$ is the discretization of a functional in $H^{-1}(\Omega)$, the linear system 
$\mathbf{B} \mathbf{w}_1 =  \mathbf{F}$ in \eqref{eq:smoothing_problem_a} is a FE element discretization of  the differential problem:
$$
\left\{
\begin{array}{ll}
\displaystyle{ -\frac{1}{\kappa_0^2} \Delta u + u = F }      &  {\rm in} \; \Omega; \\[3mm]
\displaystyle{
u\cdot \mathbf{n} = 0,
\;\;
\partial_n u \cdot \mathbf{t}_\ell = 0 ,  \;\; \ell=1,\ldots,d-1
}      &  {\rm on} \; \partial \Omega; \\
\end{array}
\right.
$$
where
the first equation should be intended in the sense of distributions and 
$\mathbf{t}_1,\ldots,\mathbf{t}_{d-1}$ are orthonormal tangent vectors on 
$ \partial \Omega$. Note that the Dirichlet boundary condition 
$u\cdot \mathbf{n} = 0$ is enforced weakly using the Nitsche's method
\cite{nitsche1971variationsprinzip}.
Similarly, 
 the linear system 
$\mathbf{B} \mathbf{u} =  \mathbf{M} \mathbf{w}_{k-1}$ in \eqref{eq:smoothing_problem_b} is a FE element discretization of  the differential problem:
$$
\left\{
\begin{array}{ll}
\displaystyle{ -\frac{1}{\kappa_0^2} \Delta u + u = w_{k-1} }      &  {\rm in} \; \Omega; \\[3mm]
\displaystyle{
u\cdot \mathbf{n} = 0,
\;\;
\partial_n u \cdot \mathbf{t}_\ell = 0 ,  \;\; \ell=1,\ldots,d-1
}      &  {\rm on} \; \partial \Omega. \\
\end{array}
\right.
$$
In conclusion, the action of $\mathbf{K}^{-1}$ can be interpreted as the solution to a Laplace problem with mixed boundary conditions followed by $s-1$ Laplacian smoothing steps. Since the Laplace operator is a linear bounded operator from $H^\alpha$ to $H^{\alpha-2}$, we can interpret the matrix $\mathbf{K}$ as the discretization of an operator 
from $H^\alpha$ to $H^{\alpha-2s}$. 
We can thus loosely  interpret the penalty function in \eqref{eq:quadratic_penalty} as a proxy of the $H^s$ norm.

We might also attempt to interpret the system 
$\mathbf{K} \mathbf{u}= \mathbf{F}$  as a mixed FE discretization (see  \cite{lamichhane2011mixed} and the references therein) of the polyharmonic operator (see, e.g., 
\cite{antonietti2022conforming}) for a suitable choice of the boundary conditions. A rigorous formalization of this intuition is, however, beyond the scope of the present work.

\subsection{Expectation-maximization procedure}
\label{sec:EM}
We remove dependence on the parameter $\mu\in \mathcal{P}$ to shorten notation; furthermore, we denote by $Q$ (as opposed to $Q_\mu$) the total number of target points $\{y_j \}_{j=1}^Q \subset \overline{\Omega}$. Since the weights $\mathbf{P}$  in \eqref{eq:pointwise_sensor} are typically unknown, we should optimize 
\eqref{eq:tractable_optimization_based_registration} with objective \eqref{eq:pointwise_sensor} and penalty \eqref{eq:quadratic_penalty} for both the velocity $v$ and the weights $\mathbf{P}$.
Towards this end, following
 \cite{myronenko2010point}, we resort to an expectation-maximization (EM) procedure.

The EM algorithm is a well-established technique in machine learning, which was proposed in \cite{dempster1977maximum} to find maximum likelihood solutions for 
statistical  models with latent (unobservable) variables.
We refer to \cite[Chapter 9]{bishop2006pattern}  and 
 \cite[Chapter 8]{hastie2009elements}    for a thorough review of the methodology and for a rigorous theoretical justification: in particular, we refer to  
\cite[Chapter 9.4]{bishop2006pattern} and 
\cite[Chapter 8.5.3]{hastie2009elements}
for the interpretation of the EM method as an alternating   minimization algorithm.

We introduce the basic EM procedure in section \ref{sec:EM_basic};
next,
in section \ref{sec:probabilistic_model}, 
we introduce the probabilistic model that is considered for the registration task;
in section \ref{sec:overviewEM}, we review the complete procedure and we comment on the termination condition;
finally, 
in section \ref{sec:optimization_problem}, we discuss the computational procedure for the optimization problem  \eqref{eq:tractable_optimization_based_registration} that is solved at each EM iteration.

\subsubsection{Abstract procedure}
\label{sec:EM_basic}
The goal of the EM algorithm is to find maximum likelihood (ML) solutions for models with latent variables. We denote by $\mathbf{Y} = [y_1,\ldots,y_Q] \in \mathbb{R}^{d\times Q}$ the observed data, and by $\mathbf{Z} = (z_1,\ldots,z_Q) \in \{1,\ldots,N\}^Q$ the latent variables for some $N\in \mathbb{N}$, and by $\theta\in \Theta$ the set of model parameters. 
We introduce the probability model   $\theta \mapsto {\rm Pr}(\mathbf{Y}, \mathbf{Z}\big| \theta)$
and the corresponding marginal distribution 
$\theta \mapsto {\rm Pr}(\mathbf{Y}, \mathbf{Z}\big| \theta)$ obtained by integrating the joint density over $\mathbf{Z}$,
\begin{subequations}
\label{eq:EM_pointdeparture}
\begin{equation}
 {\rm Pr} \left( \mathbf{Y} \big| \theta \right)
 =
 \sum_{\mathbf{Z} \in \{ 1,\ldots,N\}^Q } {\rm Pr} \left( \mathbf{Y}, \mathbf{Z} \big| \theta \right)
\end{equation}
 We can also consider continuous latent variables by replacing the sum   with an integral over a continuous measure. 
 Then, we  seek solutions to the maximization problem:
\begin{equation}
\theta^\star \in {\rm arg} \max_{\theta\in \Theta} 
\;\; 
 {\rm Pr} \left( \mathbf{Y} \big| \theta \right)
 = 
 {\rm arg} \max_{\theta\in \Theta} 
\;\; 
\log\left(
 {\rm Pr} \left( \mathbf{Y} \big| \theta \right)
\right).
\end{equation}
\end{subequations}

Direct computation of the solution to \eqref{eq:EM_pointdeparture} is computationally challenging due to the fact that the summation 
over $\mathbf{Z}$
in \eqref{eq:EM_pointdeparture} appears inside the logarithm;
however, for many probability models of interest, the estimation of the solution to 
\eqref{eq:EM_pointdeparture} becomes significantly easier    if $\mathbf{Z}$ is known.
For this reason, the EM algorithm features  an iterative procedure in which we first estimate the probability of $\mathbf{Z}$ given $\mathbf{Y}$ and the current estimate
$\theta^{\rm old}$
of the parameters $\theta$  (\emph{E-step}), and then we determine the new estimate of $\theta$ by maximizing the expected log-likelihood  (\emph{M-step}), 
$$
\mathcal{Q}(\theta \big|\theta^{\rm old})
:=
\sum_{\mathbf{Z}} {\rm Pr} \left( \mathbf{Z} \big| \mathbf{Y}, \theta^{\rm old} \right)
\log \left(
{\rm Pr} \left( \mathbf{Y}, \mathbf{Z} \big| , \theta \right)
\right).
$$
Provided that $y_1,\ldots,y_Q$ and 
$z_1,\ldots,z_Q$ are independent realizations of the random variables $Y$ and $Z$, respectively, 
we have that
${\rm Pr} \left( \mathbf{Z} \big| \mathbf{Y}, \theta^{\rm old} \right)
= \prod_{j=1}^Q 
{\rm Pr} \left( Z = z_j \big| Y = y_j, \theta^{\rm old} \right)
$ and
${\rm Pr} \left( \mathbf{Y}, \mathbf{Z} \big|  \theta^{\rm old} \right)
= \prod_{j=1}^Q 
{\rm Pr} \big(Y = y_j,$ 
$Z = z_j \big|  \theta^{\rm old} \big)
$. Therefore,
the latter reduces to 
\begin{equation}
\label{eq:abstract_Mobjective}
\mathcal{Q}(\theta \big|\theta^{\rm old})
:=
\sum_{i=1}^N \sum_{j=1}^Q 
P_{i,j}
\log \left(
{\rm Pr} 
\left(Y=y_j, Z=i \big| , \theta \right)
\right),
\quad
{\rm where} \;\;
P_{i,j} = {\rm Pr}\left( Z=i \big| Y=y_j, \theta^{\rm old} \right).   
\end{equation}

We notice that the EM method aims to improve $\mathcal{Q}(\theta \big| \theta^{\rm old})$ rather than directly improving ${\rm Pr}(\mathbf{Y} \big| \theta)$ at each iteration. 
As rigorously shown in 
\cite{dempster1977maximum}, exploiting   Jensen's inequality, we can verify that
\begin{equation}
\label{eq:theory_EM_jensen}
\log \left(
{\rm Pr}(\mathbf{Y} \big| \theta)
\right)
-
\log \left(
{\rm Pr}(\mathbf{Y} \big| \theta^{\rm old})
\right)
\geq
\mathcal{Q}(\theta \big| \theta^{\rm old})
-
\mathcal{Q}(\theta^{\rm old} \big| \theta^{\rm old}).    
\end{equation}
Therefore, improvements in $\theta \mapsto \mathcal{Q}(\theta \big| \theta^{\rm old})$ lead to improvements in
$\theta \mapsto \log \left(
{\rm Pr}(\mathbf{Y} \big| \theta)
\right)$.
Estimate \eqref{eq:theory_EM_jensen} shows that the EM procedure fits in the framework of bound optimizer algorithms, for which extensive analysis can be carried out
\cite{salakhutdinov2003adaptive}.

Algorithm \ref{alg:EM_abstract} summarizes the EM procedure.
Meng and Rubin
\cite{meng1993maximum} proposed to partition  $\theta$ into groups and then break down the M-step into multiple steps each of which involves the optimization with respect to one subgroup while the remainders are held fixed; furthermore, we can adapt the EM procedure to tackle maximum a posteriori (MAP) estimates by augmenting $\mathcal{Q}(\theta \big|\theta^{\rm old})$ with a prior  for $\theta$ in the M-step. In the remainder, we exploit both extensions of Algorithm \ref{alg:EM_abstract}.

{\renewcommand{\baselinestretch}{1.5}\selectfont
\begin{algorithm}[H]
\caption{General expectation-maximization (EM) procedure.}
\label{alg:EM_abstract}
\KwData{$\mathbf{Y}\in \mathbb{R}^{d\times Q}$ observations;
$\theta_0$  initial estimate of the parameters,
probabilistic model of \eqref{eq:abstract_Mobjective}
}
\KwResult{$\theta^\star$}

\textit{Initialization}: set 
$\theta^\star \gets \theta_0$.

\For{$k = 0,\ldots,{\rm until \;\; convergence}$}{
\textbf{E-step:}
Determine 
$P_{i,j} = {\rm Pr}\left( Z=i \big| Y=y_j, \theta^{\star} \right)$ for $i=1,\ldots,N$, $j=1,\ldots,Q$
 \;

\textbf{M-step:}
Solve  
$\displaystyle{
\theta^\star  \gets {\rm arg } \max_{\theta\in \Theta}
\mathcal{Q}(\theta \big|\theta^{\rm old})
}$
}
\end{algorithm}
}

\subsubsection{Probabilistic model}
\label{sec:probabilistic_model}

The application of the general EM algorithm critically depends on the choice of the probabilistic models for the observations.
Following \cite{myronenko2010point}, we consider the mixture model for the location of the target points
\begin{equation}
\label{eq:CPD_probabilistic_model}
Y = \sum_{i=1}^N \delta(Z, i ) Y^{(i)} +
 \delta(Z, N+1) Y^{(N+1)},
\end{equation}
where 
$\delta(i,j) = 1$ if $i=j$ and $0$ otherwise,
$Y^{(i)} \sim \mathcal{N}(m_i, \sigma^2 \mathbbm{1} )$ for $i=1,\ldots,N$,
$Y^{(N+1)} \sim {\rm Uniform}(\mathcal{D}),$ 
and $Z\sim {\rm Multinomial}(\{1,\ldots,N+1\})$ are independent random variables. Given $w\in [0,1)$, we set
\begin{equation}
\label{eq:model_multinomial}
{\rm Pr}(Z=i )
= 
\left\{
\begin{array}{ll}
\displaystyle{\frac{1-w}{N}} & i=1,\ldots,N, \\
\displaystyle{w} & i=N+1; \\
\end{array}
\right.
\quad
|\mathcal{D}|=Q.    
\end{equation}
In order to ensure that the GMM centroids move coherently
(cf. \cite{myronenko2010point})
and lie in $\Omega$, 
we propose the 
deformation model:
\begin{equation}
\label{eq:deformation_model}
m_i = F[v] (\xi_i) , \qquad
i=1,\ldots,N;
\end{equation}
where $\{\xi_i \}_{i=1}^N$ is the reference point cloud, and  $F[v]$ is the
flow of the velocity $v\in V_h$.
Note that the deformation model \eqref{eq:deformation_model} differs from the one of \cite{myronenko2010point} and is designed to enforce the bijectivity of the mapping in $\Omega$.
In the remainder, we assume that $w,\mathcal{D}$ are user-defined constants, while the velocity $v$ and the variance $\sigma^2$ are the design parameters, i.e., $\theta=(v,\sigma^2)$.

As discussed in the next section, the probabilistic model introduced above enables the application of (a variant of) Algorithm \ref{alg:EM_abstract}.
Below, we  comment   on our modeling choices.

\begin{itemize}
\item 
Eq. \eqref{eq:CPD_probabilistic_model} reads as a  GMM with a  perturbation given by the uniform random variable $Y^{(N+1)}$: the latter is intended to account for noise and outliers in the dataset. We   do not explicitly construct the domain $\mathcal{D}$: we simply assume that all the observed datapoints belong to 
$\mathcal{D}$.
\item 
The choice of the deformation model \eqref{eq:deformation_model} ensures that the GMM centroids lie in $\overline{\Omega}$; however, the mixture model \eqref{eq:CPD_probabilistic_model}
does not satisfy the constraint
${\rm Pr}(Y\in \overline{\Omega} \big| \theta) = 1$. 
Even if we did not 
experience any convergence issue for the experiments of section \ref{sec:numerics}, we envision that this shortcoming of our model might lead to convergence issues for the EM iterative procedure. The development of more sophisticated probabilistic models for point-set registration in bounded domains is the subject of ongoing research.
\item 
In several applications, we might be interested in exploiting several sensors at the same time.
In these scenarios, we should deal with multiple point clouds to register at the same time. Below, we discuss how to update the probabilistic model \emph{a posteriori} to cope with multiple point clouds.
\end{itemize}

\subsubsection{EM  procedure for point-set registration}
\label{sec:overviewEM}
Exploiting the previous hypotheses,
we can derive an explicit expression for 
$P_{i,j} =       
{\rm Pr}(Z=i \big| Y=y_j, \theta )$.   
If we recall
 the 
definition of the conditional probability and the definition of the marginal probability,
we find
$$
P_{i,j} =       
{\rm Pr}(Z=i \big| Y=y_j, \theta )   
=
\dfrac{{\rm Pr}(Z=i , Y=y_j \big|\theta )   }{{\rm Pr}(Y=y_j \big|\theta )  }
=
\dfrac{ {\rm Pr}(Z=i , Y=y_j \big|\theta )   }{\sum_k {\rm Pr}(Z=k, Y=y_j \big|\theta ) }.
$$
Next, 
we notice that, by construction,
if $Z=i$, we have $Y=Y^{(i)}$ (cf. \eqref{eq:CPD_probabilistic_model}); therefore,
${\rm Pr}(Z=i, Y=y_j \big|\theta ) = 
{\rm Pr}(Z=i, Y^{(i)}=y_j \big|\theta )$ and  the latter is equal to 
${\rm Pr}(Z=i) {\rm Pr}(Y^{(i)}=y_j \big|\theta )$ since $Z$ and $Y^{(1)},\ldots,Y^{(N)}$ are independent. Therefore,
$$
P_{i,j}
=
\dfrac{ {\rm Pr}(Z=i) {\rm Pr}(Y^{(i)}=y_j \big|\theta )   }{\sum_k {\rm Pr}(Z=k)
{\rm Pr}(Y^{(k)}=y_j \big|\theta ) }.
$$
Finally,
if we recall the probability laws for $\mathbf{Z}$ and $Y^{(i)}$ (cf. \eqref{eq:CPD_probabilistic_model} and \eqref{eq:model_multinomial}), we obtain
\begin{equation}
\label{eq:Pij_cpd}
P_{i,j} =    
\dfrac{  {\rm exp}\left( -\frac{\| y_j - m_i \|_2^2}{2\sigma^2}  \right)       }{    \sum_{k=1}^N 
 {\rm exp}\left( -\frac{\| y_j- m_k \|_2^2}{2\sigma^2}  \right)
+ 
c },
\quad
{\rm where} \;\;
c=\frac{w}{1-w}  \frac{N}{Q}  (2\pi \sigma^2)^{d/2},
\end{equation}
for $j=1,\ldots,N$ and
$i=1,\ldots,N, j=1,\ldots,Q$.

If multiple point clouds are used, we should ensure that ${\rm Pr}(Z=i, Y=y_j \big|\theta ) = 0$ if the indices $i,j$ correspond to different clouds  to ensure that each coherent structure is independently tracked: we should hence modify \eqref{eq:Pij_cpd} as follows
\begin{equation}
\label{eq:Pij_cpd_modified}
P_{i,j} =      
\dfrac{  {\rm exp}\left( -\frac{\| y_j - m_i \|_2^2}{2\sigma^2}  \right)  \overline{P}_{i,j}     }{    \sum_{k=1}^N 
 {\rm exp}\left( -\frac{\| y_j- m_k \|_2^2}{2\sigma^2}  \right) \overline{P}_{k,j}
+ 
c },
\end{equation}
where $\overline{P}_{i,j} = 0$ if $i,j$ belong to different clouds and $1$ otherwise.

Similarly, 
exploiting the same argument as in 
\eqref{eq:Pij_cpd},
we find
$$
\log \left(
{\rm Pr} 
\left(
Y=y_j, Z=i \big| \theta
\right)
\right)
=
\left\{
\begin{array}{ll}
\displaystyle{ 
-\dfrac{d}{2} \log(  \sigma^2  ) 
-\dfrac{1}{2\sigma^2} \| y_j - m_i \|_2^2 +
\log \left(
\frac{1-w}{N (2\pi)^{d/2}}
\right)
}      &  i=1,\ldots,N\\[3mm]
 \log (w) - 2 \log Q    &  i=N+1. \\
\end{array}
\right.
$$
and then
$$ 
-\mathcal{Q}\left( 
\theta \big|\theta^{\rm old} \right)
:=
-
\sum_{i=1}^N \sum_{j=1}^Q 
P_{i,j}
\log \left(
{\rm Pr} 
\left(Y=y_j, Z=i \big| , \theta \right)
\right)
\,= \,
\left(
\frac{1}{2\sigma^2}
 \sum_{i=1}^N
\sum_{j=1}^Q
P_{i,j}
\| y_j - m_i  \|_2^2
\right)
+\frac{d Q_{\rm p}}{2} \log(\sigma^2) 
+C,
$$
where $Q_{\rm p} =  \sum_{i=1}^N
\sum_{j=1}^Q P_{i,j} $, and $C$ is a constant that is independent of $\theta$ and can thus be omitted. If we express the GMM centroids 
$\{m_i \}_{i=1}^N$
using the
deformed corresponding elements of the reference point cloud through \eqref{eq:deformation_model}, we finally obtain
\begin{equation}
\label{eq:practical_Mobjective}
-\mathcal{Q}\left( 
v,\sigma^2 \big|\theta^{\rm old} \right)
\,= \,
\left(
\frac{1}{2\sigma^2}
 \sum_{i=1}^N
\sum_{j=1}^Q
P_{i,j}(\theta^{\rm old})
\| y_j - F[v](\xi_i)  \|_2^2
\right)
+\frac{d Q_{\rm p}}{2} \log(\sigma^2) 
+C,
\end{equation}

Following \cite{meng1993maximum} and \cite{myronenko2010point}, we propose to split the M-step in two parts: first, we compute the velocity field $v^\star$ based on the previous estimate of $\sigma^2$; second, we compute the variance $\sigma^2$.
Furthermore, we augment the log-likelihood with an additional term that ensures  the smoothness of the velocity field and the (approximate) satisfaction of the constraint $v\cdot \mathbf{n}\big|_{\partial \Omega} = 0$.
In more detail, we consider the optimization problem to find the velocity  $v^\star$:
\begin{equation}
\label{eq:MstepI_tmp}
\max_{v\in V_h}  \mathcal{Q}(v, \sigma^{2,\rm old} \big| \sigma^{\rm old}) - \frac{\lambda}{\sigma^{2,\rm old}}
\mathfrak{f}_{\rm pen}(v)
\end{equation}
where $\lambda>0$ is a given constant and $\mathfrak{f}_{\rm pen}$ is the penalty function of section \ref{sec:penalty}.  
If we omit the terms that do
 not depend on $v$, we obtain the minimization statement
\begin{equation}
\label{eq:MstepI}
\min_{v\in V_h} \;\; 
\mathfrak{f}^{\rm obj}(v) :=
\frac{1}{2}
\sum_{i=1}^Q
\sum_{j=1}^N
P_{i,j}(\theta^{\rm old})
\| y_j - F[v](\xi_i)  \|_2^2
\; + \; 
\frac{\lambda }{2}
\mathfrak{f}_{\rm pen}(v).    
\end{equation}
Note that the maximization step \eqref{eq:MstepI} corresponds to the variational problem introduced in \eqref{eq:tractable_optimization_based_registration}.
Given the new estimate $v^\star$, we then set $\sigma^{2,\star}$ as
\begin{equation}
\label{eq:MstepII}
\sigma^{2,\star}
= {\rm arg} \min_{\sigma^{2} \in \mathbb{R}_+}
- \mathcal{Q}(v^\star, \sigma^{2} \big| \sigma^{\rm old})
 \quad \Rightarrow 
 \sigma^{2,\star} = 
\dfrac{  \sum_{i=1}^N \sum_{j=1}^Q P_{i,j} \|y_j - F[v^\star](\xi_i) \|_2^2   }{d  \sum_{i=1}^N \sum_{j=1}^Q P_{i,j}     }.
\end{equation}

Algorithm \ref{alg:EM_cpd} summarizes the overall procedure.
In this work, we consider the termination condition
\begin{equation}
\label{eq:termination_condition}
 R_1 := \frac{1}{N} \mathfrak{f}^{\text{tg}}(v^\star) < \epsilon_1,
 \quad
  R_2 := 
\dfrac{\| \mathbf{K}^{-1}   \nabla   \mathfrak{f}^{\text{obj}}(v^\star)        \|_2     }{\| \mathbf{K}^{-1}   \nabla   \mathfrak{f}^{\text{obj}}(v^0)        \|_2}  < \epsilon_2,
\quad
R_3 := \big| \sigma^{2,\star} - \sigma^{2,\rm old}  \big|< \epsilon_3,    
\end{equation}
where $\epsilon_1,\epsilon_2,\epsilon_3$ are user-defined parameters. Note that $R_1$ measures the misfit error,
$R_2$ is a weighted norm of the objective gradient, finally $R_3$ measures the increment of the estimate of the variance $\sigma^2$.

{\renewcommand{\baselinestretch}{1.5}\selectfont
\begin{algorithm}[H]
\caption{Expectation-maximization (EM) procedure for registration in bounded domains.}
\label{alg:EM_cpd}
\KwData{$\mathbf{Y}\in \mathbb{R}^{d\times Q}$ observations;
$v_0,\sigma_0^2$  initial estimate of the parameters,
$w$ (cf. \eqref{eq:model_multinomial}), $\lambda$ (cf. \eqref{eq:MstepII}), 
$\epsilon_1,\epsilon_2,\epsilon_3$ (cf. \eqref{eq:termination_condition}).
}
\KwResult{$v^\star$}

\textit{Initialization}: set 
$v^\star \gets v_0, \sigma^{2,\star} \gets \sigma_0^2$.

\For{$k = 0,\ldots,{\rm until \;\; convergence}$}{
\textbf{E-step:}
Determine 
$P_{i,j}$ using \eqref{eq:Pij_cpd} (or   \eqref{eq:Pij_cpd_modified}) for $i=1,\ldots,N$, $j=1,\ldots,Q$
 \;

\textbf{M-step (I):}
Estimate the solution $v^\star$ to  
 \eqref{eq:MstepI} \;

\textbf{M-step (II):}
Set  $\sigma^{2,\star}$ using 
 \eqref{eq:MstepII} \;

 Check the termination condition \eqref{eq:termination_condition}. 
}
\end{algorithm}
}

\begin{remark}
As in \cite{meng1993maximum},
we augmented the log-likelihood with a regularization term that encodes prior information on the parameters.
However,as opposed to 
 \cite{meng1993maximum},
the regularization $\lambda \mathfrak{f}_{\rm pen}(v)$ 
in \eqref{eq:MstepI_tmp}
is divided by $\sigma^{2,\rm old}$: this choice is justified by the fact that  
the penalty  became less effective as $\sigma^{2,\rm old}\to 0$.
In conclusion,
as opposed to \cite{meng1993maximum},
we cannot rigorously interpret the regularization $\mathfrak{f}_{\rm pen}(v)$ as a prior for the parameters. 
\end{remark}

\subsubsection{Solution to the optimization problem}
\label{sec:optimization_problem}

The estimate of $v^\star$ (cf. M-step (I), Algorithm \ref{alg:EM_cpd}) requires the solution to a highly nonlinear non-convex minimization problem. Towards this end, we resort to a left-preconditioned gradient-descent technique:
\begin{equation}
\label{eq:gradient_descent_technique}
\mathbf{v}^{\star, (q+1)} = 
\mathbf{v}^{\star, (q)} - \eta^{(q)} \mathbf{K}^{-1} \nabla \mathfrak{f}^{\rm obj}\left(  \mathbf{v}^{\star, (q)} \right), \quad
q=1,2,\ldots,
\end{equation}
where $\mathbf{K}$ is the symmetric positive definite matrix introduced in 
\eqref{eq:quadratic_penalty} and $\eta^{(q)}$ is determined using
line search. 
The choice of $\mathbf{K}^{-1}$ as left preconditioner is consistent with the variational interpretation of the velocity field (cf. \cite{younes2010shapes}): we refer to 
\cite[section 3.1.1]{iollo2026mathematical} for a thorough discussion on the subject.

In the numerical experiments, we considered two distinct line search methods:
backtracking line search
 and
critical point line search
(see, e.g., \cite[Appendix A]{dennis1996numerical}) as implemented in  the PETSC functions \texttt{SNESLINESEARCHBT} and
\texttt{SNESLINESEARCHCP}.
The two methods exhibit comparable performance in the numerical experiments of section \ref{sec:numerics}.

Since the  the objective of \eqref{eq:MstepI} is updated at each outer loop iteration of the EM procedure,
we empirically found that an inexact minimization based on limited updates (e.g.,  a few gradient steps) is preferable in practice. We remark that our empirical finding is consistent with the extensive literature on EM-like procedures
\cite{salakhutdinov2003adaptive}.

\section{Numerical results}
\label{sec:numerics}

\subsection{Transonic inviscid flow past a NACA0012 airfoil}

\subsubsection{Model problem}
\begin{figure}[hbt!]
    \centering     
    \subfigure[]{\label{fig:naca hf t01}\includegraphics[width=0.4\textwidth]{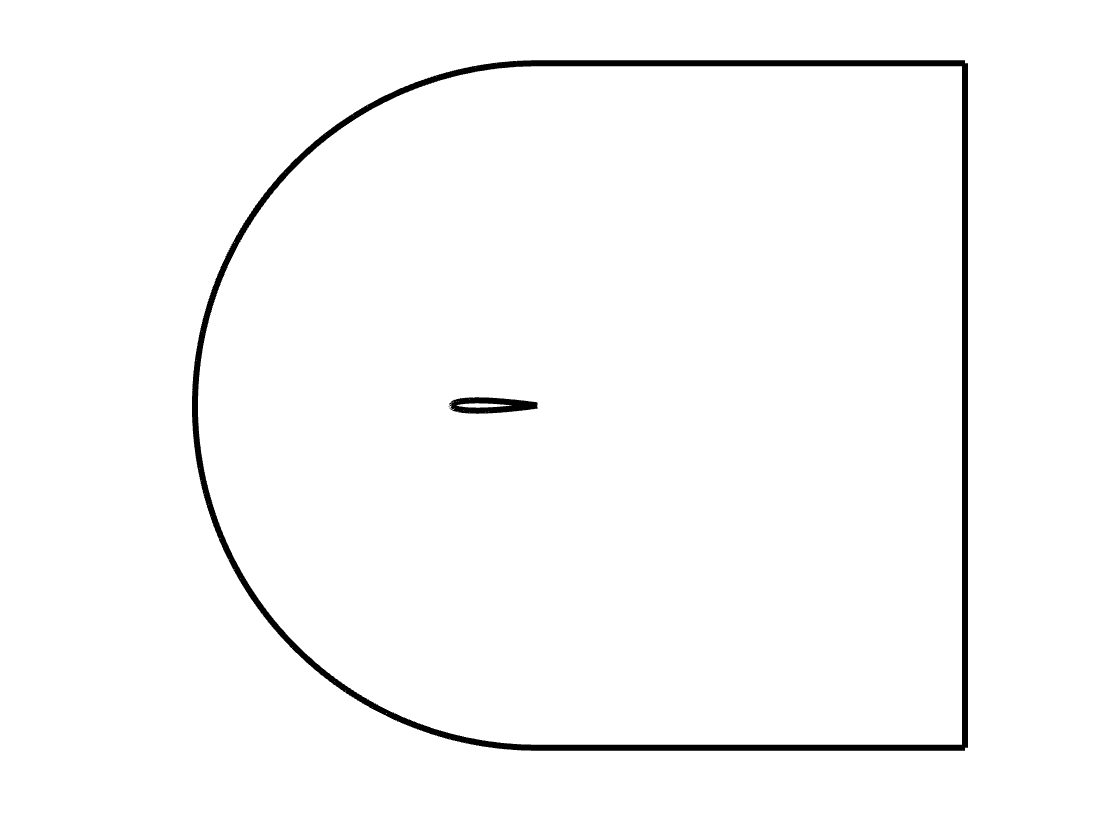}}
    \subfigure[]{\label{fig:naca hf t01}\includegraphics[width=0.4\textwidth]{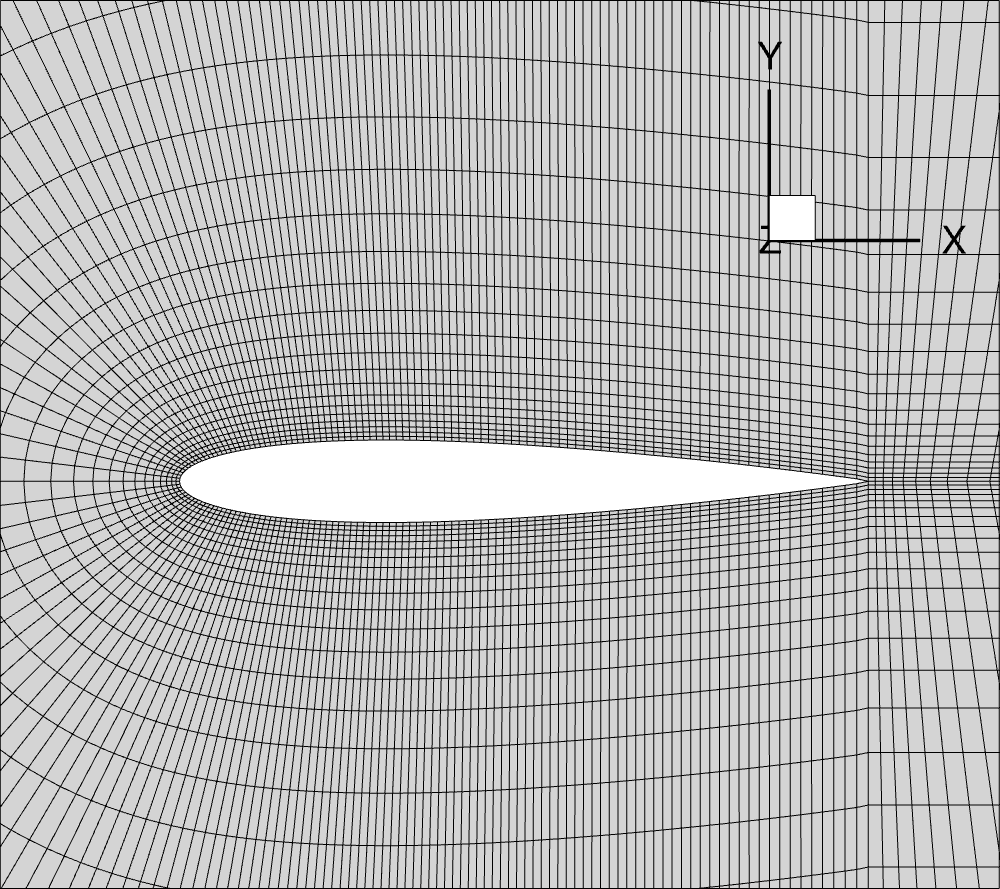}}

        \subfigure[$M_\infty=0.8$]{\label{fig:naca hf t00}\includegraphics[width=0.4\textwidth]{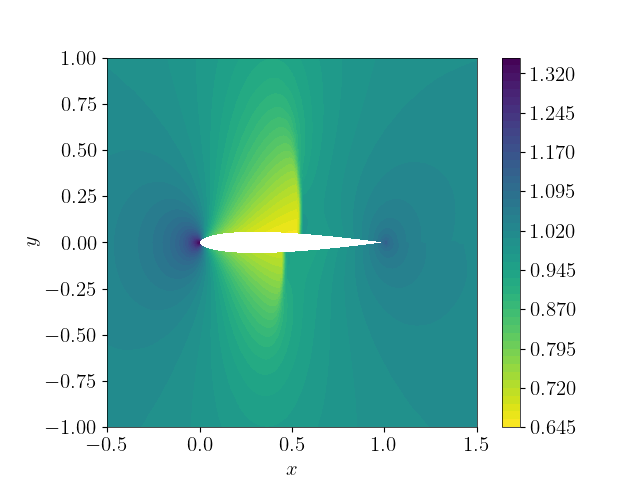}}
   \subfigure[$M_\infty=0.85$]{\label{fig:naca hf t01}\includegraphics[width=0.4\textwidth]{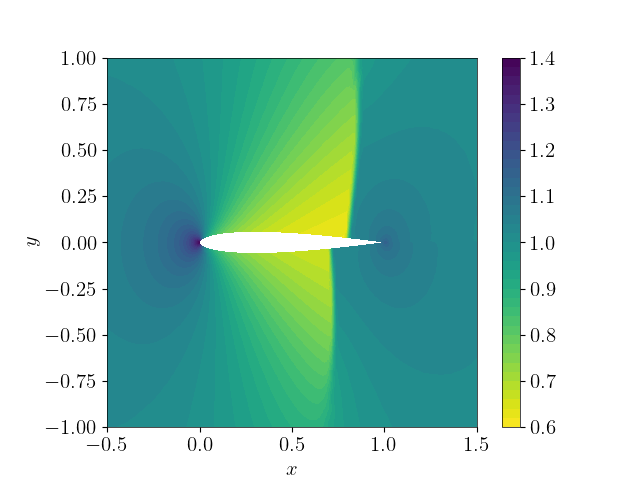}}

 \caption{transonic flow past a NACA airfoil.
    (a) geometric configuration.
    (b) computational mesh.
    (c)-(d) density profiles for two values of the free-stream Mach number and angle of attack $0.4^o$.}
     \label{fig:naca_intro}
\end{figure}

We consider the transonic inviscid flow past a NACA0012 airfoil for varying inflow Mach number
$\mu=M_\infty \in [0.8, 0.85]$ and angle of attack 
${\rm AoA}=0.4^o$. 
We introduce the vector of conserved variables 
$U= (\rho, \rho v_1,\rho v_2, \rho E )$ where $\rho$ is the flow density, $v$ is the flow velocity, $E$ is the total specific energy;  we also introduce the pressure $p$ that is related to the state variables through the equation of state
$p  = (\gamma-1) \left(\rho E - \frac{1}{2}\rho \|v  \|_2^2  \right)$, where $\gamma=7/5$ is the ratio of specific heats. 
The computational domain $\Omega = \Omega_{\rm out} \setminus \Omega_{\rm naca}$ is depicted in Figure \ref{fig:naca_intro}(a): the airfoil 
$\Omega_{\rm naca}$ has unit chord and its leading edge  is located at the origin $(0,0)$; $\Omega_{\rm out}$ is the union of a half circle of radius $100$ centered at the trailing edge $(1,0)$ and  the rectangle 
$(1,100)\times (-100,100)$.
CFD calculations are performed using the DG solver Aghora \cite{renac2015aghora} on a structured mesh with $N_{\rm e} = 14800$ quadrilateral elements (cf. Figure \ref{fig:naca_intro}(b)).
Figures \ref{fig:naca_intro}(c) and \ref{fig:naca_intro}(d) show the behavior of the density profile in the proximity of the airfoil for $M_\infty = 0.8$ and $M_\infty = 0.85$, respectively: we notice that the solution exhibits two normal shocks --- one on the upper (suction) side and one on the lower (pressure) side of the airfoil --- whose shape and location strongly depend on the value of the parameter.

\subsubsection{Setup}
We apply the 
nonlinear interpolation strategy of section \ref{sec:formulation} in combination with the 
registration method of section \ref{sec:methodology} to the two snapshots $U_0 = U(\mu_0=0.8)$ and $U_1 = U(\mu_1= 0.85)$. Towards this end, we apply Algorithm \ref{alg:EM_cpd} with initial velocity $v_0=0$, initial variance $\sigma_0^2=1$, and the hyper-parameters $w=0.1$, $\lambda=10^{-4}$;  we consider the termination condition
 $R_3 < 10^{-6}$ (that is, we set $\epsilon_1=\epsilon_2=0$ and $\epsilon_3=0$ in \eqref{eq:termination_condition}), and
 we set $s=2$ in the definition of the penalty matrix $\mathbf{K}
= \left(  
\mathbf{B}  \mathbf{M}^{-1}
\right)^{s-1} \mathbf{B}
 $ in \eqref{eq:quadratic_penalty}.
 At each M-step, we perform two iterations of the gradient descent technique
 \eqref{eq:gradient_descent_technique} and we resort to a direct solver based on Cholesky factorization to solve the $s=2$ linear systems of type $\mathbf{B} \mathbf{x} = \mathbf{b}$ --- since 
 $\mathbf{B}$ is fixed, Cholesky factorization can be applied once for all at the beginning of the procedure. In view of the assessment, we consider several structured FE grids of increasing size and
polynomial degree $p=1$ or $p=2$; furthermore, we resort to the explicit second-order Runge Kutta method \eqref{eq:RK2} for several choices of $\Delta t$ to integrate the ODE \eqref{eq:flow_diffeomorphisms}.

\begin{figure}[hbt!]
    \centering     
    \subfigure[$U_0$]{\label{fig:naca X}\includegraphics[width=0.45\textwidth]{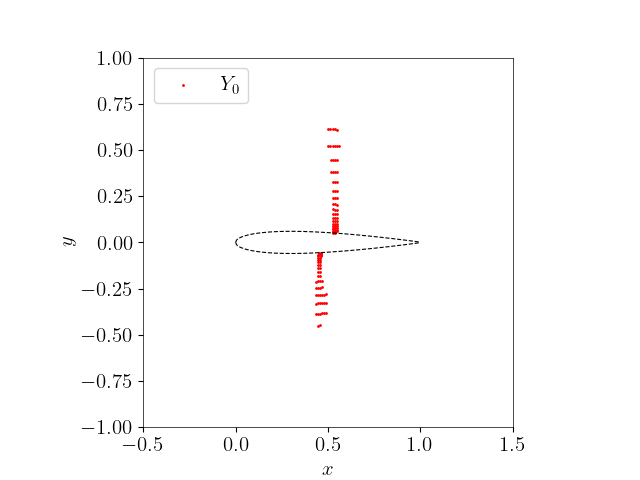}}
    \subfigure[$U_1$]{\label{fig:naca Y}\includegraphics[width=0.45\textwidth]{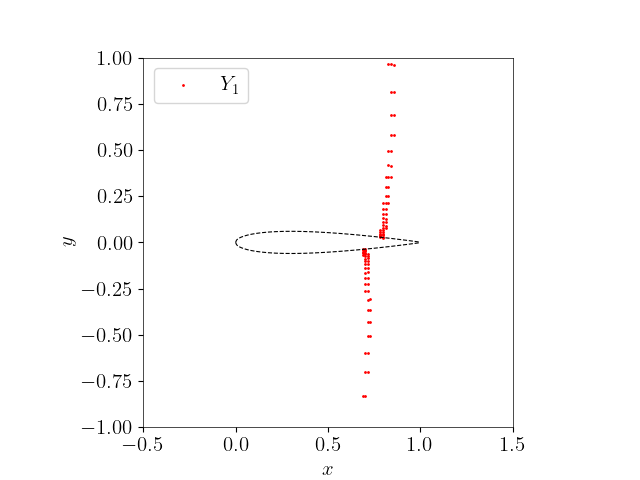}}    
    \caption{transonic flow past a NACA airfoil; selected points for the two snapshots, $U_i=U(\mu_i)$, $i=0,1$, with the sensor \eqref{eq:tecplot_sensor} ($\texttt{tol}=0.7$).}
    \label{fig:naca_sensors}
\end{figure}

Recalling the definitions of section \ref{sec:sensors}, we define the sampled point cloud as 
$$
Y_i = \left\{
x\in \Omega_{\rm discr} \, : \,
\zeta_{\rm s}(x, U_i^{\rm hf}) > 0.7
\right\},
\qquad
i=0,1.
$$
Figure \ref{fig:naca_sensors} shows the point clouds for the two snapshots. In the experiments, we   consider 
$Y_0^{\rm s}$ as reference point cloud and 
$Y_1^{\rm s}$ as target point cloud.
We also investigated the performance of the alternative sensor:
$$
Y_i^{\rm iso} = \left\{
x\in \Omega_{\rm discr} \, : \,
\zeta_{\rm iso}(x, U_i^{\rm hf}) > 0.22 \; {\rm or} \;
\zeta_{\rm iso}(x, U_i^{\rm hf}) <-0.1
\right\},
\qquad
i=0,1.
$$
where the function $\zeta_{\rm iso}$ is defined in 
\eqref{eq:iso_scalar_testing}. In our experience, we found nearly-equivalent  results to the ones obtained with the previous sensor: for this reason, we do not  further discuss this alternative.

\subsubsection{Results (I): registration problem}
\begin{figure}[h]
\centering     
\subfigure[$\Delta t=0.2$]{\label{fig:naca conv h R1}\includegraphics[width=0.32\textwidth]{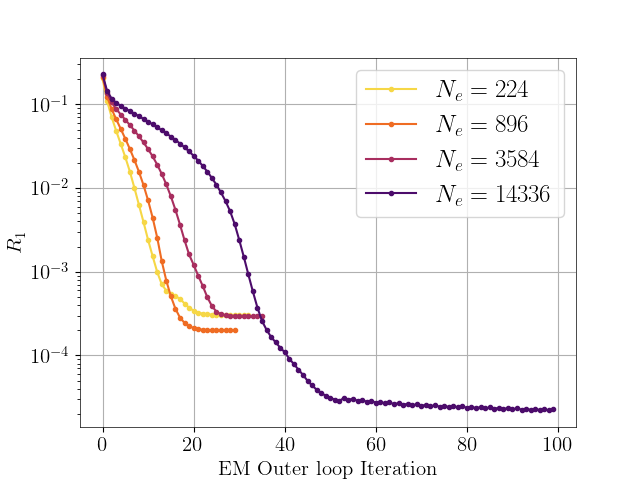}}
\subfigure[]{\label{fig:naca conv h R2}\includegraphics[width=0.32\textwidth]{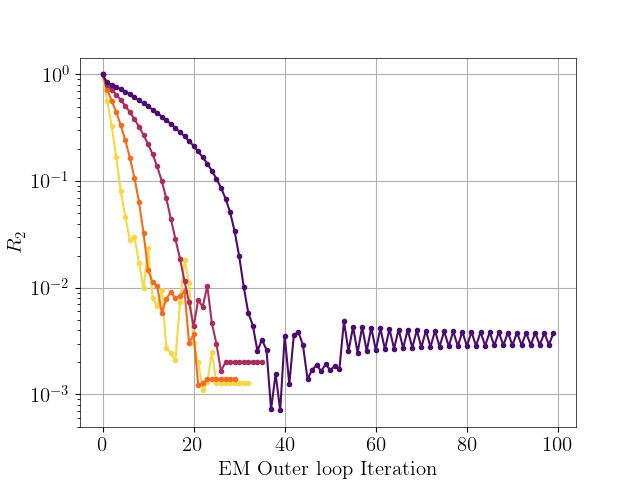}}
\subfigure[]{\label{fig:naca conv h R3}\includegraphics[width=0.32\textwidth]{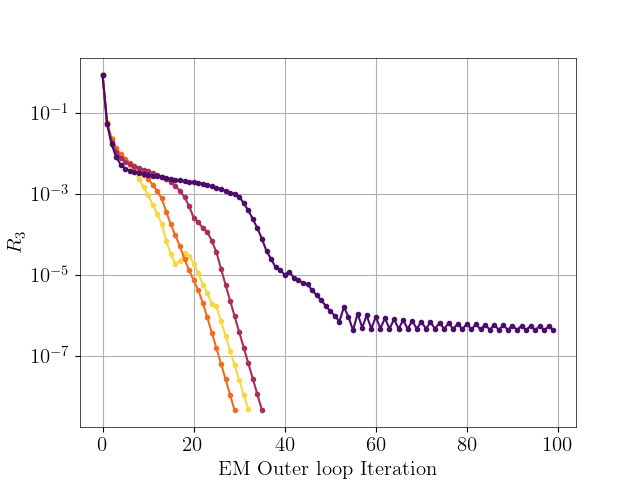}}

\subfigure[$\Delta t=0.05$]{\label{fig:naca conv h p2 R1}\includegraphics[width=0.32\textwidth]{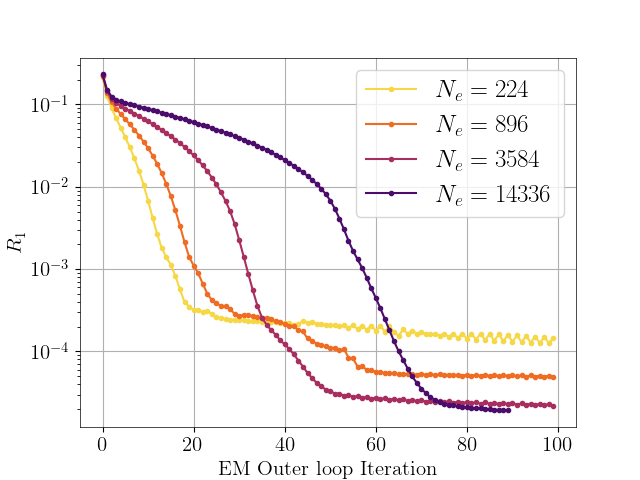}}
\subfigure[]{\label{fig:naca conv h p2 R2}\includegraphics[width=0.32\textwidth]{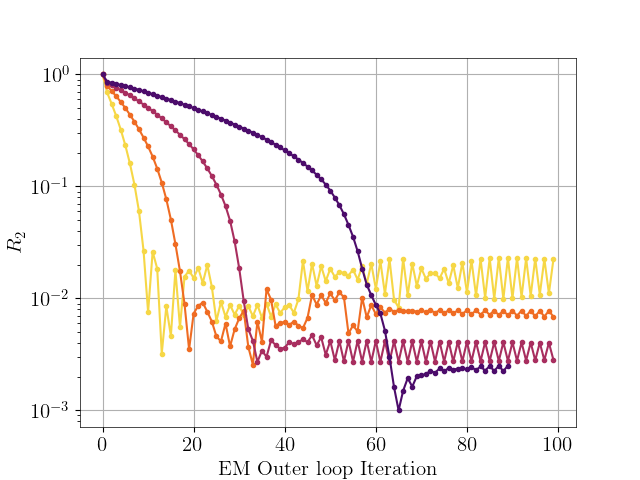}}
\subfigure[]{\label{fig:naca conv h p2 R3}\includegraphics[width=0.32\textwidth]{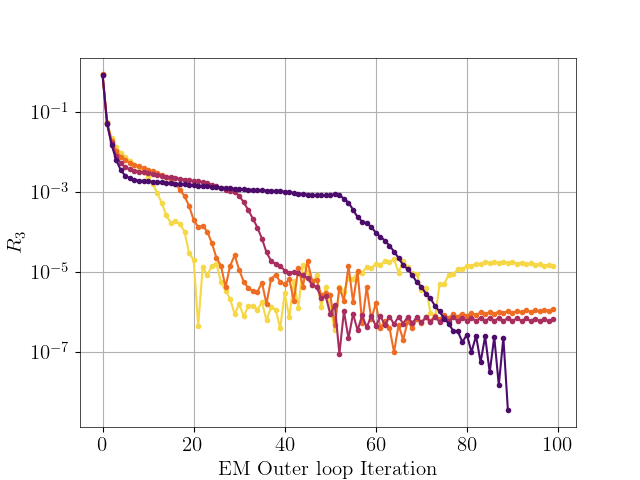}}

\caption{transonic flow past a NACA airfoil; evolution of the residuals 
$R_1,R_2,R_3$ in 
\eqref{eq:termination_condition} with respect to the EM iteration count, for different FE meshes with $N_e$.
(a)-(b)-(c) $\Delta t=0.2$, linear elements ($p=1)$.
(d)-(e)-(f) $\Delta t=0.05$, quadratic elements ($p=2)$.}
\label{fig:naca_reg_conv}
\end{figure}

Figure \ref{fig:naca_reg_conv}
shows the evolution of the residuals 
$R_1,R_2,R_3$ defined in \eqref{eq:termination_condition} with respect to the EM outer loop iteration count, for different FE meshes and two choices of the time step $\Delta t$  (polynomial degree $p=2$).
As we increase the size of the spatial mesh, we notice an improvement in the alignment performance; on the other hand, we do not observe any significant change when we reduce the time step from $\Delta t=0.2$ to $\Delta t=0.05$.

\begin{figure}[hbt!]
\centering     
\subfigure[$p=1$]{\label{fig:naca dist p1dt1}\includegraphics[width=0.4\textwidth]{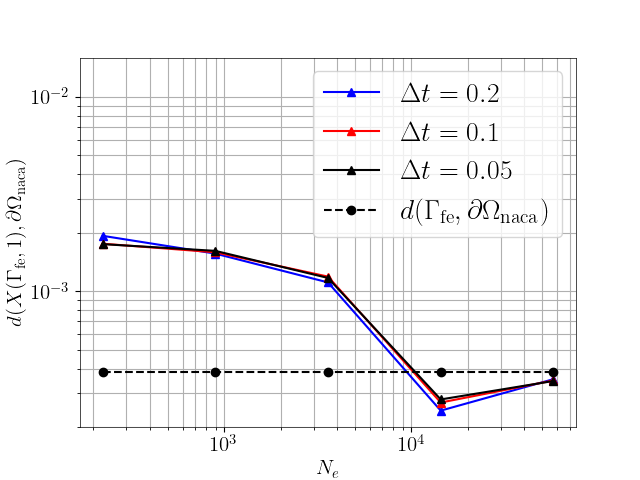}}
\subfigure[$p=2$]{\label{fig:naca dist p1dt2}\includegraphics[width=0.4\textwidth]{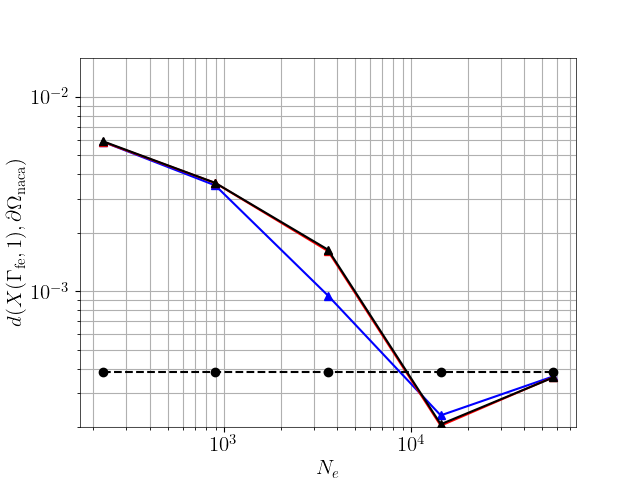}}
\caption{transonic flow past a NACA airfoil; evolution of the geometric error  \eqref{eq:geometric_error_naca}  with respect the FE mesh size for linear ($p=1$) and quadratic  ($p=2$) polynomials.}
\label{fig:naca_dist}
\end{figure}

In order to assess the geometric error,
we denote by $\Gamma_{\rm fe} : = \{ \xi_i^\Gamma \}_{i=1}^{N_{\Gamma}} \subset \partial \Omega_{\rm naca}$ the nodes of the FE mesh used for registration on the airfoil and we define the geometric error as
\begin{equation}
    \label{eq:geometric_error_naca}
    d \left( X(\Gamma_{\rm fe},1), \partial\Omega_{\rm naca} \right)= 
    \max_{i\in\{1,...,N_\Gamma\}} \min_{x\in\partial\Omega_{\rm naca}}  \big\| X(\xi_i^\Gamma,1)-x \|_2.
\end{equation}
In order to compute
$
\min_{x\in\partial\Omega_{\rm naca}}  \big\| X(\xi_i^\Gamma,1)-x \|_2$ for a given index $i\in \{1,\ldots,N_\Gamma\}$,
  we solve an optimization
problem that exploits the analytic expression of the boundary of the airfoil.
Figure \ref{fig:naca_dist} shows the behavior of \eqref{eq:geometric_error_naca} for several grid sizes and several time step sizes; for reference, we also report the geometric error of the original grid. We observe that 
for 
$N_{\rm e} \gtrsim 10^4$ 
the geometric error is of the same order
as for the original grid.
Interestingly, the time step size and the polynomial order have moderate impact on the accuracy: the latter is expected as we do not rely on curved meshes.

\subsubsection{Results (II): flow prediction}
\begin{figure}[hbt!]
    \centering     
    \subfigure[$M_\infty=0.81$]{\label{fig:naca hf t00}\includegraphics[width=0.32\textwidth]{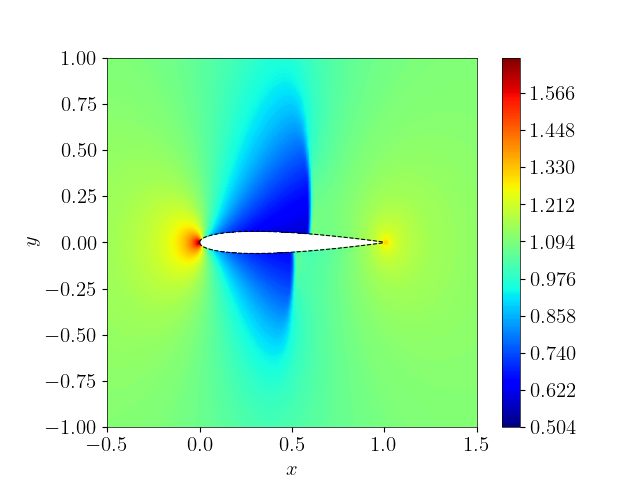}}
    \subfigure[$M_\infty=0.825$]{\label{fig:naca hf t02}\includegraphics[width=0.32\textwidth]{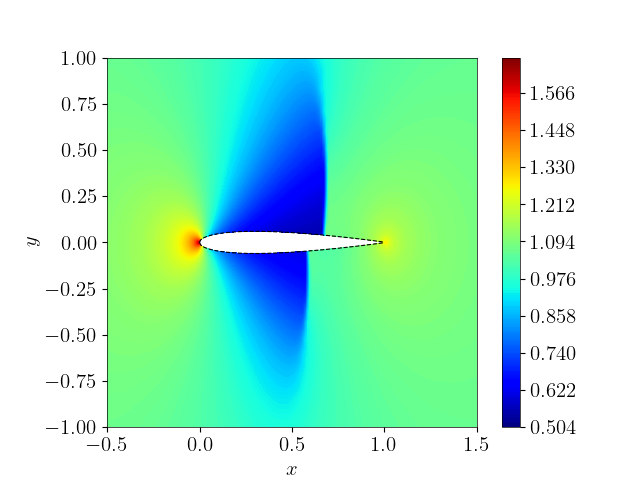}}
    \subfigure[$M_\infty=0.84$]{\label{fig:naca hf t04}\includegraphics[width=0.32\textwidth]{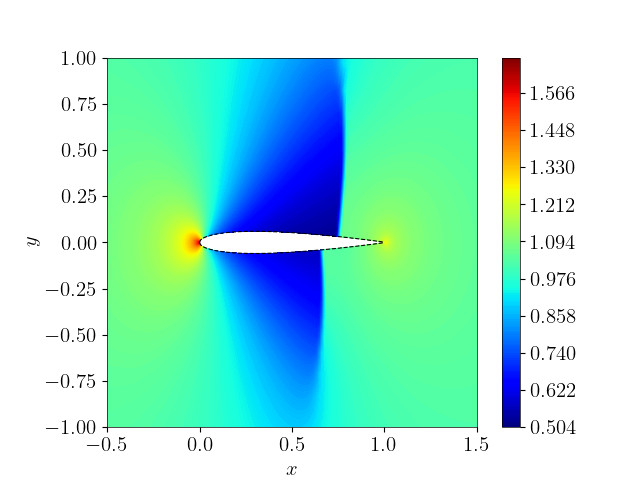}}\\
    \caption{transonic flow past a NACA airfoil; HF estimates of the pressure field for three parameter values.}
 \label{fig:naca_hf}
\end{figure}

\begin{figure}[hbt!]
\centering     
\subfigure[$M_\infty=0.81$]{\label{fig:naca cdi shock t00}\includegraphics[width=0.32\textwidth]{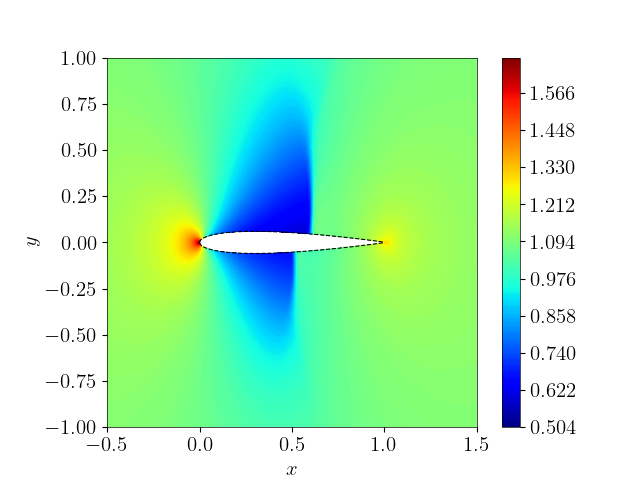}}
\subfigure[$M_\infty=0.825$]{\label{fig:naca cdi shock t02}\includegraphics[width=0.32\textwidth]{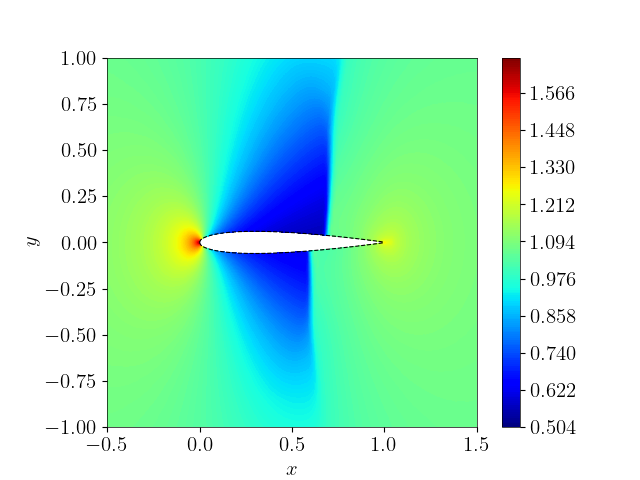}}
\subfigure[$M_\infty=0.84$]{\label{fig:naca cdi shock t04}\includegraphics[width=0.32\textwidth]{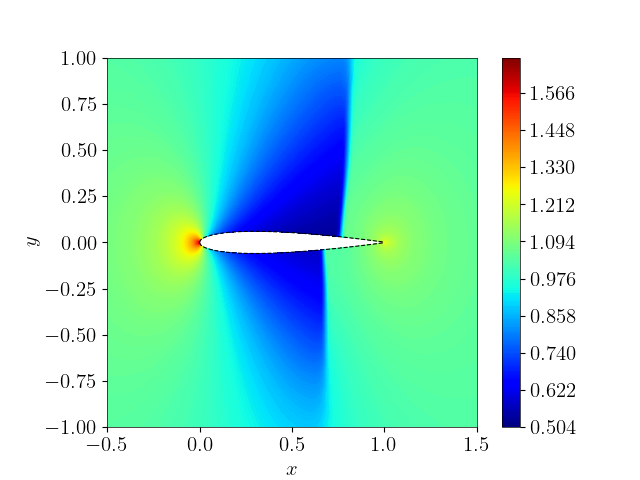}}\\
\caption{transonic flow past a NACA airfoil; CDI estimates of the pressure field for three parameter values with \texttt{shock}-based registration ($\Delta t = 0.1$ and $p=1$).}
\label{fig:naca_CDI_shock}
\end{figure}

\begin{figure}[hbt!]
\centering     
\subfigure[$M_\infty=0.81$]{\label{fig:naca ci t00}\includegraphics[width=0.32\textwidth]{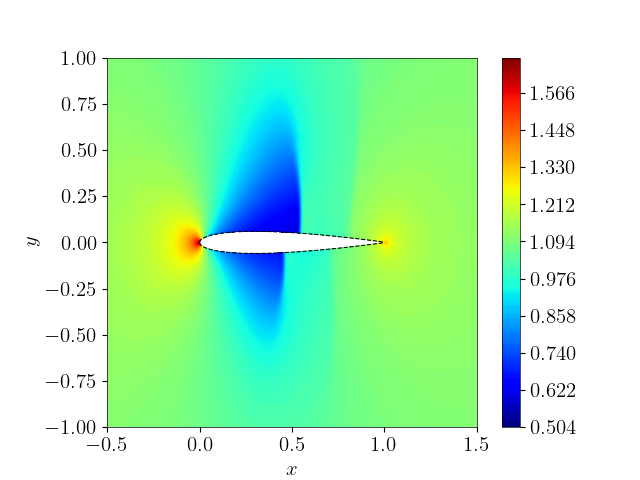}}
\subfigure[$M_\infty=0.825$]{\label{fig:naca ci t02}\includegraphics[width=0.32\textwidth]{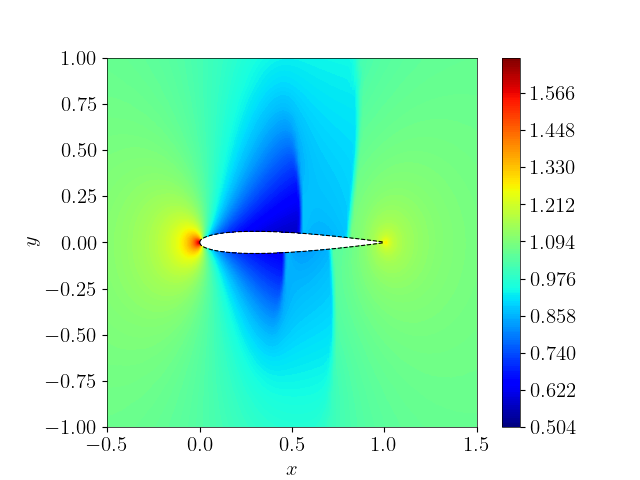}}
\subfigure[$M_\infty=0.84$]{\label{fig:naca ci t04}\includegraphics[width=0.32\textwidth]{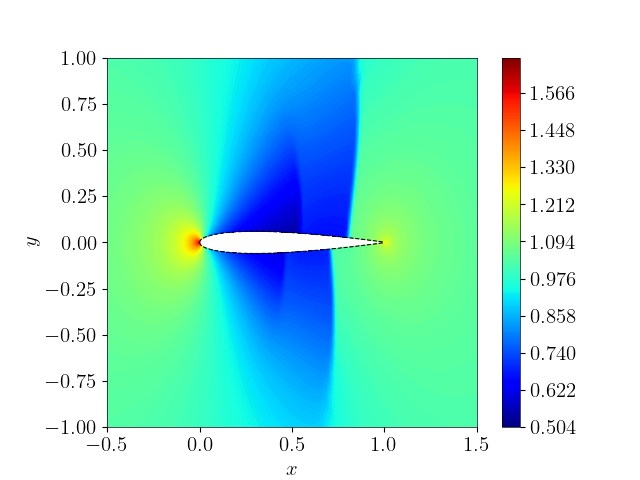}}\\

\caption{transonic flow past a NACA airfoil; CI estimates of the pressure field for three parameter values.}
\label{fig:naca_CI}
\end{figure}

We apply the CDI method \eqref{eq:CDI} to predict the flow state for $M_\infty \in [0.8,0.85]$. Figure 
\ref{fig:naca_hf} shows the truth  pressure field 
as predicted by the high-fidelity (HF) DG solver;
Figure    \ref{fig:naca_CDI_shock} shows the CDI prediction based on the registration method outlined above;
finally, Figure \ref{fig:naca_CI} shows the performance of a simple convex interpolation.
We notice that   CDI outperforms
convex interpolation, for the same amount of training data (that is, two snapshots).

\begin{figure}[hbt!]
    \centering     
    \subfigure[$M_\infty=0.81$]{\label{fig:naca cp t00}\includegraphics[width=0.32\textwidth]{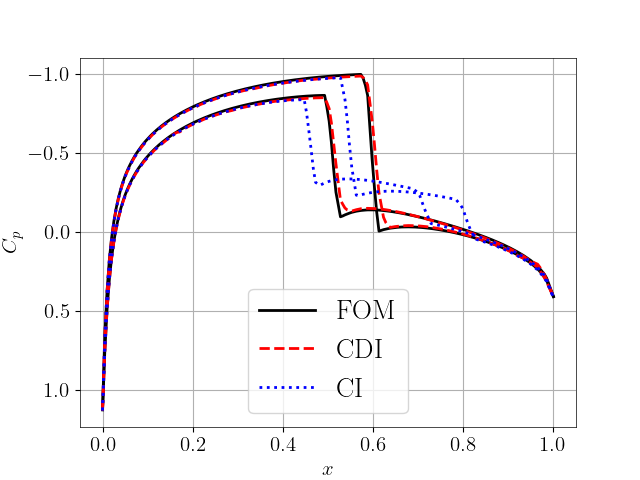}}
    \subfigure[$M_\infty=0.825$]{\label{fig:naca cp t02}\includegraphics[width=0.32\textwidth]{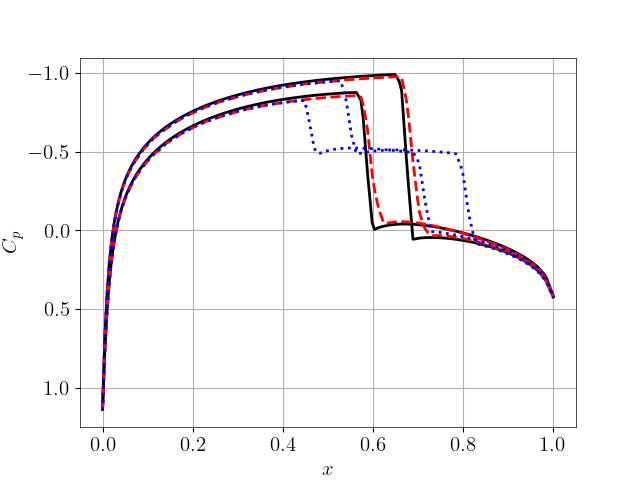}}
    \subfigure[$M_\infty=0.84$]{\label{fig:naca cp t04}\includegraphics[width=0.32\textwidth]{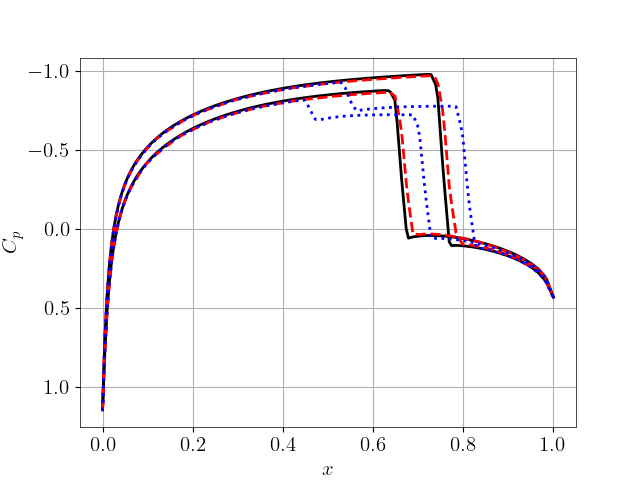}}\\

    \caption{transonic flow past a NACA airfoil; prediction of the pressure coefficients $C_p$ for several parameter values ($N_e = 14336$).}
      \label{fig:naca_Cp}
\end{figure}

\begin{figure}[hbt!]
\centering     
\subfigure[$p=1$, $\Delta t=0.1$]{\label{fig:naca cdi L2 norm p1}\includegraphics[width=0.45\textwidth]{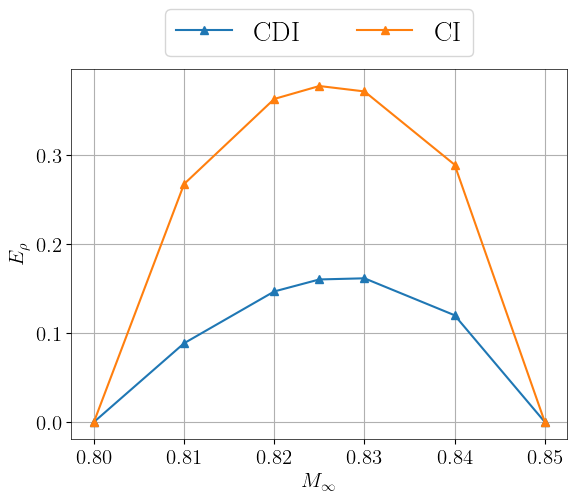}}
    \subfigure[$p=2$, $\Delta t=0.05$]{\label{fig:naca cdi L2 norm p2}\includegraphics[width=0.45\textwidth]{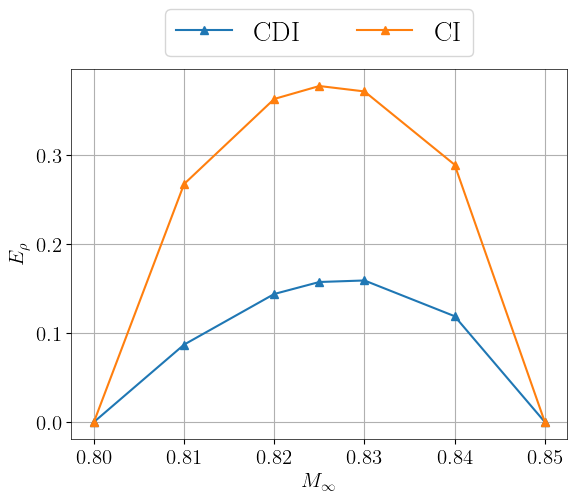}}
 
\caption{transonic flow past a NACA airfoil; 
    relative prediction $L^2$ error for the  density field $\rho$, for two different discretizations.p    ($N_e = 14336$).}
\label{fig:naca_rho_L2_norm}
\end{figure}

In order to offer a more quantitative comparison between the different methods, we report in Figure \ref{fig:naca_Cp} the behavior of the pressure coefficient
$$
C_p : = \dfrac{p-p_\infty}{\frac{1}{2} \rho_\infty \| v_\infty \|_2^2}
$$
on both the suction and the pressure sides of the airfoil for both implementations of the CDI, and we compare with convex interpolation.
Furthermore,  Figure \ref{fig:naca_rho_L2_norm} shows the relative error in the prediction of the density field
$$
E_\rho: = \dfrac{\| \rho - \hat{\rho}   \|_{L^2(\Omega)}}{\| \rho - {\rho}_\infty   \|_{L^2(\Omega)}},
$$
where $\widehat{\rho}$ denotes the CDI (or CI) estimate.
We consider two different discretizations of the registration task:
the results of Figure \ref{fig:naca_rho_L2_norm}(a) correspond to a linear ($p=1$) FE discretization and $\Delta t = 0.1$, while the  
results of Figure \ref{fig:naca_rho_L2_norm}(b) correspond to a quadratic ($p=2$) FE discretization and $\Delta t = 0.05$.
We observe that CDI clearly outperforms convex interpolation for this test case; we further notice that the polynomial degree of the FE discretization, as well as the time step $\Delta t$ have a negligible impact on the performance.

\subsubsection{Results (III): data augmentation}

We exploit the CDI model to generate artificial snapshots in the range $\mu \in [0.8, 0.85]$. These snapshots are used to build a reduced basis that can be later used for model-based (projection-based, collocated) ROMs. In more detail, we generate $N_{\rm snap} = 500$ artificial pressure field snapshot with the CDI model using  different choices of the interpolation parameter $s$ in \eqref{eq:CDI}. 
We apply proper orthogonal decomposition to  
the snapshot set of 
$N_{\rm snap}$
synthetic snapshots to 
build a reduced basis $\{\varphi_1, ..., \varphi_M\}$. We denote by $\mathcal{W}_m = \text{Span}(\varphi_1, ..., \varphi_m)$ the truncated POD subspace for the pressure field of dimension $m$.

\begin{figure}[hbt!]
    \centering     
    \subfigure[]{\label{fig:projection error decay}\includegraphics[width=0.45\textwidth]{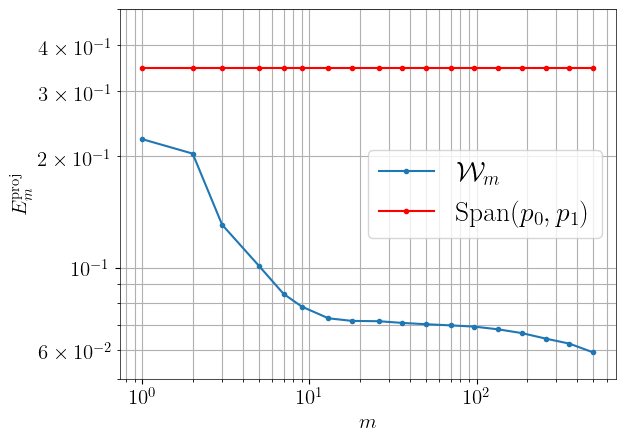}}
    \subfigure[]{\label{fig:projected snap M26}\includegraphics[width=0.45\textwidth]{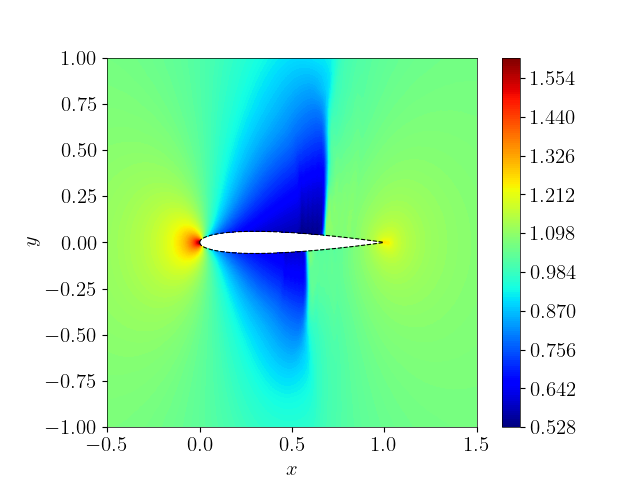}}
    \caption{(a) $L^2$ projection error of FOM snapshot $p_\mu$ with $\mu=0.825$ onto subspace $\mathcal{W}_m$, (b) Scalar field $\Pi_{\mathcal{W}_m}p_\mu$ with $m=26$.}
    \label{fig:NACA L2 proj error dataAug}
\end{figure}

Figure \ref{fig:NACA L2 proj error dataAug}(a) shows the behavior of the relative projection error
\begin{equation}
\label{eq:L2 projective error dataAug}
E_m^{\rm proj} = \frac{\|p_{\mu} - \Pi_{\mathcal{W}_m}p_\mu\|_{L^2}}{\|p_\mu- p_\infty\|_{L^2}},
\end{equation}
with respect to the size of the POD space, for the test parameter  $\mu_{\rm test} = 0.825$.
Finally, Figure \ref{fig:NACA L2 proj error dataAug}(b) shows the projected pressure field for $m=26$.
We observe that data augmentation enhances approximation properties of the reduced space: the projection error drops from $E^{\rm proj} = 3 \cdot 10^{-1} $ for ${\rm span}\{p_0,p_1\}$  to approximately 
$E^{\rm proj} = 7 \cdot  10^{-2}$ for $m \gtrsim 20$.

 
\subsection{Viscous flow past an ONERA M6 wing}

\subsubsection{Model problem}
We consider the transonic viscous turbulent flow past an Onera M6 wing 
\cite{schmitt1979pressure}
for varying angle of attack $\mu={\rm AoA}\in [3.06^o,5.1^o]$. We set $M_\infty=0.84$ and we consider the 
Reynolds number ${\rm Re}=14.6\cdot 10^6$.
We rely on the compressible Reynolds-averaged Navier Stokes (RANS) equations with Spalart-Allmaras turbulence model
\cite{allmaras2012modifications}; the vector of state variables $U$ is given by $U={\rm vec}(\rho,\rho v_1,\rho v_2,\rho v_3, \rho E, \rho \nu_{\rm T})$ where $\rho, v, E$ are the same as in the previous test case, while 
$\nu_{\rm T}$ is the turbulent viscosity.
CFD calculations are performed using the DG solver CODA \cite{stefanin2024aircraft}.
We consider a hybrid mesh for CFD calculations with $N_{\rm e}=1,351,648$ elements: prisms elements are employed in the boundary layer region near the wing, while tetrahedral elements are employed in the far field, which is set at $12$ wing root chords.  

We resort to a $p=1$ discretization for the computation of the HF snapshots;
we rely on the 
ROE    up-winding method with positivity preserving limiter  to handle the  convection term.
We consider a pseudo-transient continuation method to solve the nonlinear system of equations;
we  initialize the CFL condition to one and then we consider a SER ramp with exponent of $0.4$ to increase the time step.
At each pseudo-time iteration, 
we exploit the
Newton's method with 
finite differencing approximation of the Jacobian to solve the nonlinear system of equations,  and we use the 
restarted GMRES method for linear solves.
Finally, 
for $p=0$, we initialize the DG solver using the free-stream  solution;
for $p=1$, we   use the $p=0$ solution.
We set the convergence level to 
$10^{-10}$ for $p=0$ computation, and to  $10^{-8}$ for $p=1$ computation.

\begin{figure}[hbt!]
    \centering     
        \subfigure[]{\label{fig:M6 t1}\includegraphics[width=0.32\textwidth]{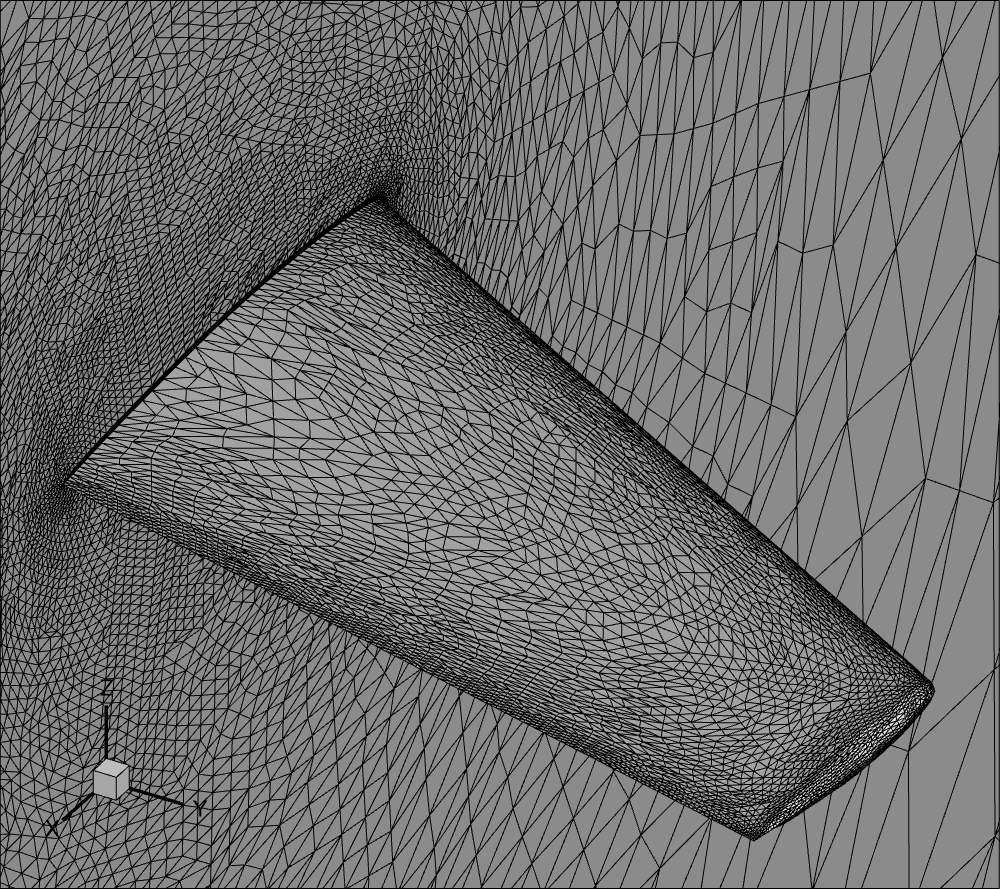}}
    \subfigure[${\rm AoA}=3.06^o$]{\label{fig:M6 t00}\includegraphics[width=0.32\textwidth]{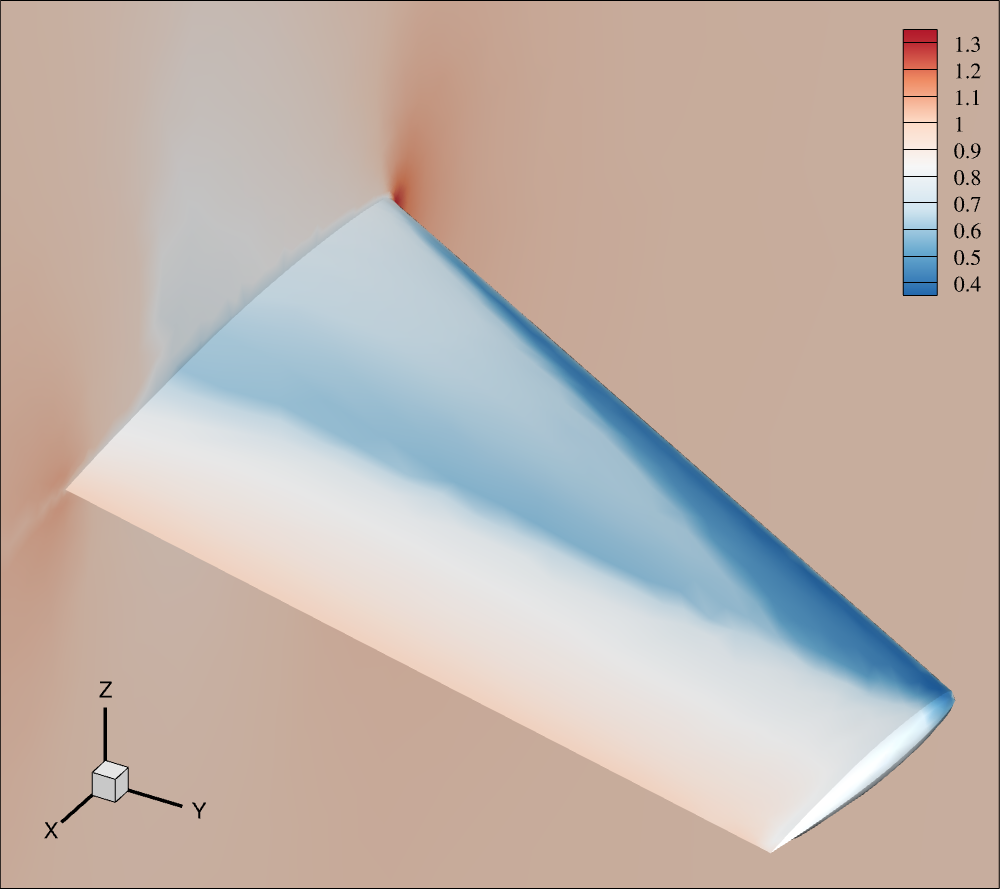}}
    \subfigure[${\rm AoA}=5.1^o$]{\label{fig:M6 t00}\includegraphics[width=0.32\textwidth]{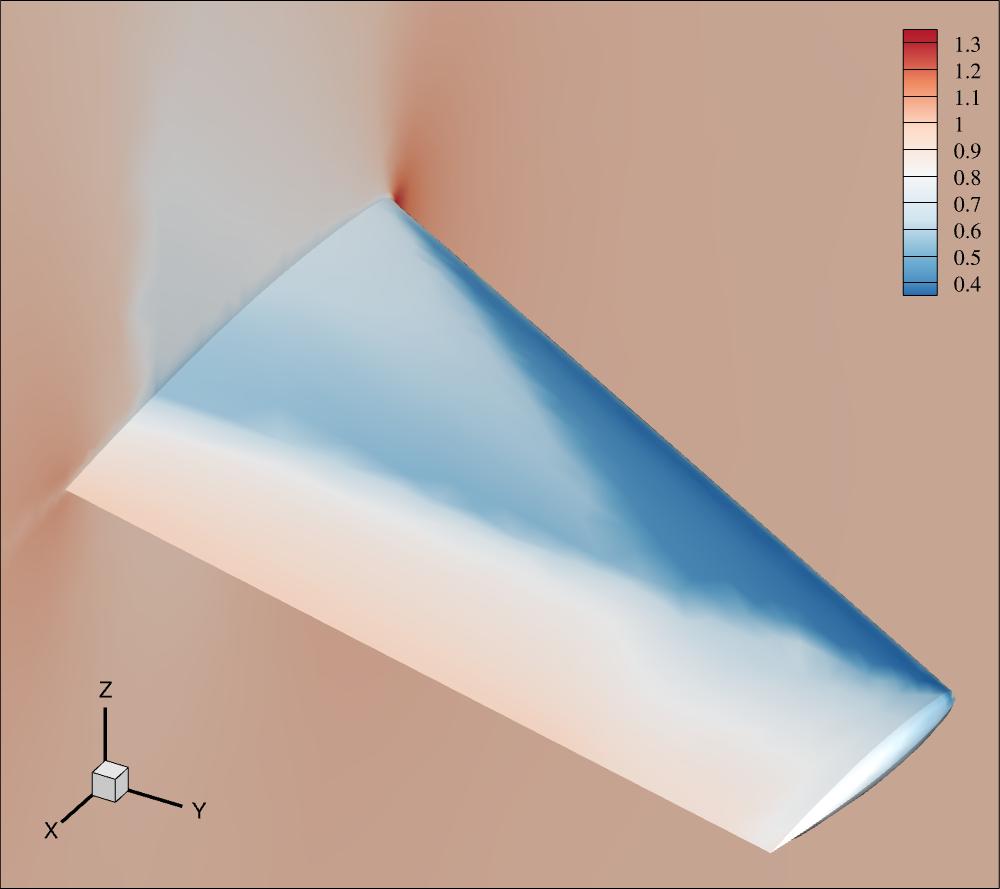}}
    \caption{Viscous flow past an ONERA M6 wing.
    (a) computational mesh.
    (b)-(c) density profiles for two values of the angle of attack.}
    \label{fig:M6_vis}
\end{figure}

Figure \ref{fig:M6_vis}(a) shows the computational mesh; 
Figures \ref{fig:M6_vis}(b) and (c) show the density profiles for two parameter values.
As extensively discussed in the literature (see, e.g., \cite{kuzmin2014lambda}),
this test  features a $\lambda$-configuration ($\lambda$-shock) of isobars on the upper surface of the wing; the location and the shape of  the supersonic regions is highly sensitive to the inflow conditions (Mach number and angle of attack).

\subsubsection{Setup}
We  extract the source and the  target point clouds from the CFD mesh. In more detail, we compute the shock sensor presented in (\ref{eq:tecplot_sensor}) on the CFD mesh nodes  as follows:
\begin{equation}
\label{eq:point_cloud_M6}
Y_i^{\rm s} = \left\{
x\in \Omega_{\rm discr} \, : \,
\zeta_{\rm s}(x, U_i^{\rm hf}) > 0.6
\right\},
\qquad
i=0,1;
\end{equation}
For the selected snapshots,
we retrieve two point sets of size $N_0=6731$ and $N_1=4250$ to represent the geometry of the lambda shock.
To visualize the point clouds in Figure 
\ref{fig:M6 reg results}, we show vertical cuts of  the indicator function  
$\chi_i(x) := \mathbbm{1}_{Y_i^{\rm s}}(x)$ for  three different choices of $x_3$.


As for the previous test case, we perform two steps of the gradient descent technique \eqref{eq:MstepI} at each outer loop iteration of 
Algorithm \ref{alg:EM_cpd}.
We use a critical point line search method with an initial search interval of $[0, 10^{-3}]$, and maximum step  $\eta_{\rm max}=10^3$ to ensure   smooth convergence of the EM procedure.
  We resort to the conjugate gradient method with 
  ILU  preconditioner to solve \eqref{eq:gradient_descent_technique}.
We initialize Algorithm \ref{alg:EM_cpd} with the velocity 
$v_0=0$;  we set 
the initial covariance of the GMM  to $\sigma_0^2=2$;  we use $w=0.1$ for the noise of the probability model;
furthermore,
we consider the regularization parameter 
$\lambda=10^{-4}$ in \eqref{eq:MstepI} and 
we set $s=3$ 
in  \eqref{eq:quadratic_penalty};
we consider the 
cut-off frequency $\kappa_0=1$.

\subsubsection{Results (I): registration problem}

\begin{figure}[hbt!]
    \centering     
    \subfigure[$x_3=0.3$]{\label{fig:M6 reg 0 cut03}\includegraphics[width=0.3\textwidth]{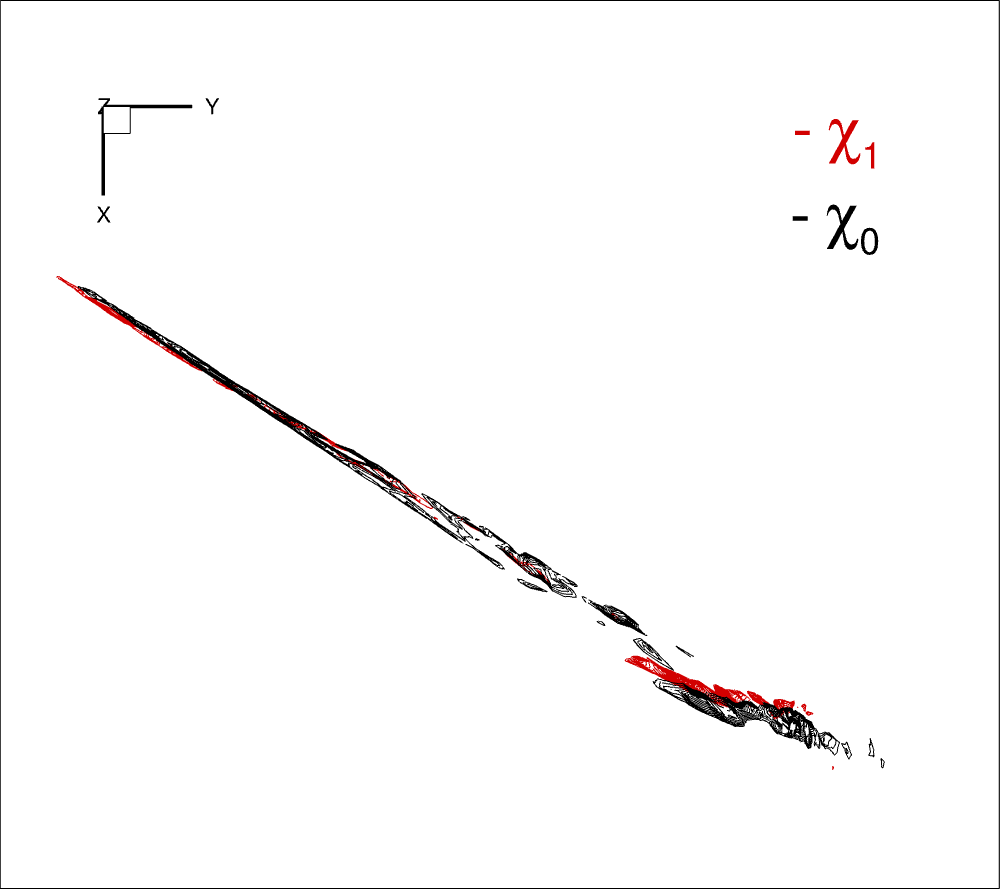}}
    \subfigure[$p=1$, $\Delta t = 0.1$]{\label{fig:M6 map shock p1}\includegraphics[width=0.3\textwidth]{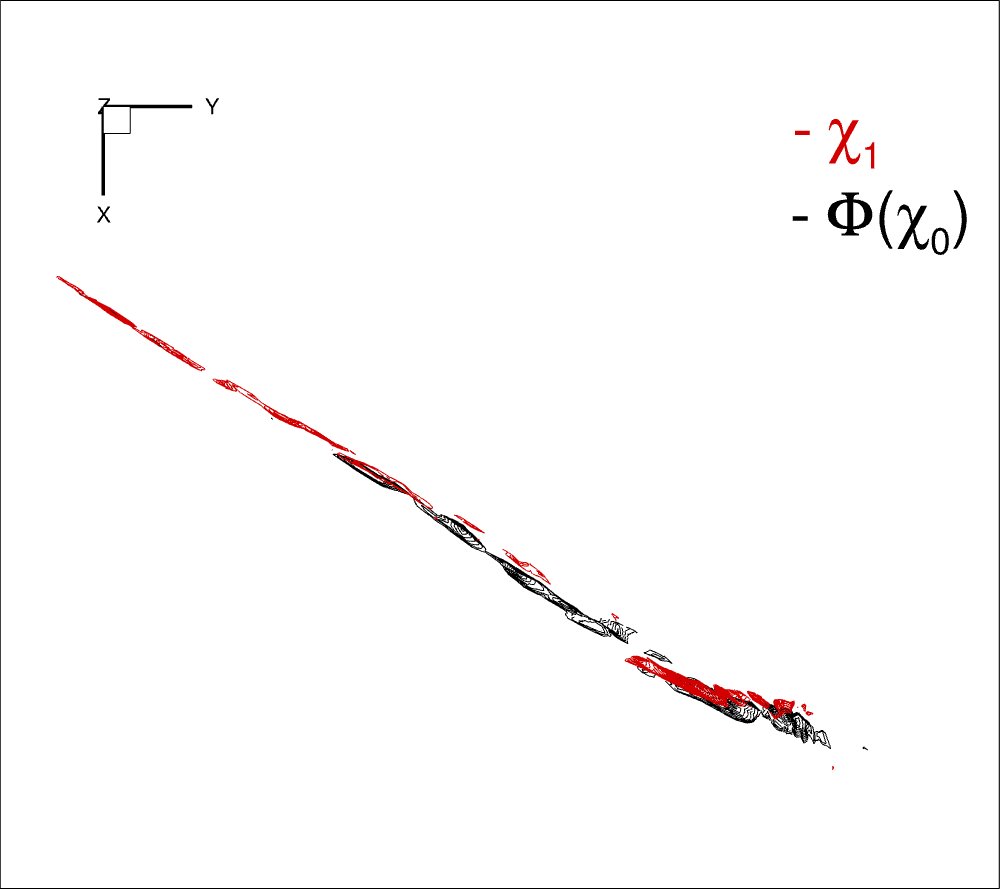}}
    \subfigure[$p=2$, $\Delta t = 0.05$]{\label{fig:M6 map shock p2}\includegraphics[width=0.3\textwidth]{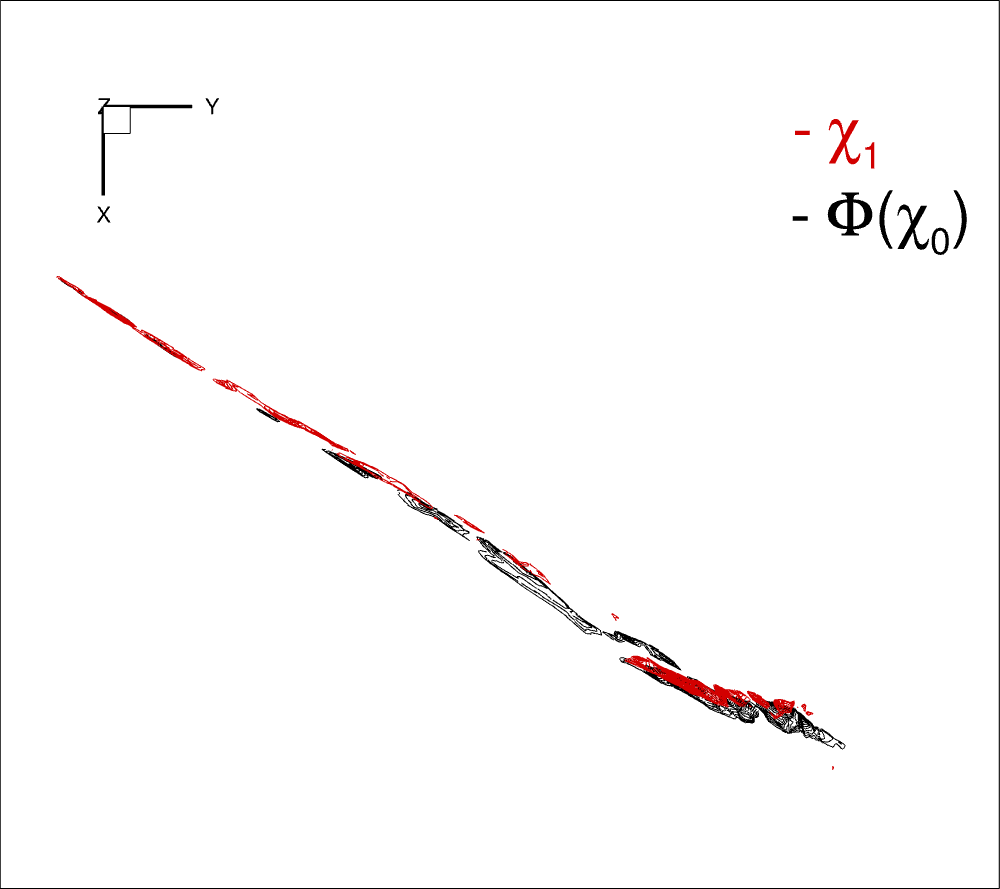}} \\
    \subfigure[$x_3=0.5$]{\label{fig:M6 reg 0 cut05}\includegraphics[width=0.3\textwidth]{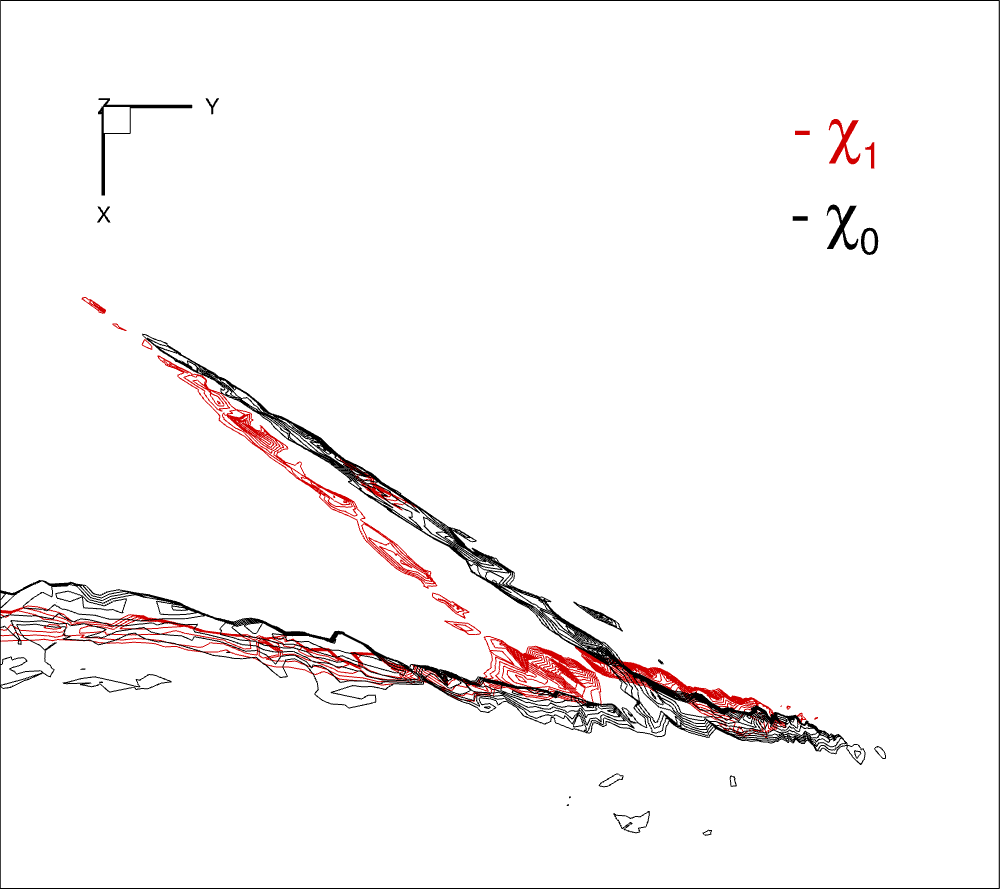}}
    \subfigure[]{\label{fig:M6 map shock p1}\includegraphics[width=0.3\textwidth]{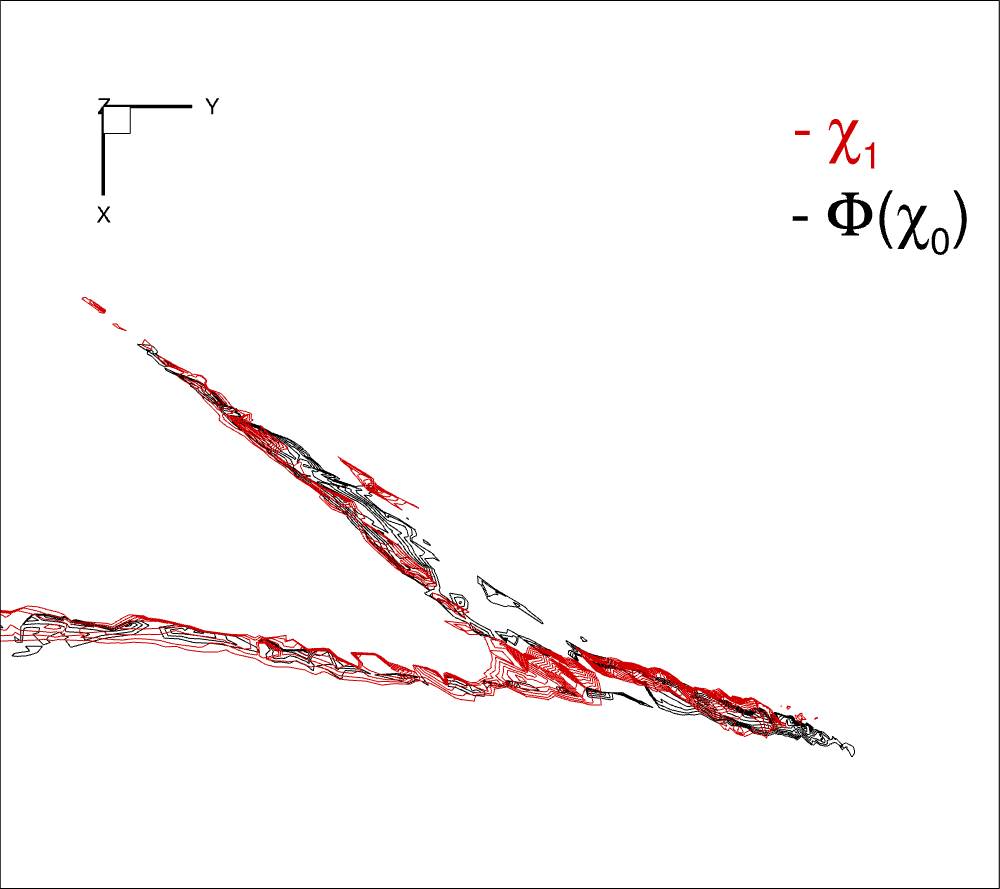}}
    \subfigure[]{\label{fig:M6 map shock p2}\includegraphics[width=0.3\textwidth]{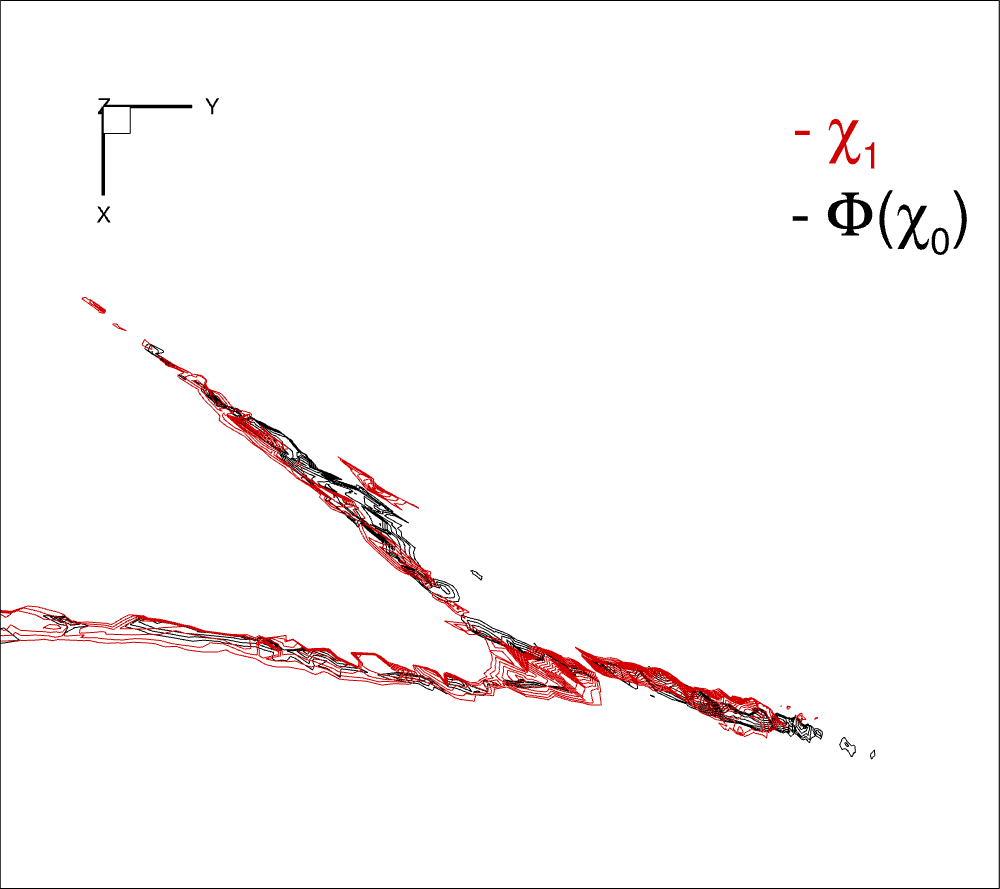}} \\
    \subfigure[$x_3=1.7$]{\label{fig:M6 reg 0 cut17}\includegraphics[width=0.3\textwidth]{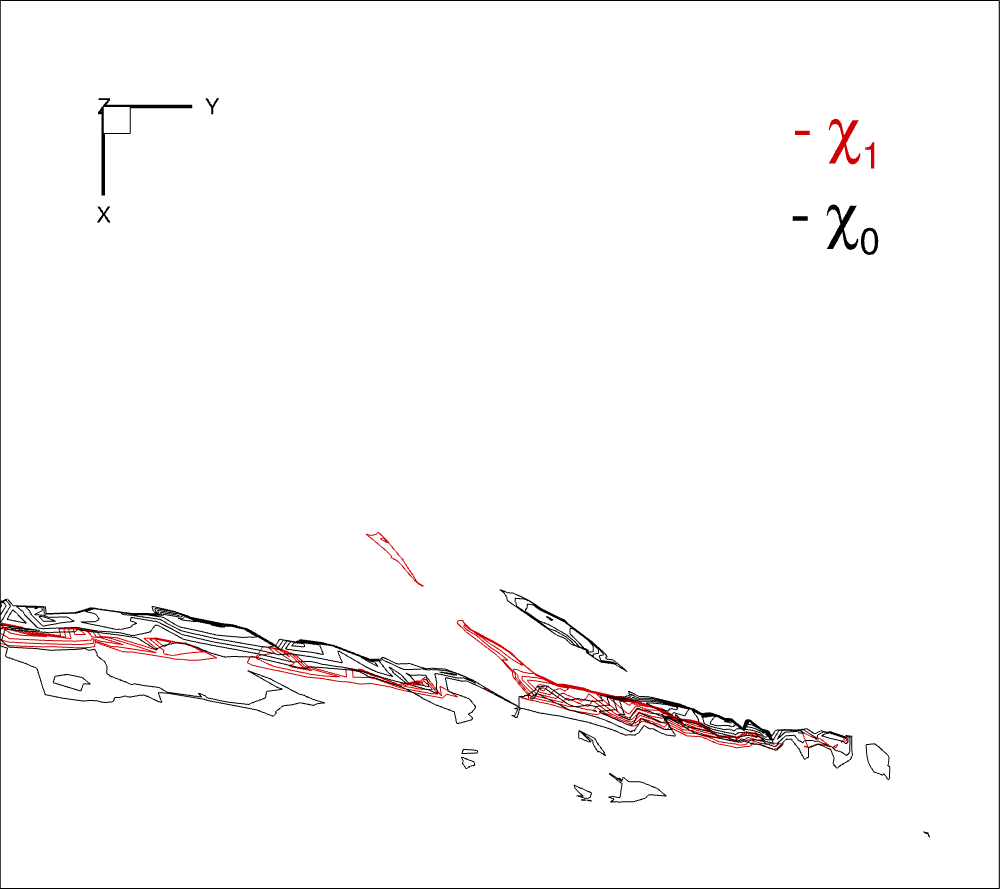}}
    \subfigure[]{\label{fig:M6 map shock p1}\includegraphics[width=0.3\textwidth]{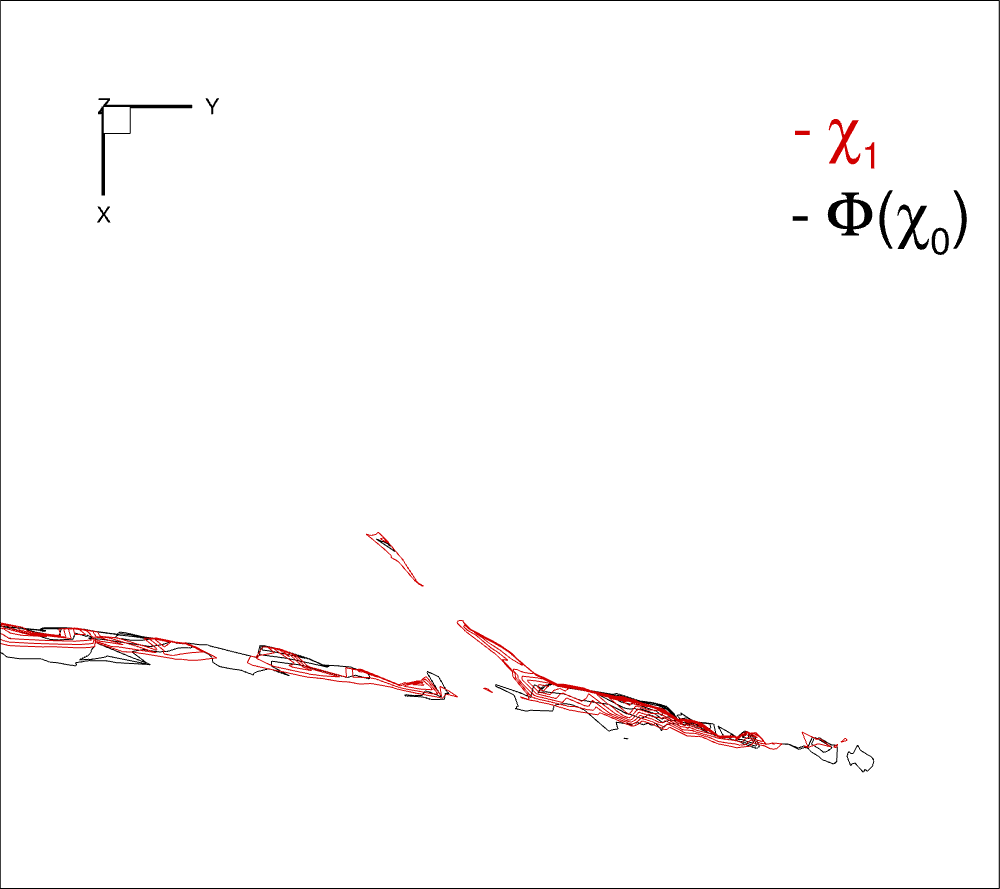}}
    \subfigure[]{\label{fig:M6 map shock p2}\includegraphics[width=0.3\textwidth]{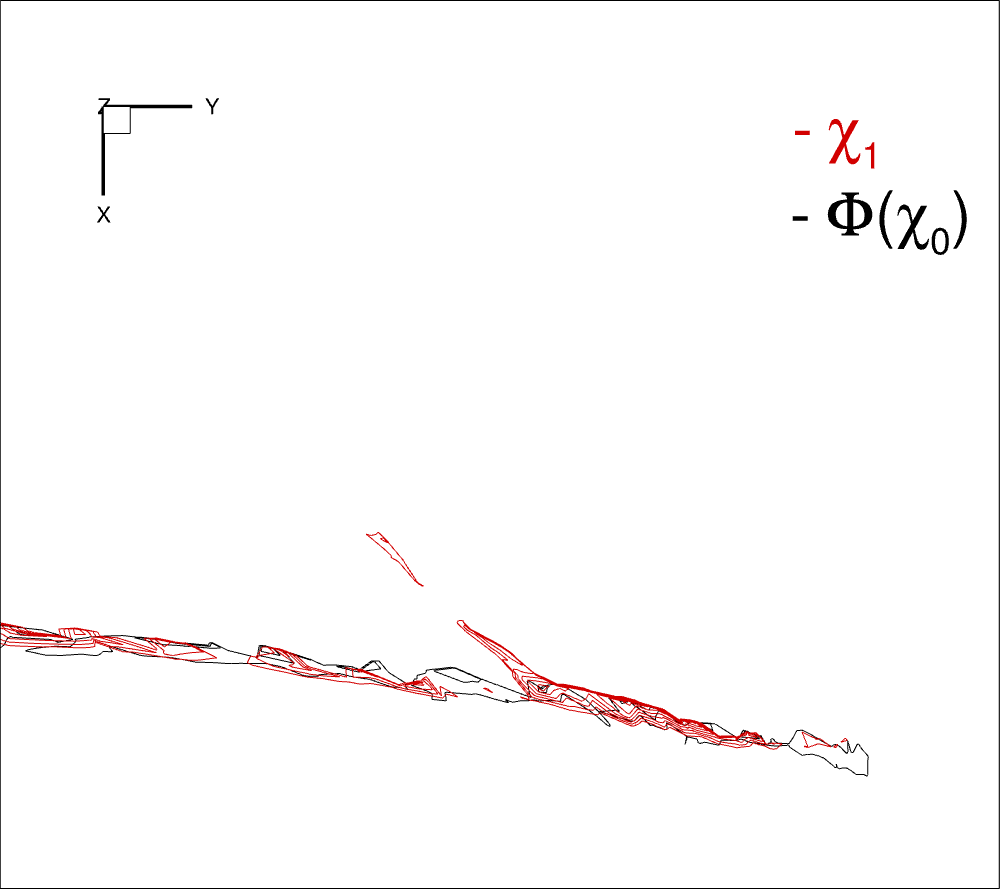}}
    \caption{viscous flow past an ONERA M6 wing;    
    vertical cuts of the indicator function 
$\chi_i(x) := \mathbbm{1}_{Y_i^{\rm s}}(x)$    associated with 
    \eqref{eq:point_cloud_M6}.
Each row corresponds to a different vertical cut;
the  left column corresponds to 
 the  reference and the target point clouds;
 the middle and the right columns compare the indicator functions associated with the deformed point clouds 
 for two choices of the temporal and the spatial discretizations.
}
\label{fig:M6 reg results}
\end{figure}

Figures 
\ref{fig:M6 reg results} illustrate the effectiveness of the
registration procedure to properly deform the point clouds.
Each row corresponds to a different vertical cut ($x_3=0.3$, $x_3=0.7$ and $x_3=1.7$)
 of the indicator function 
$\chi_i(x) := \mathbbm{1}_{Y_i^{\rm s}}(x)$    associated with 
    \eqref{eq:point_cloud_M6};
the  left column shows the indicator functions of 
 the  reference and the target point clouds on each cut;
 the middle and the right columns compare the indicator functions associated with the deformed point clouds 
 for two choices of the temporal and spatial discretization ($p=1$, $\Delta t = 0.1$ and $p=2$, $\Delta t = 0.05$).
We observe that our procedure is effective to 
deform the point clouds and is insensitive to the choice of the discretization.

\begin{figure}[hbt!]
\centering     
\subfigure[$\Delta t=0.2$ , $p=1$]{\label{fig:M6 R1 p1}\includegraphics[width=0.3\textwidth]{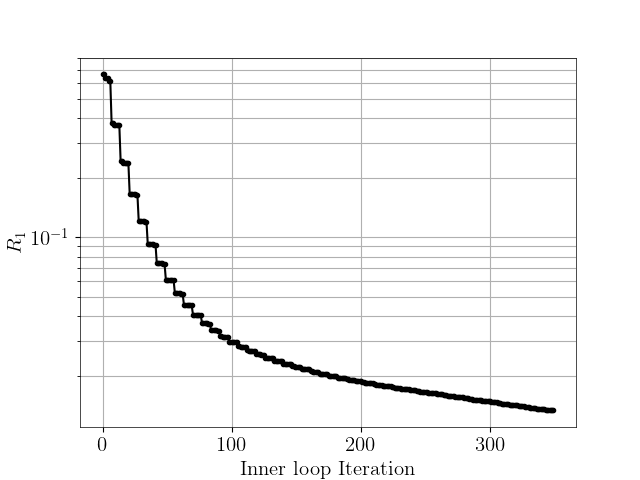}}
    \subfigure[ ]{\label{fig:M6 R2 p1}\includegraphics[width=0.3\textwidth]{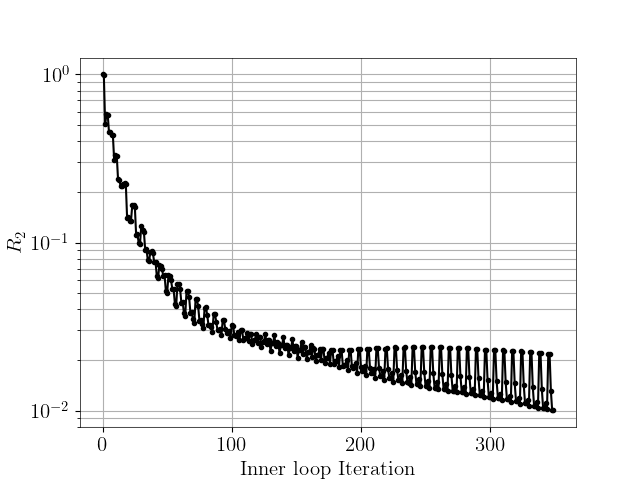}}
    \subfigure[]{\label{fig:M6 R3 p1}\includegraphics[width=0.3\textwidth]{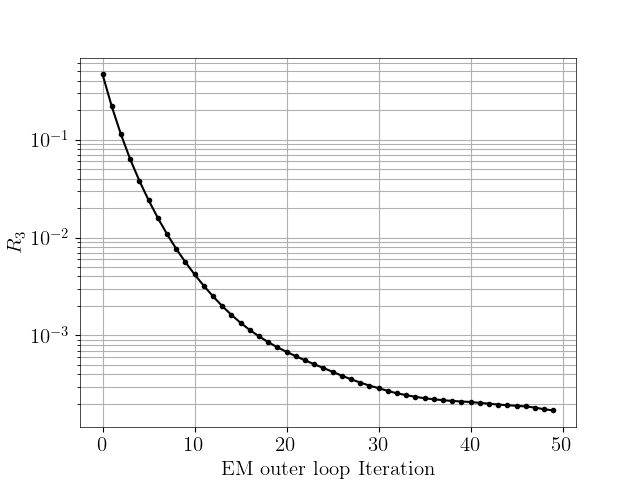}}
    
    \subfigure[$\Delta t=0.05$, $p=2$]{\label{fig:M6 R1 p2}\includegraphics[width=0.3\textwidth]{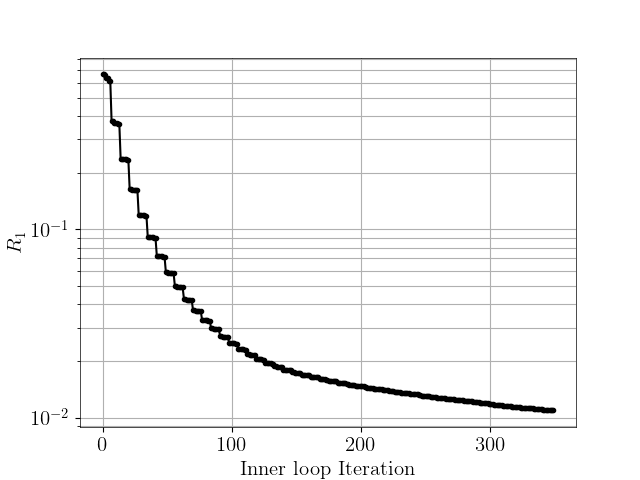}}
    \subfigure[]{\label{fig:M6 R2 p2}\includegraphics[width=0.3\textwidth]{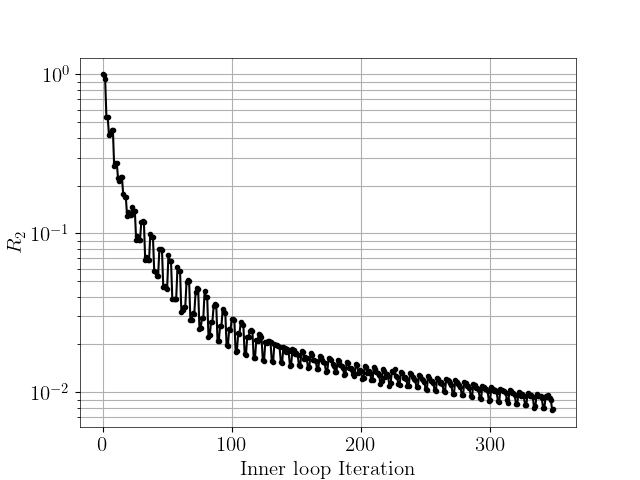}}
    \subfigure[]{\label{fig:M6 R3 p2}\includegraphics[width=0.3\textwidth]{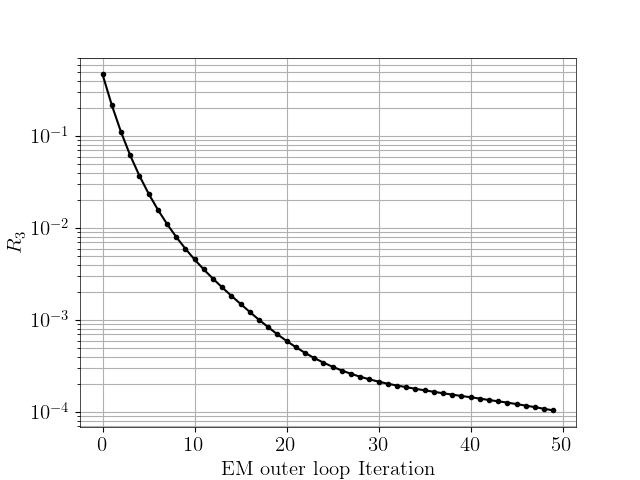}}    
    
    \caption{viscous flow past an ONERA M6 wing; evolution of the residuals 
$R_1,R_2,R_3$ in 
\eqref{eq:termination_condition} with respect to the EM iteration count, for two  different discretizations.
(a)-(b)-(c) $\Delta t=0.2$, linear elements ($p=1)$.
(d)-(e)-(f) $\Delta t=0.05$, quadratic elements ($p=2)$.}
    \label{fig:M6_reg_all}
\end{figure}

Figure \ref{fig:M6_reg_all}
shows the evolution of the residuals 
$R_1,R_2,R_3$ in 
\eqref{eq:termination_condition} with respect to the EM iteration count, for two  different discretizations.
We remark that we plot the residuals $R_1$ and $R_2$  against the inner loop iterations, which include the iterations of the M-step. 
We notice a modest improvement in the alignment performance (cf. $R_1$) as we increase the  polynomial order.
Interestingly, we note that the line search is more stable with the finer discretization: this 
can be explained by observing that the 
effectiveness of the line search procedure is linked to the accuracy of the gradient estimate, which increases as we 
 reduce the time step.

\subsubsection{Results (II): flow prediction and data augmentation}
\begin{figure}[hbt!]
    \centering     
    \subfigure[$AoA=3.06^\circ$]{\label{fig:M6 t00}\includegraphics[width=0.18\textwidth]{Images/TC3/M6_rho0_3D.png}}
    \subfigure[$AoA=3.59^\circ$]{\label{fig:M6 cdi t025}\includegraphics[width=0.18\textwidth]{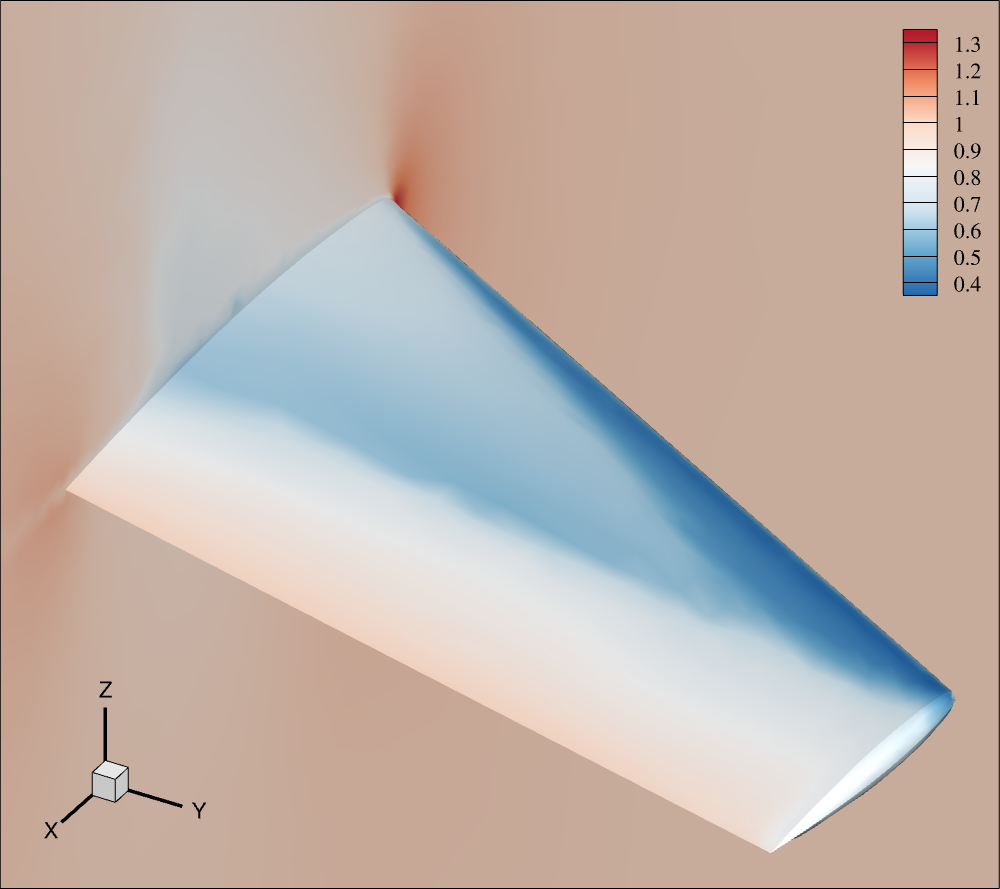}}
    \subfigure[$AoA=4.08^\circ$]{\label{fig:M6 cdi t05}\includegraphics[width=0.18\textwidth]{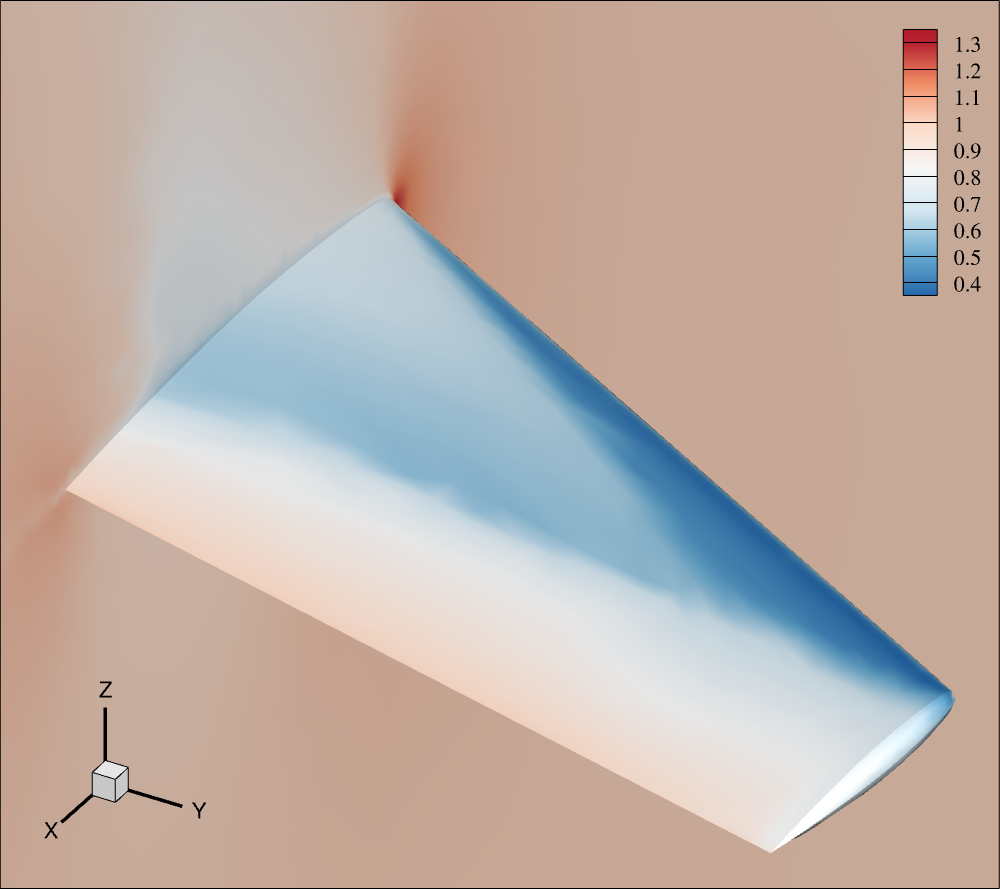}}
    \subfigure[$AoA=4.59^\circ$]{\label{fig:M6 cdi t075}\includegraphics[width=0.18\textwidth]{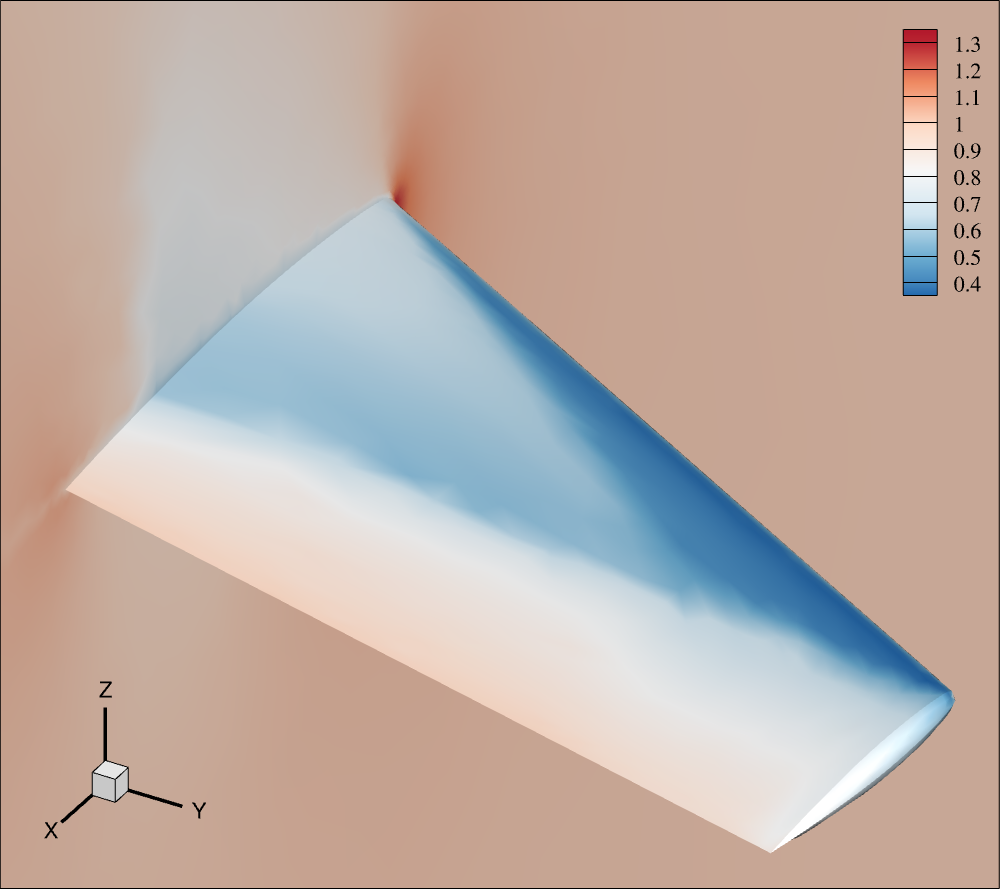}}
    \subfigure[$AoA=5.1^\circ$]{\label{fig:M6 t1}\includegraphics[width=0.18\textwidth]{Images/TC3/M6_rho1_3D.png}}
    \caption{viscous flow past an ONERA M6 wing;  
    CDI density prediction $\tilde \rho_{CDI}(.,\mu)$ over range $AoA \in [3.06, 5.10]$.}
    \label{fig:M6 cdi predictions}
\end{figure}

\begin{figure}[hbt!]
    \centering     
    \subfigure[$AoA=3.06^\circ$]{\includegraphics[width=0.18\textwidth]{Images/TC3/M6_rho0_3D.png}}
    \subfigure[$AoA=3.59^\circ$]{\label{fig:M6 ci t025}\includegraphics[width=0.18\textwidth]{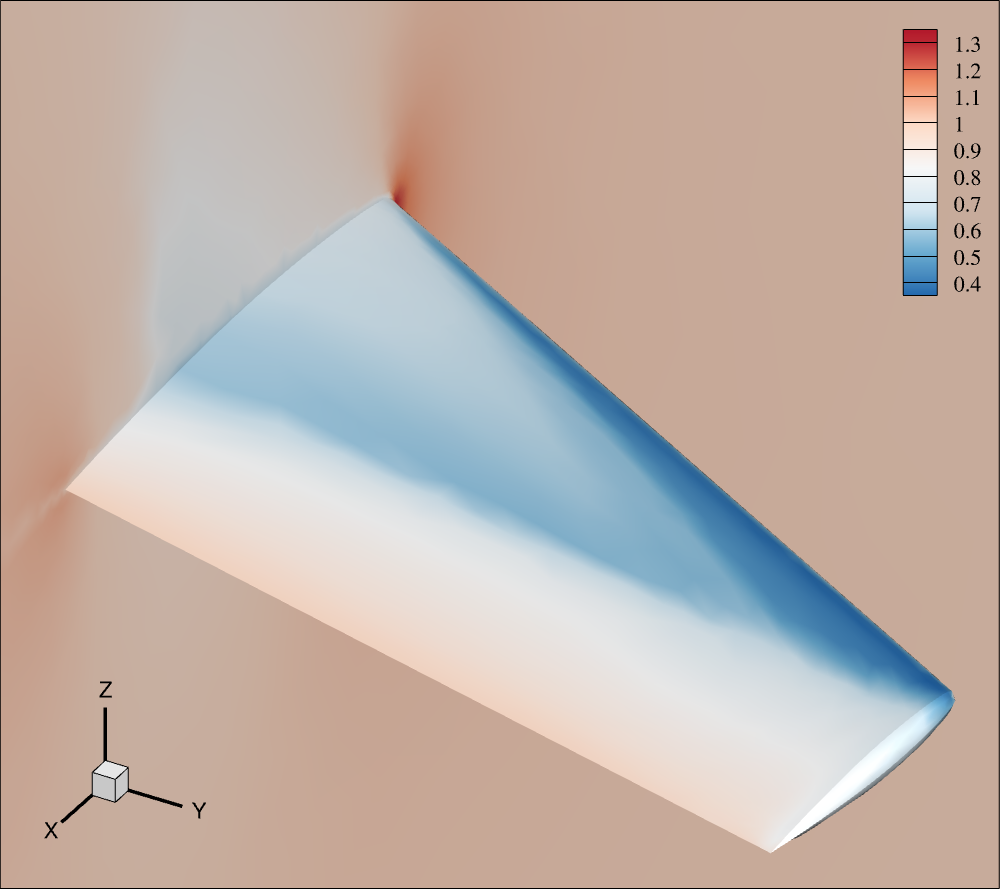}}
    \subfigure[$AoA=4.08^\circ$]{\label{fig:M6 ci t05}\includegraphics[width=0.18\textwidth]{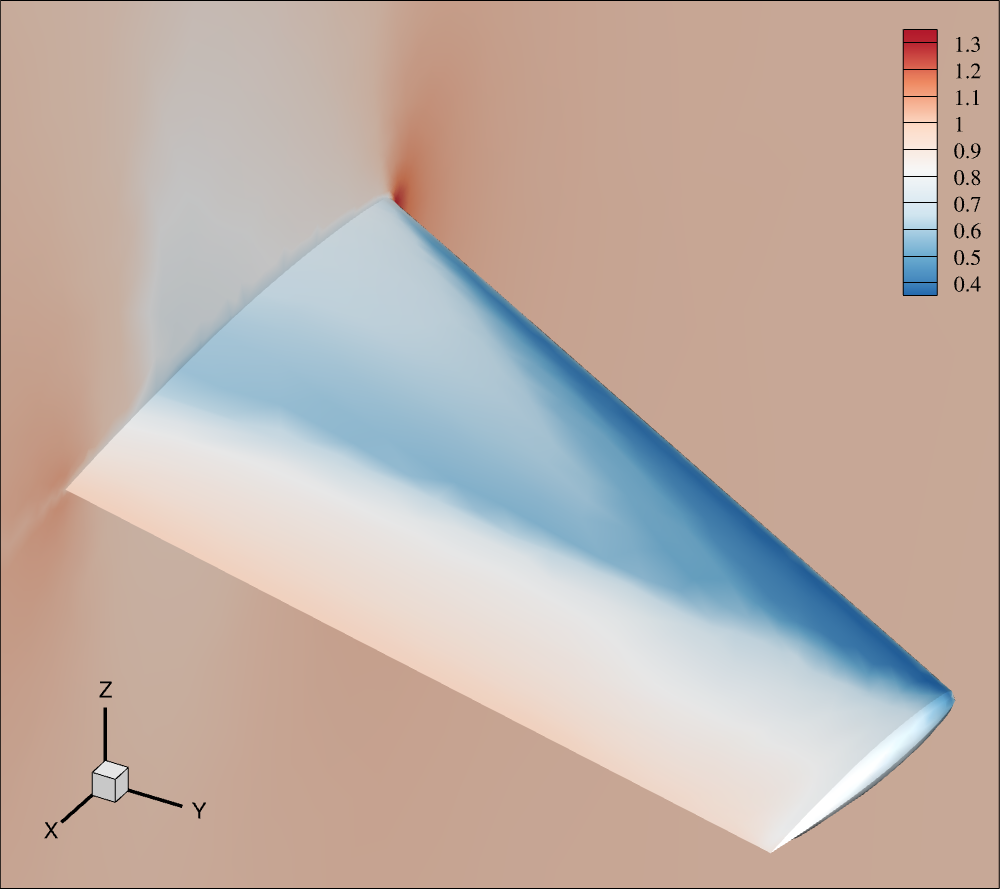}}
    \subfigure[$AoA=4.59^\circ$]{\label{fig:M6 ci t075}\includegraphics[width=0.18\textwidth]{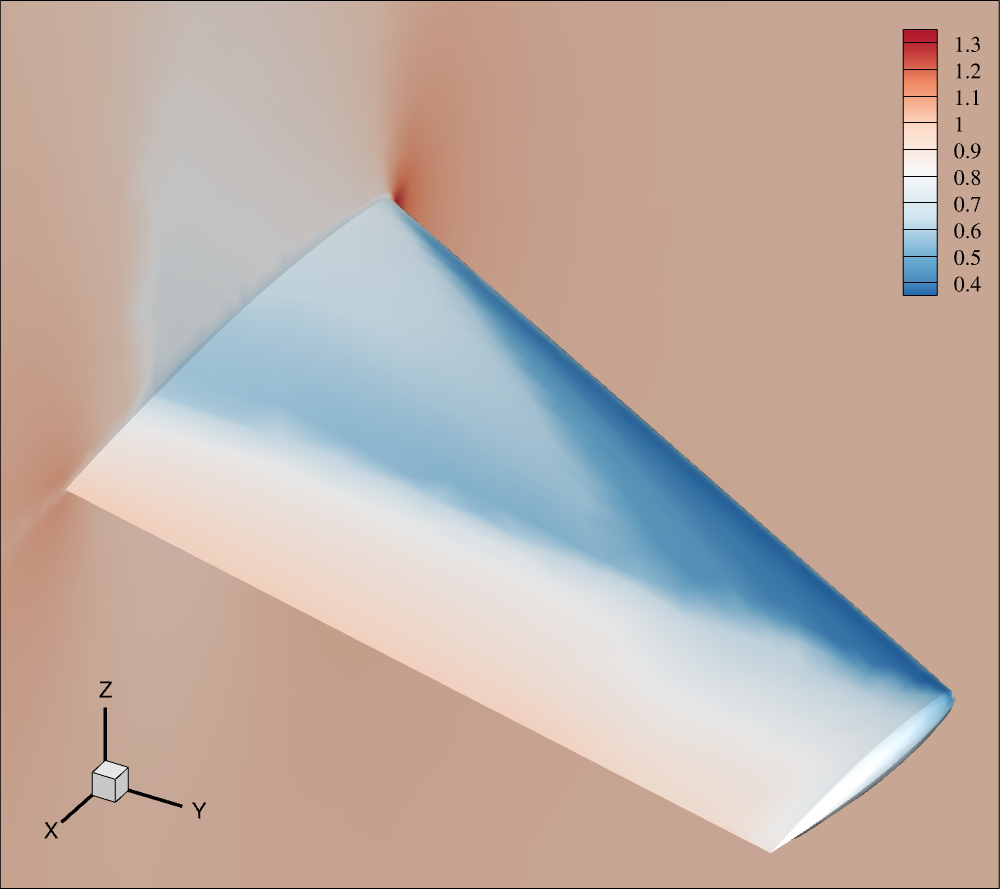}}
    \subfigure[$AoA=5.1^\circ$]{\includegraphics[width=0.18\textwidth]{Images/TC3/M6_rho1_3D.png}}
    \caption{viscous flow past an ONERA M6 wing; CI density prediction $\tilde \rho_{CI}(.,\mu)$ on range $AoA \in [3.06, 5.10]$.}
    \label{fig:M6 ci predictions}
\end{figure}

\begin{figure}[hbt!]
    \centering     
    \subfigure[CDI]{\label{fig:M6 cdi cut}\includegraphics[width=0.3\textwidth]{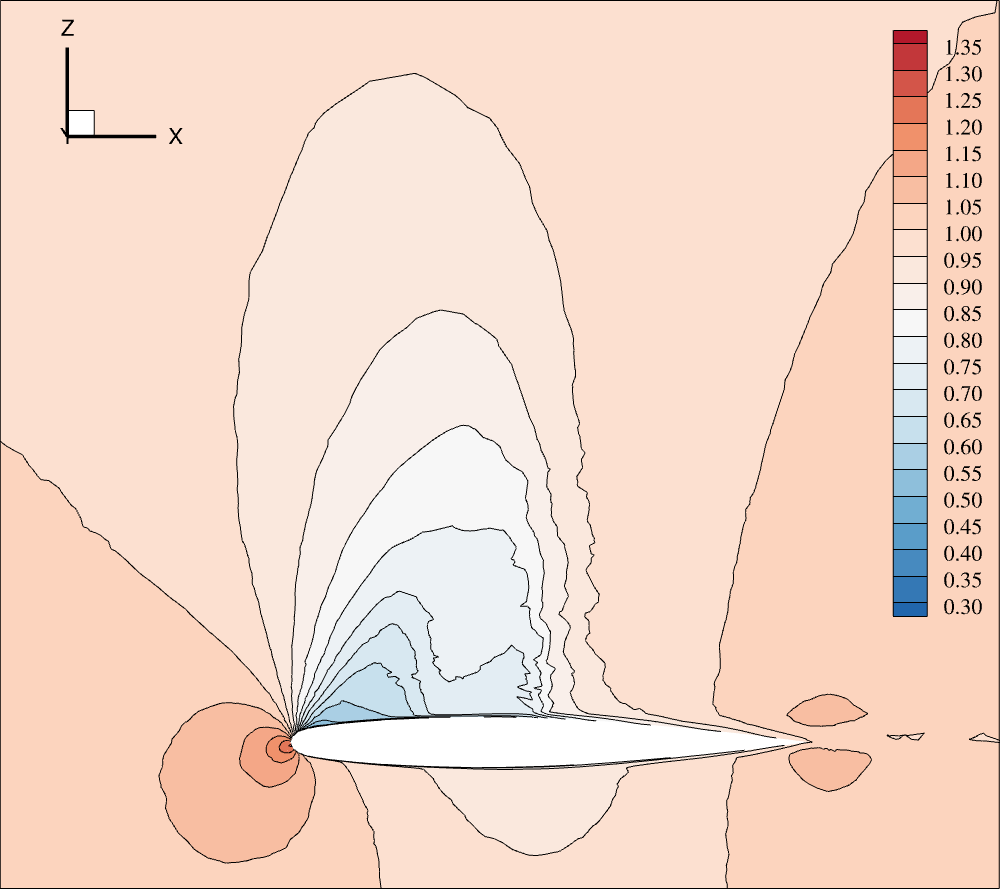}}
    \subfigure[FOM]{\label{fig:M6 fom cut}\includegraphics[width=0.3\textwidth]{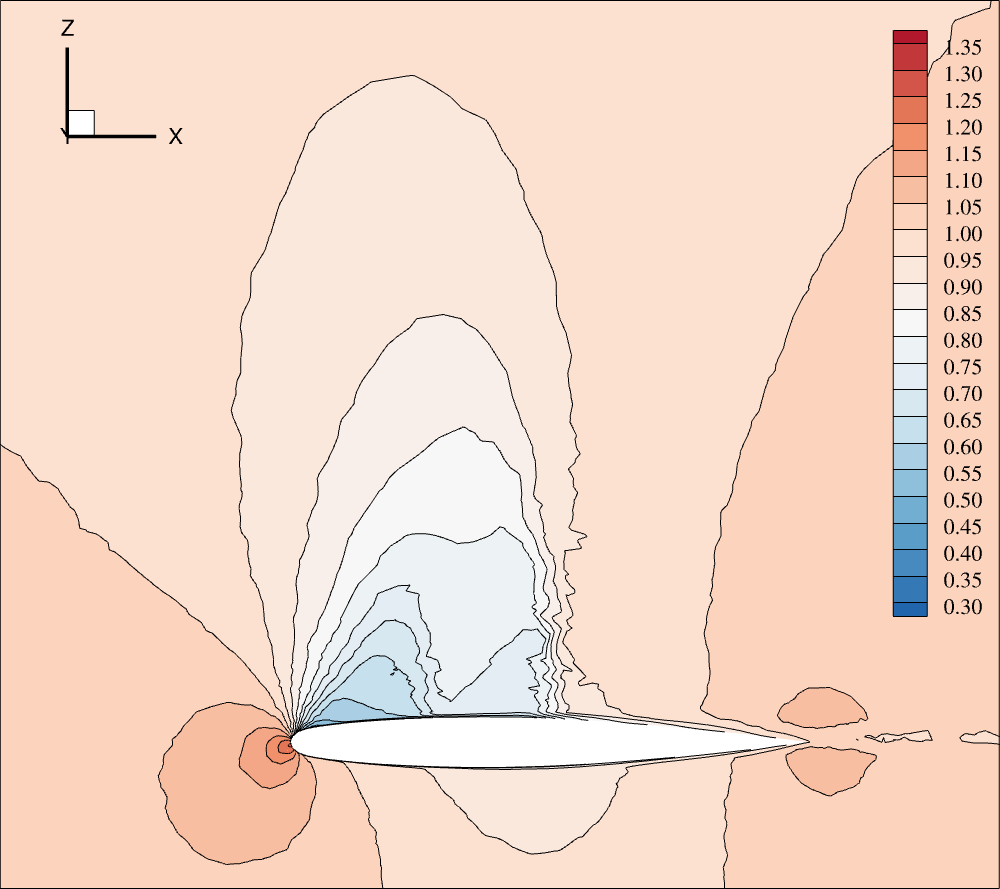}}
    \subfigure[CI]{\label{fig:M6 ci cut}\includegraphics[width=0.3\textwidth]{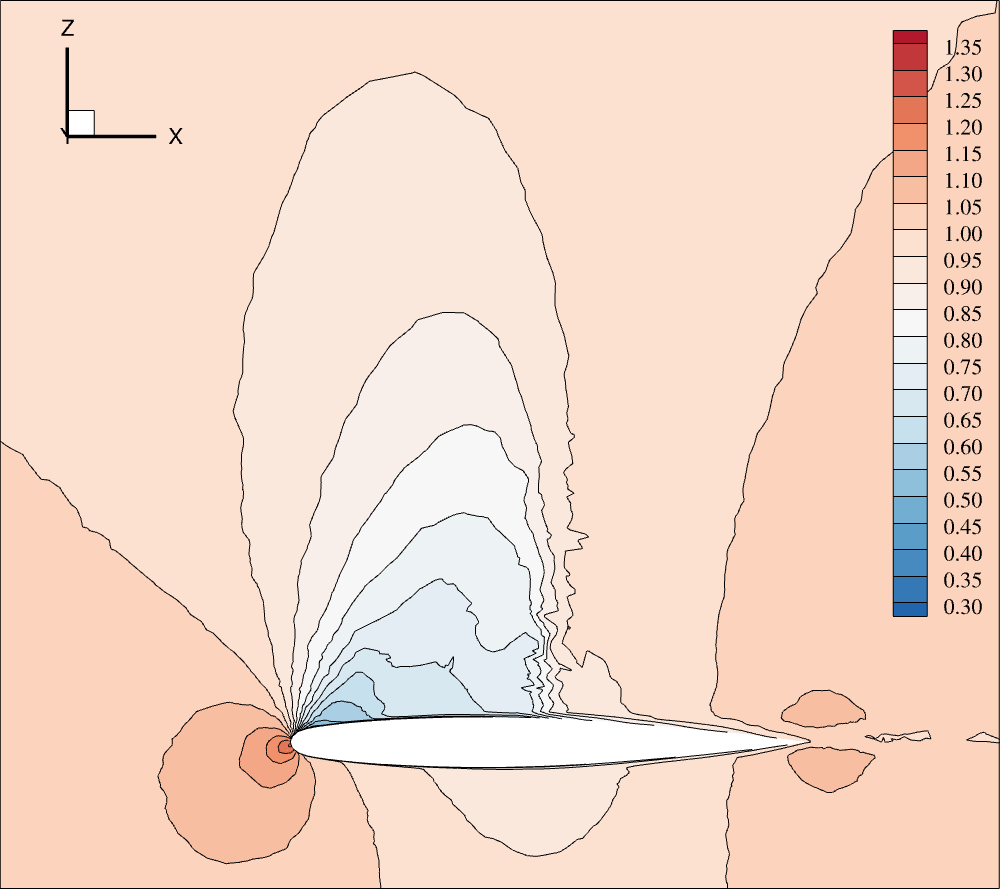}}
    \caption{viscous flow past an ONERA M6 wing; cut of the full field at $Y=8.8$ as predicted by CDI, FOM and CI.}
    \label{fig:M6 cut predictions}
\end{figure}

We exploit the solution to the registration algorithm
 to perform the nonlinear interpolation 
\eqref{eq:flowsCDI} 
  of the flow state.
  We here consider the results obtained for the coarser discretization (
  $p=1$ and $\Delta t=0.2$).
Figure \ref{fig:M6 cdi predictions} shows the   CDI predictions of the density field, while 
  Figure \ref{fig:M6 ci predictions} shows the corresponding CI predictions:
  we observe that CDI coherently transports the shock over the wing, while  CI only mixes the two snapshots.
This observation is  confirmed by the results of  Figure \ref{fig:M6 cut predictions}, which depicts 
a volume cut at  $x_2=8.8$: the CDI estimate features two distinct peaks of the density profile that are consistent with the   FOM solution.

We investigate the effectiveness of CDI and CI to initialize the DG solver for the RANS equations for three choices of the parameter value.
Towards this end, we compare the convergence history of the density residual for three out-of-sample parameters associated with three different initialization of the DG solver:
(i) CDI, (ii) $p=0$ DG solution, (iii) CI.
We observe that CDI enables a $2-3 \times $ reduction over CI of the total number of iterations that are required to meet the required tolerance. We justify this fact by observing that CDI is effective to convect the  fronts where the solution is discontinuous:
since Newton's method  is significantly more effective    at adjusting amplitudes than at at moving discontinuities, the ability of CDI to adequately predict the shock location greatly reduces the required number of iterations to reach convergence and contributes to robustify the nonlinear solver.

\begin{figure}[hbt!]
    \centering     
    \subfigure[$AoA = 3.57 \ (s=0.25)$]{\includegraphics[width=0.3\textwidth]{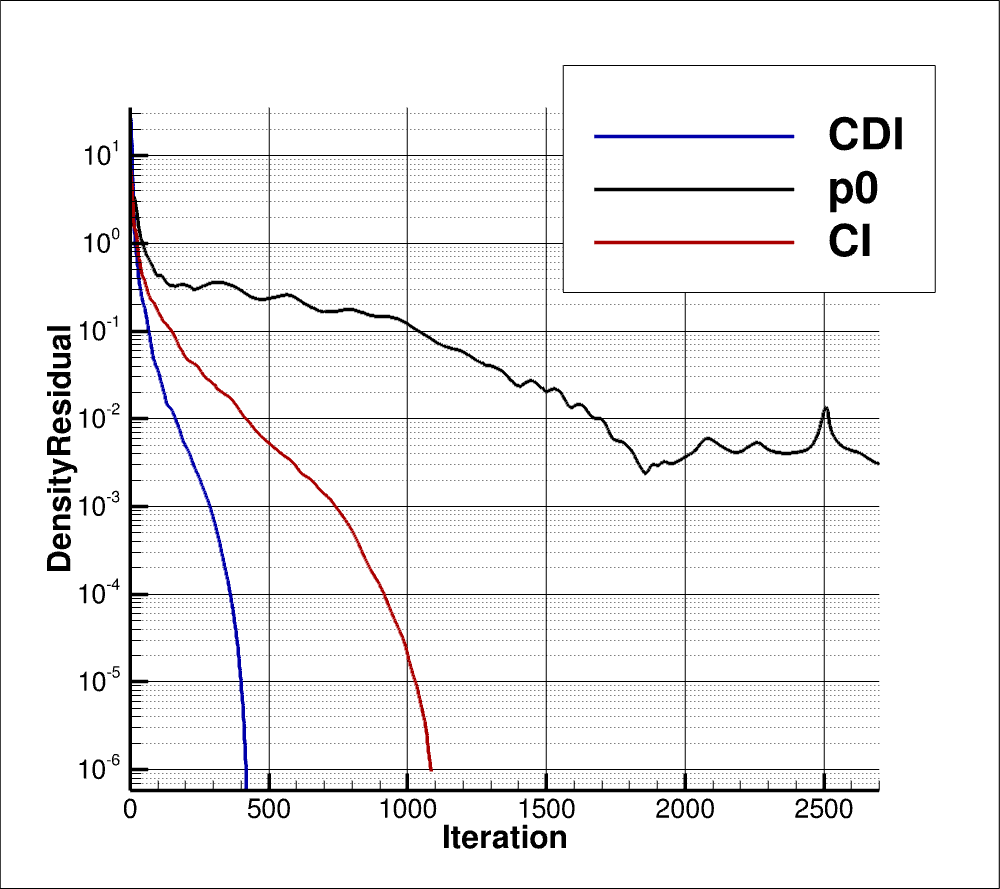}}
    \subfigure[$AoA = 4.08 \ (s=0.5)$]{\label{fig:res convergence 0.5}\includegraphics[width=0.3\textwidth]{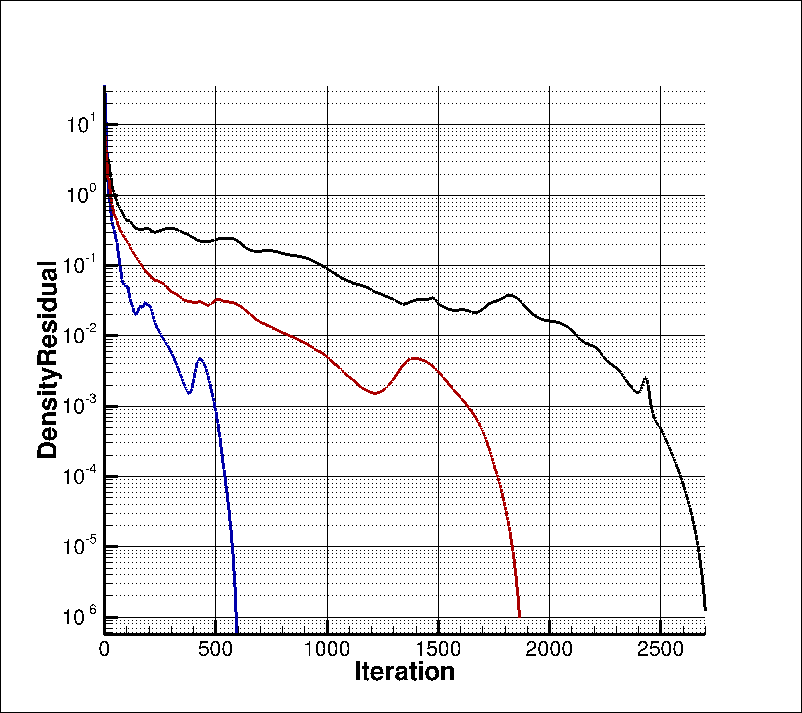}}
    \subfigure[$AoA = 4.59 \ (s=0.75)$]{\label{fig:res convergence 0.75}\includegraphics[width=0.3\textwidth]{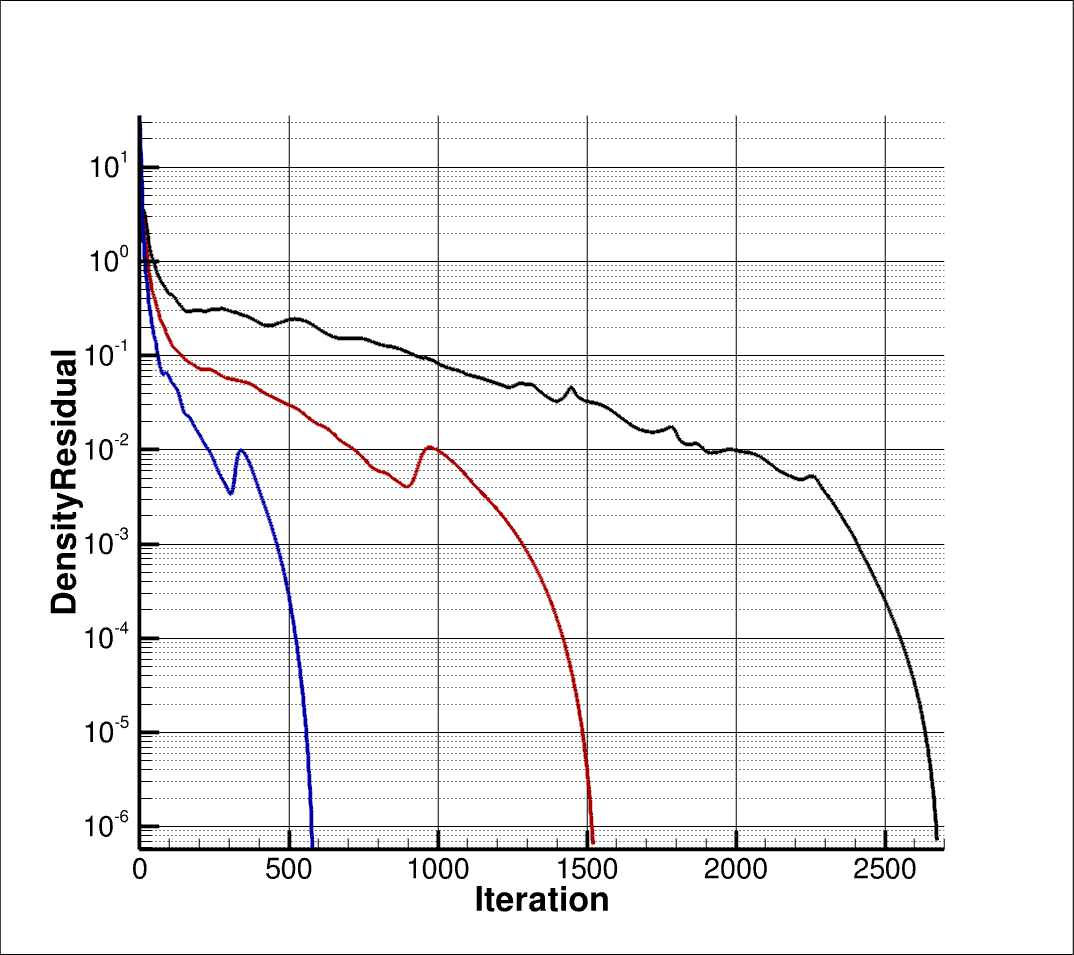}}
       \caption{viscous flow past an ONERA M6 wing;  convergence history of the  density residuals for  different initializations of the  DG FOM.}
        \label{fig:res_convergence}
\end{figure}

\begin{figure}[hbt!]
    \centering     
\includegraphics[width=0.5\textwidth]{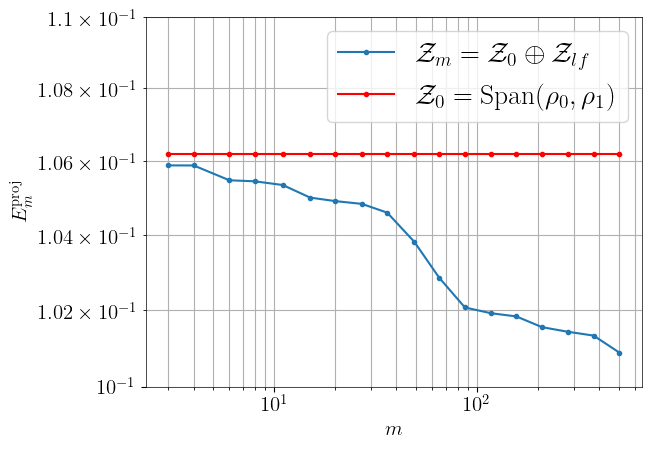}
    \caption{viscous flow past an ONERA M6 wing;  data augmentation based on CDI.}
    \label{fig:data_augmentation_3D}
\end{figure}

Figure \ref{fig:data_augmentation_3D} shows the performance of CDI to generate meaningful artificial snapshots. 
We apply CDI to the density profile to generate $500$ artificial snapshots; next, we define the ($m-2$)-dimensional POD space 
$\mathcal{Z}_{\rm lf}$
associated with the projected snapshots
$\{ \widehat{\rho}_{\rm CDI}( s_i ) - \Pi_{\mathcal{Z}_0} \widehat{\rho}_{\rm CDI}( s_i ) \}_{i=1}^{500} $ where $\mathcal{Z}_0 = {\rm span}\{ \rho_0, \rho_1\}$; finally, we show the projection error \eqref{eq:L2 projective error dataAug} associated with the space $\mathcal{Z}_m = \mathcal{Z}_0 \oplus \mathcal{Z}_{\rm lf}$ and we compare it with the error of $ \mathcal{Z}_0$.
We observe that the addition of synthetic snapshots yields only a modest improvement in the POD accuracy; we conjecture that this limited gain is due to the choice of sensors, which may not be sufficiently informative for effective CDI interpolation and lack of grid refinement in the region of the shocks.

\section{Conclusions}
\label{sec:conclusions}

We have presented a registration procedure for parametric model order reduction in two and three-dimensional bounded domains. Our method combines vector flows with an expectation-maximization algorithm and  can be interpreted as a   CPD approach adapted to bounded geometries.
Even if the procedure is general (i.e., independent of the underlying PDE model), in this work we specifically target  nonlinear interpolation of aerodynamic fields.

We focused on the deployment of  CDI to handle parameter-dependent discontinuities. The numerical results for transonic inviscid and viscous flows demonstrated that CDI effectively reconstructs flow features such as shocks, and significantly outperforms standard linear interpolation techniques.
However, the results for the ONERA M6 wing illustrate that the accuracy of CDI is highly-dependent on the appropriateness of the sensor: this observation motivates further research on the development of effective sensors for shock-dominated flows of aeronautic interest.

Several additional perspectives arise from this work. 
First, we plan to rigorously incorporate the tools developed in this work into the framework of 
projection-based model reduction: in this respect, we plan to extend the strategies of 
\cite{bernard2018reduced} and \cite{barral2024registration} to a broader class of parametric problems.
Second, we plan to investigate the use of second-order methods for registration 
\cite{mang2016constrained} 
 to improve the computational efficiency of the optimization procedure.

\bibliographystyle{abbrv} 
\bibliography{all_references}

\end{document}